\documentclass[twocolumn]{aastex701}
\usepackage{CJK}
\usepackage{rotating}
\usepackage{booktabs}
\usepackage{amsmath}

\usepackage{booktabs}

\shortauthors{C. E. Williams et al.}

\begin{document}
\begin{CJK*}{UTF8}{bsmi}

\title{Wandering intermediate-mass black holes and off-nuclear tidal disruption events from early star clusters}

\author[orcid=0000-0003-2369-2911,sname='Williams']{Claire E. Williams}
\affiliation{Department of Physics and Astronomy, UCLA, Los Angeles, CA 90095}
\affiliation{Mani L. Bhaumik Institute for Theoretical Physics, Department of Physics and Astronomy, UCLA, Los Angeles, CA 90095, USA}
\email[show]{clairewilliams@astro.ucla.edu}  

\author[orcid=0000-0002-9802-9279,sname='Naoz']{Smadar Naoz}
\affiliation{Department of Physics and Astronomy, UCLA, Los Angeles, CA 90095}
\affiliation{Mani L. Bhaumik Institute for Theoretical Physics, Department of Physics and Astronomy, UCLA, Los Angeles, CA 90095, USA}
\email[]{snaoz@astro.ucla.edu}  

\author[orcid=0000-0002-8828-8461,sname='Zhou']{Yihao Zhou 
(周亦豪)}
\affiliation{McWilliams Center for Cosmology, Department of Physics, Carnegie Mellon University, Pittsburgh, PA 15213, USA}
\email[]{yihaoz@andrew.cmu.edu}

\author[orcid=0000-0002-0647-718X, sname='Akiba-Lewis']{Tatsuya Akiba-Lewis}
\affiliation{Department of Physics and Astronomy, UCLA, Los Angeles, CA 90095}
\email[]{astroakiba@gmail.com}

\author[0000-0001-8593-7692]{Mark Vogelsberger}
\affil{Department of Physics and Kavli Institute for Astrophysics and Space Research, Massachusetts Institute of Technology, Cambridge, MA 02139, USA}
\affil{Fachbereich Physik, Philipps Universit\"at Marburg, D-35032 Marburg, Germany}
\email[]{mvogelsb@mit.edu} 

\author[0000-0001-5817-5944]{Blakesley Burkhart}
\affiliation{Department of Physics and Astronomy, Rutgers, The State University of New Jersey, 136 Frelinghuysen Rd, Piscataway, NJ 08854, USA \\}
\affiliation{Center for Computational Astrophysics, Flatiron Institute, 162 Fifth Avenue, New York, NY 10010, USA \\}
\email[]{bburkhart@flatironinstitute.org}

\author[0000-0001-7925-238X]{Naoki Yoshida}
\affiliation{Department of Physics, The University of Tokyo, 7-3-1 Hongo, Bunkyo, Tokyo 113-0033, Japan}
\affiliation{Kavli Institute for the Physics and Mathematics of the Universe (WPI), UT Institute for Advanced Study, The University of Tokyo, Kashiwa, Chiba 277-8583, Japan}
\affiliation{Research Center for the Early Universe, School of Science, The University of Tokyo, 7-3-1 Hongo, Bunkyo, Tokyo 113-0033, Japan}
\email[]{naoki.yoshida@ipmu.jp}

\author[0000-0002-8859-7790]{Avi Chen}
\affiliation{Department of Physics and Astronomy, Rutgers, The State University of New Jersey, 136 Frelinghuysen Rd, Piscataway, NJ 08854, USA \\}
\email[]{avi.chen832@gmail.com} 

\author[0000-0002-4086-3180]{Kyle Kremer}
\affil{Department of Astronomy \& Astrophysics, University of California, San Diego; La Jolla, CA 92093, USA}
\email[]{clairewilliams@astro.ucla.edu}

\author[0000-0002-4227-7919]{William Lake}
\affil{Department of Physics and Astronomy, Dartmouth College, Hanover, NH 03755, USA \\}
\email[]{William.Lake@dartmouth.edu} 

\author[0000-0003-3816-7028]{Federico Marinacci}
\affiliation{Dipartimento di Fisica e Astronomia “Augusto Righi”, Università di Bologna, via Piero Gobetti 93/2, I-40129 Bologna, Italy\\}
\affiliation{INAF, Osservatorio di Astrofisica e Scienza dello Spazio di Bologna, via Piero Gobetti 93/3, I-40129 Bologna, Italy\\}
\email[]{federico.marinacci2@unibo.it}

\author[0000-0001-5944-291X]{Shyam H. Menon}
\affiliation{Center for Computational Astrophysics, Flatiron Institute, 162 Fifth Avenue, New York, NY 10010, USA \\}
\affiliation{Department of Physics and Astronomy, Rutgers, The State University of New Jersey, 136 Frelinghuysen Rd, Piscataway, NJ 08854, USA \\}
\email[]{smenon@flatironinstitute.org}

\begin{abstract}
Off-nuclear tidal disruption events (TDEs) reveal evidence for massive black holes outside galactic centers, but the origins of this wandering population remain uncertain. 
We investigate wandering intermediate-mass black holes (IMBHs) formed through runaway stellar collisions in dense, high-redshift star clusters and predict their TDE signatures across cosmic time. Our semi-analytic model samples clusters associated with low-mass subhalos at $7<z<20$ and uses merger trees from the IllustrisTNG TNG50 simulation to follow the assembly of 15 elliptical galaxies with stellar masses of $10^9-10^{11}~M_\odot$. 
We compare two models for high-redshift cluster formation: a maximal model taken from a high-resolution simulation and a conservative model calibrated from local cluster relations. 
Off-nuclear TDEs produced through this mechanism occur in bursts within 10-100 Myr of IMBH formation as the loss-cone surrounding the black hole is depleted.
We estimate the rate of off-nuclear TDEs expected across time per galaxy, and characterize these events as typically occurring between $10-10^3$ kpc from their host galaxy centers. 
Event rates increase with galaxy mass; in the maximal model, the progenitors of the most massive galaxies reach peak off-nuclear TDE rates of approximately $5\times10^{-4}-4\times10^{-2}~\mathrm{yr}^{-1}$ per galaxy at $7\lesssim z\lesssim13$. 
This work provides a framework for constraining the early population of star clusters that form IMBHs through runaway collapse via their contribution to the rate of off-nuclear TDEs in the high redshift Universe.
\end{abstract}

\keywords{\uat{Intermediate-mass black holes}{816} ---\uat{Star clusters}{1567} --\uat{Tidal disruption}{1696} --- \uat{Galaxy Formation}{595} --- \uat{Elliptical Galaxies}{456}}


\section{Introduction} 


Off-nuclear tidal disruption events (TDEs) offer a direct way to uncover massive black holes outside galactic centers, where weak persistent accretion can make them difficult to detect \citep{ricarte_unveiling_2021,lin_luminous_2018,lin_multiwavelength_2020,jin_intermediate-mass_2025,yao_massive_2025,stein_tde_2026}. The four candidates analyzed by \citet{guolo_demographics_2026} are associated with massive, early-type parent galaxies with stellar masses $M_\star\sim10^{11}$~M$_\odot$. 
These events bring a long-standing theoretical prediction into observational reach: hierarchical galaxy assembly can leave massive black holes outside galactic nuclei when the orbital decay of black holes accreted during mergers is inefficient \citep[e.g.,][]{governato_fate_1994,volonteri_assembly_2003,islam_massive_2004,bellovary_wandering_2010}. Furthermore, off-nuclear TDEs may provide the opportunity to probe intermediate-mass black holes formed in early star clusters, provided their formation and subsequent stellar dynamics translate into off-nuclear TDE activity as galaxies assemble.

Dense star clusters in the early Universe provide a promising environment for forming intermediate-mass black holes (IMBHs), here defined as black holes with masses $10^2$--$10^5$~M$_\odot$. Observations at $z\gtrsim8$ reveal compact, young stellar systems with masses and sizes comparable to present-day globular clusters, or with even higher stellar densities \citep{Adamo+24,Mowla+24}. In sufficiently dense clusters, runaway stellar collisions can build a very massive star that subsequently collapses into an IMBH \citep[e.g.,][]{zwart_runaway_2002,glebbeek_evolution_2009,Sakurai17,gonzalez_prieto_intermediate-mass_2024,vergara_rapid_2025,williams_cosmological_2026}. The surrounding cluster also supplies stars that may be tidally disrupted after the black hole forms. The TDE activity of this population therefore depends on both IMBH production and the subsequent evolution of the stellar orbits that populate their loss cones.

In the local universe, wandering black holes are expected to accrete extremely weakly \citep{ricarte_unveiling_2021}, and the dynamical detection is a promising method \citep[with some candidates in globular clusters and ultra-compact dwarfs:][]{seth_supermassive_2014,taylor_supermassive_2025, pechetti_detection_2022}, alongside gravitational waves \citep{abbott_gw190521_2020}. 
However, stars in the vicinity of a wandering IMBH may be tidally disrupted, producing a bright flare that reveals the presence of the off-nuclear black hole \citep{lin_luminous_2018, lin_multiwavelength_2020, jin_intermediate-mass_2025, yao_massive_2025,stein_tde_2026, grotova_population_2025, chang_multiwavelength_2025, maksym_tidal_2013, donato_tidal_2014}.
\citet{guolo_demographics_2026} identifies a sample of four TDE candidates that are all offset from their host by 0.9" to 53": 3XMM J215-05 \citep{lin_luminous_2018,lin_multiwavelength_2020}, EP20240222a \citep{jin_intermediate-mass_2025}, AT2024tvd \citep{yao_massive_2025}, and AT2025abcr \citep{stein_tde_2026}. 
These events are all hosted by massive, early-type galaxies with $M_*\sim 10^{11}M_\odot$. 
An additional three luminous, fast-evolving optical transients observed at 4-10 kpc from elliptical host galaxies \citep[e.g.,][]{vinko_luminous_2014,nicholl_at_2023} could originate from IMBH-driven TDEs \citep[e.g.,][]{perley_fast_2019}.

Cosmological simulations that include black hole seeding prescriptions and allow black holes to evolve on independent orbits provide a framework for studying wandering populations during galaxy assembly. Examples include {\tt ROMULUS} \citep{tremmel_romulus_2017} and {\tt ASTRID} \citep{ni_astrid_2022, zhou_astrid_2026}. Studies based on these simulations predict substantial off-nuclear populations: {\tt ROMULUS} finds that wandering black hole abundance increases approximately linearly with halo mass, while {\tt ASTRID} shows that mergers can redistribute black holes away from galactic centers, causing central and wandering occupation fractions to evolve differently \citep{ricarte_origins_2021, ricarte_unveiling_2021,di_matteo_vast_2023,guolo_demographics_2026,weller_black_2026} . Their adopted seed masses, however, do not sample the full IMBH regime, and the stellar collisions and loss-cone dynamics that govern cluster-born IMBH formation and TDE production remain unresolved. 
Furthermore, the halo-based MBH seeding model limits these simulations to one seed per halo, which may cause an underestimation of the wandering IMBH population \citep{zhou_acorn_2026}. 
Quantifying the contribution of this formation channel therefore requires connecting the formation and internal dynamics of early star clusters to their subsequent evolution during hierarchical galaxy assembly. The abundance of wandering IMBHs alone does not determine their TDE rate: the rate also depends on the evolving supply of stars on orbits that lead to disruption.

In this work, we unravel the interplay between hierarchical structure formation and the formation of the Universe's first star clusters, connecting IMBH birth in dense stellar environments to off-nuclear TDE activity across cosmic time. Predicting this activity requires following both the assembly of the wandering population and the evolving supply of stars on orbits that lead to disruption. We therefore use a semi-analytic framework that derives IMBH formation and TDE production from the same modeled cluster populations while following their evolution during galaxy assembly. This approach allows us to determine how early cluster conditions and subsequent loss-cone depletion shape the rates and spatial distributions of TDEs from wandering IMBHs. 

Specifically, we use merger trees from Illustris TNG50-1 \citep{nelson_illustristng_2019} to get the hierarchical growth histories of 15 elliptical galaxies of  $z=0$ stellar mass between $10^9-10^{11}M_\odot$. 
We then seed the initial halos with star cluster populations, testing two possible methods--a ``maximal" model for initial star cluster seeding based on high-resolution cosmological simulations \citep{Williams+25}, and a ``conservative" model for early seeding based on empirical scaling relations in the local Universe. 
We use the analytic model presented in \citet{williams_cosmological_2026}, to follow the formation of IMBHs through the runaway collapse of the star cluster and formation of a Very Massive Star (VMS). 
Each cluster hosting an IMBH uses up the stellar reservoir within 10-100Myr of black hole formation, resulting in a burst of TDEs over that time. 
We predict the rates of off-nuclear TDEs produced by this process. 

The paper is organized as follows: \S~\ref{sec:methods} describes the semi-analytic modeling used to populate the TNG merger tree with star clusters, seed black holes in the clusters, integrate cluster orbits to $z=0$, and predict TDE rates. 
\S~\ref{sec:results} describes the results of this analysis. 
\S~\ref{sec:discussion} discusses the implications and limitations of this study. 
The findings are summarized in \S~\ref{sec:summary}. 

\section{Methods}
\label{sec:methods}
We combine high-resolution, high-redshift cosmological hydrodynamics simulations \citep{Williams+25} with the merger tree history from Illustris TNG50 \citep{nelson_illustristng_2019}. 
Star clusters are seeded into the progenitor halos of the tree using a population sampler either calibrated on a high-resolution simulation or empirical relations, and their orbits are followed through the tree to $z=0$. Finally, the potential formation of an IMBH in each cluster through runaway stellar collisions and the collapse of a very massive star (VMS) is computed with an analytic model.

\subsection{TNG50 Galaxy selection and merger tree}

The preference for off-nuclear TDEs for massive, early-type hosts \citep[e.g.,][]{guolo_demographics_2026} motivates us to consider TDE rates in elliptical galaxies of various masses. 
We select galaxies from the TNG50-1 simulation \citep{pillepich_simulating_2018, nelson_illustristng_2019}, which has a comoving box size of 51.7~Mpc, a dark matter particle mass of $4.5\times10^5\,M_\odot$ and a mean baryonic mass resolution of  $8.5\times10^4\,M_\odot$. 
We identify elliptical galaxies in the $z=0$ subhalo 
catalog (snapshot 99) in three steps. First, we restrict to central galaxies, i.e.\ the 
most massive \textsc{subfind} subhalo of their friends-of-friends (FOF) group 
\citep{springel_populating_2001}. 
We sample the total bound stellar mass of these galaxies as identified by \textsc{subfind} in three mass bins: 
$10^9 M_\odot \leq M_* \leq 10^{10} M_\odot$, $10^{10} M_\odot \leq M_* \leq 10^{11} M_\odot$, and $M_* \geq 10^{11}\,M_\odot$. 
Next, we require these  galaxies to be quenched, with an instantaneous star formation rate 
$\mathrm{SFR} < 0.1\,M_\odot\,\mathrm{yr}^{-1}$. 
Then, because being quenched does not guarantee a spheroidal  morphology, we classify each candidate by its stellar kinematics. 
We use the star particles within 30~kpc of the subhalo center, measured relative to their  mass-weighted mean velocity, and compute $\kappa_{\rm rot}$, the fraction of 
stellar kinetic energy in ordered rotation about the net stellar angular  momentum axis \citep{Sales+12}. 
We treat systems with $\kappa_{\rm rot} < 0.5$ as 
dispersion-dominated, and we select five candidates with the lowest $\kappa_{\rm rot}$ as our prototypical ellipticals. 
For each star we also compute the orbital circularity $\epsilon = j_z/j_{\rm circ}(E)$ and record the mass fraction with $\epsilon > 0.7$. 
We then find the counterpart dark matter halo of the 
selected galaxy in TNG50-1-Dark, a dark-matter-only run with the same initial conditions, using the bijective subhalo matching catalog provided with the data release \citep{nelson_illustristng_2019}. 
We use the full \textsc{SubLink} merger tree of the matched halo \citep{rodriguez-gomez_merger_2015}. 
Its earliest snapshots are at $z=20$ and $z=15$. 
Because \textsc{subfind} requires at least 20 particles per  subhalo, the tree includes halos down to $M_{\rm DM} \approx 1.1\times10^7\,M_\odot$.

Because the cluster population of each galaxy is seeded into the progenitor branches of its \textsc{SubLink} tree, if any of the assembly history were missing from the tree, the corresponding clusters would be removed from our rate estimates. 
We therefore check the
TNG50-1-Dark tree of every candidate before accepting it, and replace any
candidate that fails with the next most dispersion-dominated galaxy in the same stellar-mass bin.
We require that the matched dark-matter subhalo must be the central of its FoF group at $z=0$.
Second, the tree must account for the mass growth of its main
branch. At each main-branch step we compare the mass of the descendant with
the total mass of all subhalos in the tree at the preceding one or two
snapshots (allowing for \textsc{SubLink} descendant links that skip a
snapshot), and reject the tree if the unaccounted-for mass at any step exceeds
10\% of the $z=0$ subhalo mass.
The indices of all the considered halos are provided in \autoref{tab:ellipticals} in the appendix.

Some trees also contain snapshots at which the main branch has no subhalo,
because \textsc{SubLink} links two main-branch subhalos across an intervening
snapshot. At such snapshots we construct a stand-in main progenitor whose
position and velocity are interpolated linearly in cosmic time between the
neighbouring main-branch subhalos, whose mass is interpolated logarithmically,
and whose $R_{200c}$ follows from that mass.
The stand-in is used only as the
reference for the main-progenitor background potential and for the separations
and $R_{200c}$ in \S\ref{sec:sam} (in other words, we don't assign any clusters to it). 

\subsection{Star cluster initial conditions and distribution}
\label{sec:scseeding}

We test two cases for seeding star clusters in early halos. 
The first is to directly draw the properties of star clusters from a high-resolution cosmological simulation that resolves star clusters with $M_*\gtrsim10^4 M_\odot$. 
This simulation does not include feedback, so it produces a maximal case within $\Lambda$CDM for early star cluster formation. 
For a more conservative estimate, we sample star cluster properties from empirical scaling relations. 
This results in fewer star clusters overall and lower densities. 
However, the choice of initial conditions is more ad-hoc, and the observed scaling relations do not provide constraints for the low mass halos that represent the  majority of branches in the progenitor merger tree. 
At early times $z>7$, we expect that the true rates lie  between these two estimates. 

\subsubsection{High resolution simulation and star cluster catalog selection}
\label{sec:cosmoseeding}
We use a high-resolution cosmological hydrodynamics simulation \citep{Lake+23b, Lake+24a,Williams+24, Williams+25, williams_observing_2026} to explore the population of star clusters associated with the progenitor halos. 
This simulation represents the maximal cluster formation case possible, and is used to set an upper limit on the resulting TDE rate.
However, because simulation includes thousands of clusters and their positions, masses, sizes, and velocities with respect to the host halo are known exactly, this method has the advantage of physically motivated initial conditions. 

The simulation is run with the {\tt AREPO} code, from $z=200$ to $z=12$, with resolution $768^3$, DM particle mass $M_{DM} = 1.1\times10^3 M_\odot$ and gas cell mass $M_{B} = 200 M_\odot$. 
As described in \citet{Lake+23b}, the simulation includes primordial atomic and molecular hydrogen cooling through {\tt GRACKLE}, with 15 primordial species and the rotational and vibrational transitions of molecular hydrogen \citep{smith_grackle_2017, Chiaki+19}, and a star formation prescription similar to SMUGGLE \citep{Marinacci+19}. 
In short, when a gas cell exceeds the Jeans mass, the cell is converted into a star particle on the free-fall time.
The simulation does not include feedback from the stars (radiation, SN, winds, etc) or metal enrichment and cooling. 
The simulation's assumed $\Lambda$CDM cosmology differs slightly from TNG--$\Omega_\Lambda = 0.73, \Omega_M = 0.27, \Omega_B = 0.044, \sigma_8= 1.7 ,$ and $h = 0.71$, in order to create a large statistical population of small scale baryonic structures. These parameters are assumed in the subsequent modeling. 

\begin{figure*}
    \centering
    \includegraphics[width=0.95\linewidth]{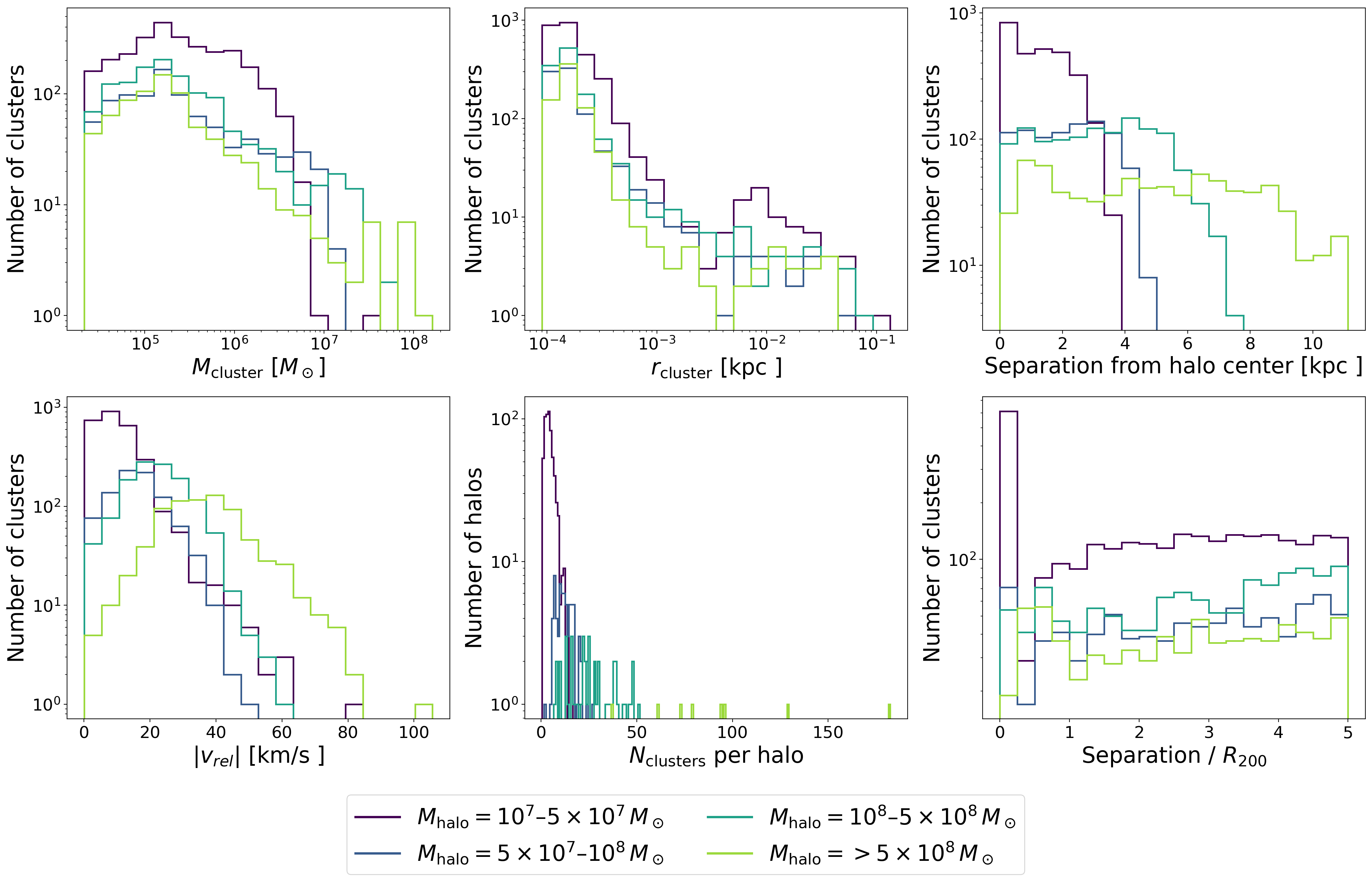}
    \caption{Distributions of star cluster properties, split by host halo mass. Each color corresponds to one halo mass bin (legend).  The top row shows cluster stellar mass, cluster half-mass radius, and three-dimensional distance from the halo center in physical kpc. The bottom row shows the magnitude of each cluster's velocity relative to its host halo, the number of associated clusters per halo, and the cluster–halo separation in units of the halo virial radius, $R_{\rm vir}$.}
    \label{fig:cluster_dist}
\end{figure*}

Following the methods introduced in \cite{Williams+25}, we combine two catalogs to ensure that we find all the star clusters in the high resolution simulation. The first is the standard FOF/\textsc{Subfind} algorithm native to {\tt AREPO}, with DM as the primary particle type and gas and star particles secondary. The second is a catalog produced with the same algorithm, but with star and gas particles as the primary type, and no secondary. This allows us to identify all clusters-not just those co-located at halo centers. For each halo in the high-resolution catalog of mass greater than the minimum mass in the TNG50 catalog, we search for all star clusters that meet the following five criteria: 
\begin{enumerate}
    \item Located within $5R_{200}$.
    \item Individually gravitationally bound (i.e., $U+K<0$).
    \item Gravitationally bound to the halo in question.
    \item Closer to the halo in question than any other halo in the simulation of equal or greater mass. 
    \item Contain at least 100 star particles. 
\end{enumerate}
This ensures that we find well resolved, bound star clusters that are truly associated with the halo of interest. 
We thus assemble a calibration catalog from which we can draw star cluster properties based on an initial halo mass.
A summary of the star cluster catalog is presented in \autoref{tab:mass_bins}. 
There are 757 halos in the catalog, and 5692 star clusters. 
\autoref{fig:cluster_dist} shows the distributions of star cluster properties represented by the catalog, split by host halo mass. 

\begin{table}
    \caption{Number of halos and associated star clusters in each halo mass bin in the high resolution simulation.}
    \label{tab:mass_bins}
    \begin{tabular}{lrrr}
        \toprule
        Halo mass range & $N_{\rm halo}$ & $N_{\rm cl}$ & $\langle N_{\rm cl} \rangle$ \\
         & & & per halo \\
        \midrule
        $10^{7}$--$5\times10^{7}\,M_\odot$ & 631 & 2806 & 4.4 \\
        $5\times10^{7}$--$10^{8}\,M_\odot$ & 72  & 897  & 12.5 \\
        $10^{8}$--$5\times10^{8}\,M_\odot$ & 46  & 1238 & 26.9 \\
        $>5\times10^{8}\,M_\odot$          & 8   & 751  & 93.9 \\
        \midrule
        Total                              & 757 & 5692 & 7.5 \\
        \bottomrule
    \end{tabular}
\end{table}
From this calibration catalog we build a kernel-weighted bootstrap sampler that resamples the catalog with a Gaussian kernel in $\log_{10}(M_{\rm halo})$, with a bandwidth of $0.3$~dex.
For a TNG host of mass $M_{200c}$ and radius $R_{200c}$ we draw a cluster population in two stages.
First, we pick one calibration halo $i$ with probability $w_i \propto \exp[-(\log M_i - \log M_{200c})^2/(2\times0.3^2)]$ and adopt its number of clusters, $N_i$, rescaled for the boosted $\sigma_8$ as described below.
Second, for each of these clusters we independently draw one calibration cluster $j$, with replacement, weighted by the same kernel evaluated at the mass of the halo that hosts $j$.
We adopt its stellar mass $M_{\rm cl}$, half-mass radius $r_{\rm h}$, host-centric distance in units of its host's $R_{200}$, speed in units of its host's $V_{200}=\sqrt{GM_{200}/R_{200}}$, and the angle $\theta$ between its position and velocity vectors.
These five quantities are drawn together, so correlations between mass, size, position and velocity within each cluster are preserved.
Clusters in the same halo, and the number of clusters, are drawn independently of each other.
The distance and speed are rescaled to the TNG host's $R_{200c}$ and $V_{200c}$. The position vector is given an isotropically random direction, and the velocity vector is placed at angle $\theta$ to it with a random azimuth.
(For computational speed, only calibration points within $6$ bandwidths of the query mass, and at most the 3000 nearest in log-mass, are considered; the neglected kernel weights are negligible.)
The host mass and radius are the FOF-group $M_{200c}$ and $R_{200c}$ of the branch's first subhalo.
Because clusters are drawn by bootstrap resampling a finite simulated catalog, the same simulated cluster can be selected multiple times; to avoid duplicate clusters we use a smoothed bootstrap, perturbing each resampled cluster's mass and half-mass radius by independent Gaussian offsets in $\log_{10} M_{\rm cl}$ and $\log_{10} r_{\rm h}$ of width $\sigma = 0.1$\,dex, which preserves the median of each quantity.

Because the largest halo in the simulation is $\sim10^9 M_\odot$, and the simulation is only run to high-redshift with Pop III stars, we cannot apply the calibration catalog to halos with very large masses, or to halos that formed at significantly lower redshift without extrapolating well outside of the calibration. 
For this reason, we only attach clusters to progenitor halos that formed at redshift $z>7$, and have total DM mass $M_{DM}<10^9 M_\odot$.

The high-resolution simulation was run with an increased $\sigma_8=1.7$ to boost the number of halos. To correct for this, we divide the number of clusters of the drawn calibration halo by a factor of $f_{ST}=5$, where $f_{ST}$ is the ratio of the halo mass function including the boosted $\sigma_8$ compared to the halo mass function with the reduced value of 0.8 at the minimum halo mass of TNG-50, and round down, $N_{\rm cl}=\lfloor N_i/5 \rfloor$ (see the appendix of \cite{Williams+24}).
With this correction, 3491 of the 10\,082 seeded branches host at least one cluster, for a total of 4834 clusters.

\subsection{Observationally calibrated cluster populations}
\label{sec:obs_sampler}
 
As an alternative model to drawing star clusters from the high-resolution cosmological simulation (Section~\ref{sec:cosmoseeding}), that allows us to make a more conservative estimate, we construct cluster populations directly from empirical scaling relations. 
For this, each progenitor halo in the TNG50 merger tree is assigned a number of clusters from the globular cluster (GC) number--halo mass relation, cluster masses from the GC mass function, half-mass radii from the birth mass--radius relation of embedded clusters, and initial positions and velocities from an analytic model of a GC system embedded in an NFW halo.
 Because these relations are not tied to the halo masses and redshifts resolved by the cluster-formation simulation, we do not apply cuts on halo mass or formation redshift in this model. 
Once drawn, clusters are assigned formation times and evolved
exactly as in the simulation-based model.

First, following \citet{burkert_high-precision_2020} we estimate the number of clusters for each halo using the observed $M_{\rm vir}-N_{\rm GC}$ relation. 
\citet{burkert_high-precision_2020} find that the number of GCs is proportional to halo
virial mass, $M_{\rm vir} = 5\times10^{9}\,{\rm M_\odot}\,N_{\rm GC}$, for
$M_{\rm vir}=10^{10}$--$2\times10^{15}\,{\rm M_\odot}$ at $z=0$
\citep[see also][]{harris_galactic_2017}. They interpret this linearity as the outcome
of hierarchical assembly from a high-redshift population of seed halos that
hosted on average one GC per $5\times10^{8}\,{\rm M_\odot}$ of dark matter,
with most of the present-day halo mass subsequently accreted smoothly and
bringing no additional GCs \citep{boylan-kolchin_globular_2017}. Since our hosts are
early progenitor halos rather than $z=0$ systems, we adopt this seed
normalization, so that the expected number of clusters in a host is
\begin{equation}
  \langle N_{\rm cl}\rangle = \frac{M_{\rm h}}{M_{\rm seed}}, \qquad
  M_{\rm seed} = 5\times10^{8}\,{\rm M_\odot}.
  \label{eq:ncl}
\end{equation}
For most hosts $\langle N_{\rm cl}\rangle \ll 1$, so we draw the number of
clusters in each host from a Poisson distribution with mean
$\langle N_{\rm cl}\rangle$. We add no further halo-to-halo scatter:
\citet{burkert_high-precision_2020} suggest that the intrinsic scatter of the relation is
consistent with Poisson statistics, and that the larger observed scatter is
dominated by uncertainties in $M_{\rm vir}$. 
We note that Equation~\eqref{eq:ncl} is applied to halos of $\sim10^{7}$--$10^{9}\,{\rm M_\odot}$, below the mass range over which the $z=0$ relation is calibrated, and that its normalization is inferred from the assembly argument above rather than measured directly.

We then draw the mass of each cluster from a log-normal distribution in $\log_{10}M_{\rm cl}$ following \citep{jordan_acs_2007},
\begin{equation}
  \frac{{\rm d}N}{{\rm d}\log_{10}M_{\rm cl}} \propto
  \exp\!\left[-\frac{\left(\log_{10}M_{\rm cl}-\log_{10}M_{\rm TO}\right)^{2}}{2\sigma_{M}^{2}}\right],
  \label{eq:gclf}
\end{equation}
with turnover mass $M_{\rm TO}=2.2\times10^{5}\,{\rm M_\odot}$ and width $\sigma_{M}=0.5$\,dex, truncated to
$10^{3}\le M_{\rm cl}/{\rm M_\odot}\le10^{8}$. Equation~\eqref{eq:gclf}
describes surviving GC populations; using it for the clusters at formation
neglects the low-mass clusters that are subsequently disrupted, which are
the least likely to form massive IMBHs.

\citet{marks_inverse_2012} constrain the initial three-dimensional half-mass radii of
embedded clusters from the binary populations of young clusters, finding
\begin{equation}
  r_{\rm h} = 0.10^{+0.07}_{-0.04}\,{\rm pc}\left(\frac{M_{\rm cl}}{{\rm M_\odot}}\right)^{0.13\pm0.04}.
  \label{eq:mk12}
\end{equation}
We use this relation to assign each cluster's radius (which feeds into the semi-analytic model of IMBH formation).
We apply a log-normal scatter of 0.2\,dex about Equation~\eqref{eq:mk12},
reflecting the uncertainty in its normalization. For the median cluster mass of our population this gives $r_{\rm h}\approx0.5$\,pc. 
 
Finally, we must initialize the clusters for the time integration at some position and  velocity. 
Observed GC systems follow de Vaucouleurs (S\'ersic index $n=4$) profiles in projection \citep{hudson_correlation_2018}, with effective radii of roughly 1--5 per cent
of $R_{200}$ \citep{lim_next_2024}. 
Therefore, we distribute clusters according to a projected S\'ersic profile with $n=4$ and effective radius $R_{\rm e}=0.03\,R_{\rm h}$, deprojected with the approximation of
\citet{prugniel_fundamental_1997},
\begin{equation}
  \rho(r)\propto s^{-p_n}\exp\!\left(-b_n s^{1/n}\right), \qquad s=r/R_{\rm e},
  \label{eq:ps}
\end{equation}
where $p_n = 1-0.6097/n+0.05463/n^{2}$ \citep{lima_neto_specific_1999} and
$b_n\approx 2n-1/3+0.009876/n$ \citep{prugniel_fundamental_1997}. 
Radii beyond $R_{\rm h}$ are redrawn, and directions are isotropic. 
Each cluster's velocity relative to its host is drawn from an isotropic Gaussian with one-dimensional dispersion $\sigma(r)=v_{\rm c}(r)/\sqrt{2}$, where $v_{\rm c}(r)$ is the circular velocity of an NFW halo \citep{NFW-97}
with mass $M_{\rm h}$, radius $R_{\rm h}$ and concentration $c=4$, the same profile used for the host in the orbit integration. Velocities at or above
the local escape speed of that profile are redrawn, so every cluster starts bound to its host.

\subsection{Runaway collapse and IMBH population}
The star clusters in the high resolution simulation are evolved through the analytic, dynamical model of \cite{williams_cosmological_2026}. 
This model is a fast, radially-resolved method that follows the buildup of a VMS in dense star clusters. 
The model includes binary heating, relaxation, dynamical friction, metallicity-dependent VMS winds, and computes the fractional mass loss involved in stellar collisions with the local velocity dispersion as a proxy for collision energy. 
We use the fiducial parameters from \citet{williams_cosmological_2026}, including a collider mass of $1M_\odot$, $\alpha=1.2$, $Z=0.1Z_\odot$, and $\phi_{\rm bin}=0.2$, and stellar eccentricity $e=0.5$
We assume that any VMS with mass $M_{\rm VMS}>200M_\odot$ formed via this channel is converted to an IMBH with a mass conversion factor of 1 (i.e., no mass is lost in the transition). 
The clusters are assumed to have an initial power-law density profile, $\rho(r)=\rho_0(r/r_0)^{-\alpha}$ with $\alpha=1.2$, truncated at $r_0$, which encloses the full cluster mass $M_{\rm cl}$.
Given the sampled half-mass radius, this gives $r_0 = 2^{1/(3-\alpha)}\,r_{\rm h}$.
The model is run in the cosmology of the high-resolution simulation.
The VMS grows through runaway stellar collisions until one of several conditions is met that terminates the static initial conditions. This can be the evolution of the $1M_\odot$ that are assumed to make up the cluster off the main sequence, or the half-mass core collapse timescale (signifying that the cluster has likely significantly restructured its internal density profile and the initial conditions are no longer valid). 
Additionally, if the star cluster reaches  $0.01R_{200c}$  of its current parent halo in the merger tree, we terminate the collisional growth of the VMS/IMBH. 
This replaces the semi-analytic ``merger timescale'' of \cite{williams_cosmological_2026} with the actual inspiral time of each cluster.
Clusters that are born within $0.01R_{200c}$ of their host centre are assumed to remain there (see \S\ref{sec:sam}), so their growth is limited only by the stellar-evolution and core-collapse conditions.

\subsection{Semi-analytic model and dynamical evolution}\label{sec:sam}
For each cluster-hosting progenitor halo, we draw a cluster population with the method described above, placing it at the branch's first snapshot relative to the position and velocity of the host subhalo.
We then follow each cluster forward through the TNG50-1-Dark merger tree one snapshot at a time until $z=0$.
At every snapshot the cluster is assigned to the descendant of its current host. When the host becomes a secondary progenitor, the cluster is handed to the halo it merges into.
Physical velocities are obtained from the tree's \texttt{SubhaloVel}.
Between snapshots, the orbit is integrated in the frame of the current host.
Because each cluster's position and velocity are measured relative to its host, they are carried over unchanged when the host continues along its own branch, since the cluster has moved and accelerated with it.
At a merger, the cluster is integrated in its old host up to the next snapshot and then transferred to the new host's frame, with the old host's position and velocity carried to that snapshot along a ballistic orbit (including the Hubble flow) in the NFW potential of the new host's progenitor, rather than by differencing two subhalos recorded at different times.
Background halos act only through their tidal field, $\mathbf{\dot{v}}_{\rm bg}(\mathbf{r})-\mathbf{\dot{v}}_{\rm bg}(\mathbf{0})$, since the current host falls freely in the same field.
A cluster that becomes unbound from a host other than the main-branch progenitor is reassigned to the subhalo at that snapshot, at least as massive as its current host, to which it is most tightly bound in units of $V_{200}^2$, or to the main-branch progenitor if it is bound to none of them and lies beyond $R_{200c}$ of its host.
Finally, one- or two-snapshot drops of more than a factor of five in the main-branch mass, where \textsc{SubLink} briefly follows a satellite during a major merger, are treated as tracking errors and stepped over, and same-branch steps with a larger mass change than this are treated as mergers.

The equation of motion includes the \cite{NFW-96a} (NFW) potential from the local host halo, dynamical friction (following \citealt{chandrasekhar_dynamical_1943}), and the background potential of the eventual main-branch galaxy (at that same snapshot, also assuming an NFW potential).
The equation of motion is:
\begin{equation}
    \mathbf{\dot{v}} = \mathbf{\dot{v}}_{\rm grav}+ \mathbf{\dot{v}}_{\rm df} +\mathbf{\dot{v}}_{\rm bg}, 
\end{equation}
where the gravitational acceleration due to the host halo is:
\begin{equation}
    \mathbf{\dot{v}}_{\rm grav} = -\frac{G M_{\rm host}(r)}{|\mathbf{r}|^2} \hat{r} \ ,
\end{equation}
the acceleration due to dynamical friction is \citep{chandrasekhar_dynamical_1943}: 
\begin{eqnarray}
  \mathbf{\dot{v}}_{\rm df} &=&  -\frac{4 \pi G^2 M_{\rm cluster} \rho(r) \ln{\Lambda}}{|\mathbf{v}|^3} \nonumber \\ 
   &\times& \left({\rm erf}(X) -\frac{2X}{\sqrt{\pi}}e^{-X^2}\right)\mathbf{v}
\end{eqnarray}
where
\begin{equation}
    X=\frac{v}{\sqrt{2}\sigma(r)} \ ,
\end{equation}
and:
\begin{equation}
    \mathbf{\dot{v}}_{\rm bg} = - \frac{G M_{\rm bg}(r)}{|\mathbf{r}|^2} \hat{r}
\end{equation}
is the background galaxy's gravitational acceleration. 
In these equations, G is the gravitational constant, $\rho$ is the DM density, $\mathbf{v}$ is the velocity of the star cluster, and $\ln(\Lambda)$ is the Coulomb logarithm. For $\rho(r)$, $M(r)$ and $\sigma(r)$ we use the NFW profile. 
This halo is the main progenitor of the $z=0$ halo at the same snapshot, centered at its position $\mathbf{r}_{\rm mb}$ relative to the current host.
It contributes gravity only, with no dynamical friction, and it is switched off when the current host is itself on the main branch, to avoid counting the same mass twice.
Here $\mathbf{v}$ and $\mathbf{r}$ are the cluster's velocity and position relative to the current host.
 
For central subhalos, the local host is the FOF group's $M_{200c}$ and $R_{200c}$, while for satellites we instead use the subhalo's own bound mass (\texttt{SubhaloMass}) with the radius enclosing a mean density of $200\rho_{\rm crit}(z)$, and add the satellite's FOF group as a second, gravity-only NFW background centered on the group center (omitted when the group's central is the main-branch progenitor, which is already included as the background halo); the same host mass and radius are used when drawing the cluster population.
$\rho(r)$ and $M(r)$ are the NFW density and enclosed mass.
For the velocity dispersion we use the isothermal approximation $\sigma(r)=V_c(r)/\sqrt{2}$, where $V_c$ is the NFW circular velocity.
The Coulomb logarithm is $\ln\Lambda=\ln(1+M_{200c}/M_{\rm cl})$, and the cluster mass $M_{\rm cl}$ is held fixed; we neglect tidal mass loss.

To avoid the central singularity, all forces are evaluated at the softened radius $\tilde r=\max(|\mathbf{r}|,r_{\rm soft})$, with $r_{\rm soft}=0.01R_{200c}$ of the current host.
If the cluster crosses inward through $r_{\rm soft}$, we stop the integration and tag it as ``inspiraled''.
We do not test for escape from intermediate hosts, since the cluster is still embedded in the larger assembling structure.
Only on the final step of the tree is a cluster that moves beyond $3R_{200c}$ tagged as ``escaped''.
Clusters that neither inspiral nor escape before $z=0$ are tagged as potential off-nuclear TDE hosts.
Clusters whose sampled initial position lies within $r_{\rm soft}$ are not integrated.
We assume they stay at the center of whichever halo hosts them, and follow them through the tree's mergers only.
They can contribute off-nuclear TDEs while their host is at a distance greater than $r_{\rm soft}$.
For each cluster we also record the Roche radius $r_{\rm Roche}=r_{\rm p}\,[M_{\rm cl}/3M_{\rm host}(r_{\rm p})]^{1/3}$ at every pericenter $r_{\rm p}$ with respect to the current host. For pinned clusters it is evaluated at $r_{\rm soft}$ in each host. We keep the minimum value and the time at which it occurs.

The snapshots in TNG50's merger tree are not uniformly spaced and can span hundreds of Myrs, and thus the host and background halos may grow significantly between snapshot outputs.
We therefore divide each snapshot interval $\Delta t$ into $N_{\rm sub}=\lceil \Delta t/1\,{\rm Gyr}\rceil$ sub-steps, and into at least four sub-steps whenever a background halo is present.
Within each sub-step, the host and background masses, radii, and relative offset, are held fixed at values linearly interpolated between the two snapshots.
During a merger step, the old host’s mass and radius are held fixed.
When the host is about to join, or has just left, the main branch, the background mass is faded linearly to or from zero across the interval.

\subsection{TDE rates}
Unlike typical nuclear TDE scenarios, the stellar population that may produce TDEs in these wandering star cluster IMBH systems does not replenish over time, resulting in a burst-like rate. 
Once a given cluster forms an IMBH, we calculate an  estimate of the time over which the system will create a burst of TDEs. 
We note that we've defined ``off-nuclear" TDEs as occurring far from the host halo center of the $z=0$ elliptical galaxy, while some of our star clusters are located at the center of a progenitor subhalo, which has not yet merged with the main branch at high redshift. 
Because these subhalo-hosted clusters are very low mass ($\sim10^4-10^7 M_\odot$), the actual low-mass host is likely not detectable, and the event appears as an off-nuclear event relative to the bright galaxy. 
For this reason, we categorize TDEs based on their distance from the host halo center at the time of the TDE. 
Distant TDEs, defined here as occurring beyond $3 R_{200c}$ of the $z=0$ host, may not be associated with the host galaxy when observed.

We compute TDE rates only for clusters whose IMBH exceeds $M_{\rm BH}=500\,M_\odot$, and because the IMBH is assembled from the cluster's own stars, we rescale the stellar density profile to contain the remaining mass, $\rho_0\rightarrow\rho_0(1-M_{\rm BH}/M_{\rm cl})$, before continuing.
Relaxation-driven TDEs supplies stars to the loss cone: 
\begin{equation}
    t_{\rm relax}= \frac{0.34 \sigma^3 }{G^2 \rho m_* \ln(\Lambda)},
    \label{eq:trelax}
\end{equation}
where the Coulomb log is $\ln (\Lambda) = \ln[M_{BH}/((1+\alpha)m_*)]$.
The size of the loss cone for a star orbiting in the cluster at semimajor axis $a$ is:
\begin{equation}
    \mathcal{R}_{\rm lc}= \frac{J_{\rm lc}^2}{J_c^2}\approx \frac{2 r_T}{a},
\end{equation}
where $J_{c}$ is the angular momentum of the orbit at the same energy and $J_{\rm lc}$ is the size of the loss cone in angular momentum, and
\begin{equation}
    r_T = R_*\left(\frac{M_{BH}}{m_*}\right)^{1/3},
\end{equation}
is the tidal radius and $R_*$ is the stellar radius. 
Given $q$, the ratio of how much relaxation changes $J^2$ in one orbit to the size of the loss cone, 
\begin{equation}
    q(a) = \frac{P}{\mathcal{R}_{\rm lc} t_{\rm relax}},
\end{equation}
where $P$ is the Keplerian orbital period, and the time for the depletion at a distance is given by: 
\begin{equation}
    t_{\rm dep}(r) = t_{\rm  relax} [\ln (1/\mathcal{R}_{\rm lc}) +q].
\end{equation}
Two cases are possible, the ``full loss cone" regime, where $q>>1$ and 
relaxation keeps the loss cone populated and 
and approximately a fraction \(\mathcal R_{\rm lc}\) of the population is disrupted per orbital period. The other regime is the ``empty loss cone" where $q<<1$ and relaxation slowly diffuses stars into the loss cone. 
Within the black hole's sphere of influence $r_{\rm infl}$, defined by $M_\star(<r_{\rm infl}) = M_{\rm BH}$, we take
\begin{equation}
    \sigma^2 \approx \frac{GM_{\rm BH}}{(1+\alpha) r},
\end{equation}
so that $t_{\rm relax}\propto r^{\alpha-3/2}$.

We start our model at the cluster's formation. 
The IMBH is formed at $t_{\rm IMBH}$, which comes from the semi-analytic model.
We assume the stellar distribution around the newly formed IMBH needs about one cluster half-mass relaxation time to adjust before it can feed the loss cone, so TDEs begin after a delayed start of one half-mass relaxation time, and a cluster produces no TDEs if its orbit ends or its Roche-limited time passes before then.

We  split clusters into thirty log-spaced spherical shells with radii $r_k$ between $r_{\rm in}= \max{[R_*, r(M* = 2m_*)]}$ and $r_{\rm out} = \min[r_{\rm infl}, r_0, r_{\rm Roche, min}]$.
 $r_{\rm Roche, min}]$ is the global minimum of the cluster's Roche radius (i.e., during the closest pericenter passage of the cluster's orbit) and is only included if the Roche radius drops below the radius of the star cluster.
Every shell contributes TDEs at a constant rate
\begin{equation}
    \dot{N}_{{\rm TDE},k} = \frac{\Delta M_k}{m_*\, t_{{\rm dep},k}},
\end{equation}
where $t_{{\rm dep},k}$ is evaluated at the shell's geometric midpoint.
The rate begins at $t_{\rm on}=t_{\rm IMBH}+t_{\rm rh}$ and lasts for
$\min(t_{{\rm dep},k},\,t_{\rm cluster})$, where
\begin{equation}
    t_{\rm cluster} = \min\left(t_{\rm Roche},\, t_{\rm inspiral},\, t_{z=0}\right) - t_{\rm on}
    \label{eq:tcluster}
\end{equation}
is the time available for TDEs. Here $t_{\rm Roche}$ is the time of the pericenter
with the minimum Roche radius (included only if $r_{\rm Roche,min}$ falls inside the
cluster), $t_{\rm inspiral}$ is the time at which the cluster inspirals (for clusters
that do), and $t_{z=0}$ is the present day, all measured from the cluster's formation.
A cluster with $t_{\rm cluster}\le0$ produces no TDEs. Shells with
$t_{{\rm dep},k}<t_{\rm cluster}$ are fully depleted, while the rest are only partly
depleted. The cluster's TDE rate is $\dot{N}_{\rm TDE}=\sum_k \dot{N}_{{\rm TDE},k}$,
and over the whole integration it produces
\begin{equation}
    N_{\rm TDE} = \sum_k \frac{\Delta M_k}{m_*}\,
    \min\left(1,\frac{t_{\rm cluster}}{t_{{\rm dep},k}}\right)
\end{equation}
TDEs, so the number of TDEs cannot exceed the number of stars in each shell.

\subsection{Wandering black holes in the ASTRID simulations}
ASTRID \citep{ni_astrid_2022,zhou_astrid_2026} is one of the largest cosmological hydrodynamic simulations that evolves to $z=0$. It has a box size of 250 ${\rm Mpc}\,h^{-1}$ per side and host $2\times 5500^{3}$ particles. 
The resolution is $m_{\rm gas} = 1.3\times 10^{6}\,M_{\odot}h^{-1}$ for gas and $m_{\rm DM} = 6.7\times 10^{6}\,M_{\odot}h^{-1}$ for DM. 
In addition to gas cooling, star formation, and stellar feedback, ASTRID models the formation, growth, and mergers of the MBHs and their AGN feedback. 
In the following, we briefly summarize the MBH prescriptions relevant to this work and refer readers to \citet{ni_astrid_2022} and \citet{bird_astrid_2022} for further details. 

ASTRID adopts a halo-based MBH seeding model: when an FOF halo achieves a total mass $M_{\rm halo}\geqslant 5\times10^9\,h^{-1}M_\odot$ and stellar mass $M_{\rm *}\geqslant 5\times10^6\,h^{-1}M_\odot$, an MBH particle is placed at its potential minimum, whose seed mass is stochastically drawn from a power-law distribution between $M_{\rm BH}=3\times10^{4}h^{-1}M_\odot$ and 
$M_{\rm BH}=3\times10^{5}h^{-1}M_\odot$ with the power-law index $-1$. 
Gas accretion onto the MBH is given by the Bondi rate \citep{DiMatteo2005} capped at twice the Eddington limit. 
AGN feedback is implemented in a two-fold model: thermal feedback for high-accretion MBHs and kinetic feedback for low-accretion MBHs \citep{Weinberger2017}. 


An important feature of ASTRID is that it explicitly follows the MBH orbital evolution by incorporating a subgrid dynamical friction model rather than artificially repositioning MBHs to galaxy centers. 
This treatment allows MBHs to remain off-center following hierarchical assembly or dynamical interactions, and naturally produces a larger population of wandering IMBHs \citep{dimatteo_vast_2023}. Comparing this population with predictions from our semi-analytical model will provide valuable insight into how different MBH seeding channels influence the wandering IMBH population in the local Universe. 

We select wandering black holes in ASTRID based on a catalog of galaxies that match our TNG criteria. 
Specifically, we choose galaxies in the $z=0$ snapshot of ASTRID for which: 
\begin{enumerate}
    \item Are the central galaxy of their host halo,
    \item Have star formation rate $<0.1 M_\odot$ yr$^{-1}$ at $z=0$, 
    \item Have stellar mass within the three mass bins considered, and 
    \item Have $\kappa_{\rm rot}<0.5$ among stars within $30$ kpc.
\end{enumerate}
Within these galaxies, we searched for wandering IMBHs by removing the most massive BH in each galaxy. We also apply a mass cut for wandering BHs $M_{\rm BH}<3\times10^5 M_\odot\,h^{-1}$, corresponding to the maximum seed mass in ASTRID. We apply this cut to match the MBHs in the semi-analytical model, which do not include the subsequent growth after formation. 
A summary of the ASTRID data investigated in this work is provided in \autoref{app:data}, \autoref{tab:astrid_population}.

\section{Results}
\label{sec:results}
\begin{figure}
    \centering
    \includegraphics[width=0.95\linewidth]{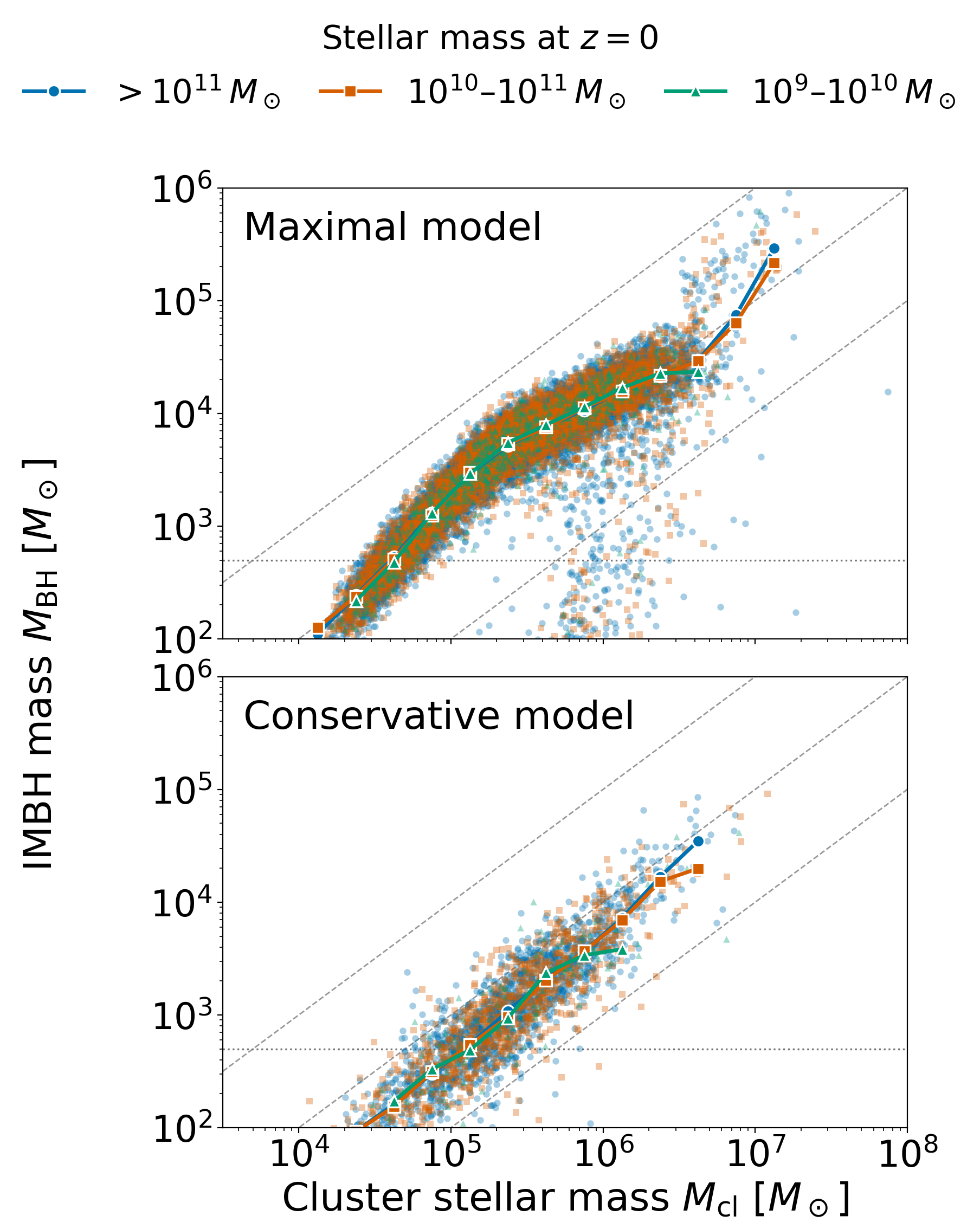}
    \caption{Black hole mass versus cluster mass for the maximal model (top) and the conservative model (bottom). Points are colored by stellar mass of the eventual host galaxy at $z=0$. From top to bottom, the the grey dashed lines show $M_{\rm BH}/M_{\rm cl}=0.1, 0.01,\text{and } 0.001$. }
    \label{fig:imbh-mass}
\end{figure}

\autoref{fig:imbh-mass} shows the IMBH mass--cluster mass relation for all of our runs. 
Because our IMBH formation model does not depend on the properties of the host galaxy, the relations for the different mass bins lie on top of one another. 
The plot shows the central difference between our models--the increased number of dense clusters seeded in the maximal model leads to a boosted IMBH mass function, driving additional TDEs. 
In \autoref{app:modelchoices}, we show that the dominant impact on the number of IMBH-hosting clusters per galaxy is simply the number of clusters seeded per subhalo, rather than the increased density of clusters in the maximal model. 
However, the cluster density does control the overall relation between $M_{\rm BH}-M_*$--as seen in the Figure, clusters in the conservative model typically host an IMBH that is $0.1-1\%$ of their mass, whereas the maximal model clusters produce IMBHs with $1-5\%$ of the cluster mass and even up to $10\%$ for the most massive clusters. 
\begin{figure*}
    \centering
    \includegraphics[width=0.95\linewidth]{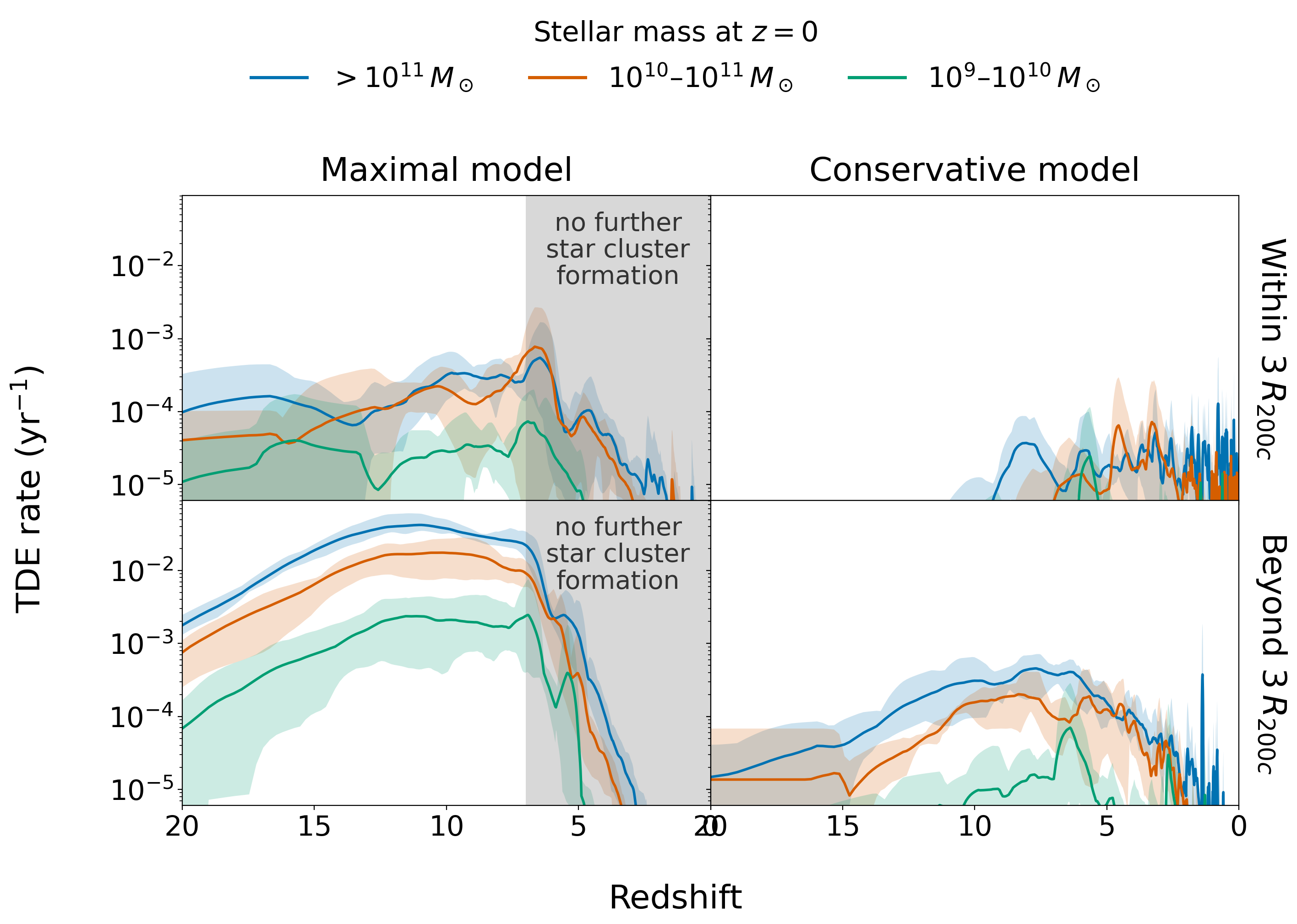}
    \caption{ Average off-nuclear TDE rate per galaxy among the sample, smoothed over 100 Myr, as a function of redshift for the maximal model (left) and the conservative model (right). The colored lines correspond to stellar mass bins (stellar mass of the Illustris TNG-50 host galaxy at $z=0$). The top panels show the TDE rate of events occurring within $3R_{ 200 c}$ of the host, and the bottom shows the rate occurring outside. Shaded regions show the spread across the five galaxies in each mass bin.}
    \label{fig:rates}
\end{figure*}

\autoref{fig:rates} shows the average TDE rate per galaxy in each of the three mass bins as a function of redshift in the maximal and conservative models. 
The rate per galaxy is the sum over every cluster seeded in every branch of one z = 0 galaxy's merger tree, averaged over the five galaxies in each mass bin. 
As described in \autoref{sec:methods}, we have split clusters in this figure by their distance from the host galaxy, with the top row showing TDE events that occur within $3R_{ 200 c}$ at the time of the TDE and the bottom row shows the events that occur outside this radius. 
At early times, $R_{ 200 c}$ of the $z=0$ host halo is relatively small, so the majority of TDEs occur outside this radius. 
For a TDE to occur inside $3R_{ 200 c}$, the cluster needs to form relatively close to the initial host, such that the burst of TDEs is still ongoing when it crosses into this region. 

We investigate the sensitivity of these results to multiple random draws of the model in \autoref{app:randomdraw}.
We find that within $3R_{ 200 c}$, the impact of random iterations on the TDE rate causes a scatter of $15-20\%$ in the upper two mass bins, while the lowest mass bin is dominated by random fluctuations, since the actual number of clusters is much smaller. 
The rates outside of $3R_{ 200 c}$ are very robust to a random draw, with the draw contributing just a few percent to the scatter. 
The scatter shown by the shaded regions in \autoref{fig:rates} is the variation between the five halos in each mass bin, and their different merger histories are generally the dominant source of scatter, especially in the highest mass bins. 
We highlight that our model does not track TDEs produced by the central SMBH of the galaxy, if it exists, and thus nuclear TDEs are not included in the rates inside  $3R_{ 200 c}$. 

The TDE rate is proportional to the host halo mass at $z=0$.
The averaged rates are smoothed over 100 Myr. 
In the maximal model, the TDE rate peaks at $z\sim 9-13$ at $\sim4\times10^{-2}$ yr$^{-1}$ for galaxies with $M_*>10^{11}M_\odot$ at $z=0$. 
In the conservative model, the maximum is roughly 2 dex lower, around $10^{-3}$ yr$^{-1}$, and peaks closer to $z=8$. 
In both scenarios the rate of TDEs decreases at late times: in the maximal model, where no new clusters are formed after $z=6$, the drop-off falls gradually, eventually  falling below $10^{-5}$ yr$^{-1}$ at $z<3$. 
In the conservative model, we do not enforce any redshift restriction on star cluster formation, so the rates continue, gradually decreasing, until $z=0$. At late times they primarily occur within $3R_{ 200 c}$. 

\begin{figure*}
    \centering
    \includegraphics[width=0.9\linewidth]{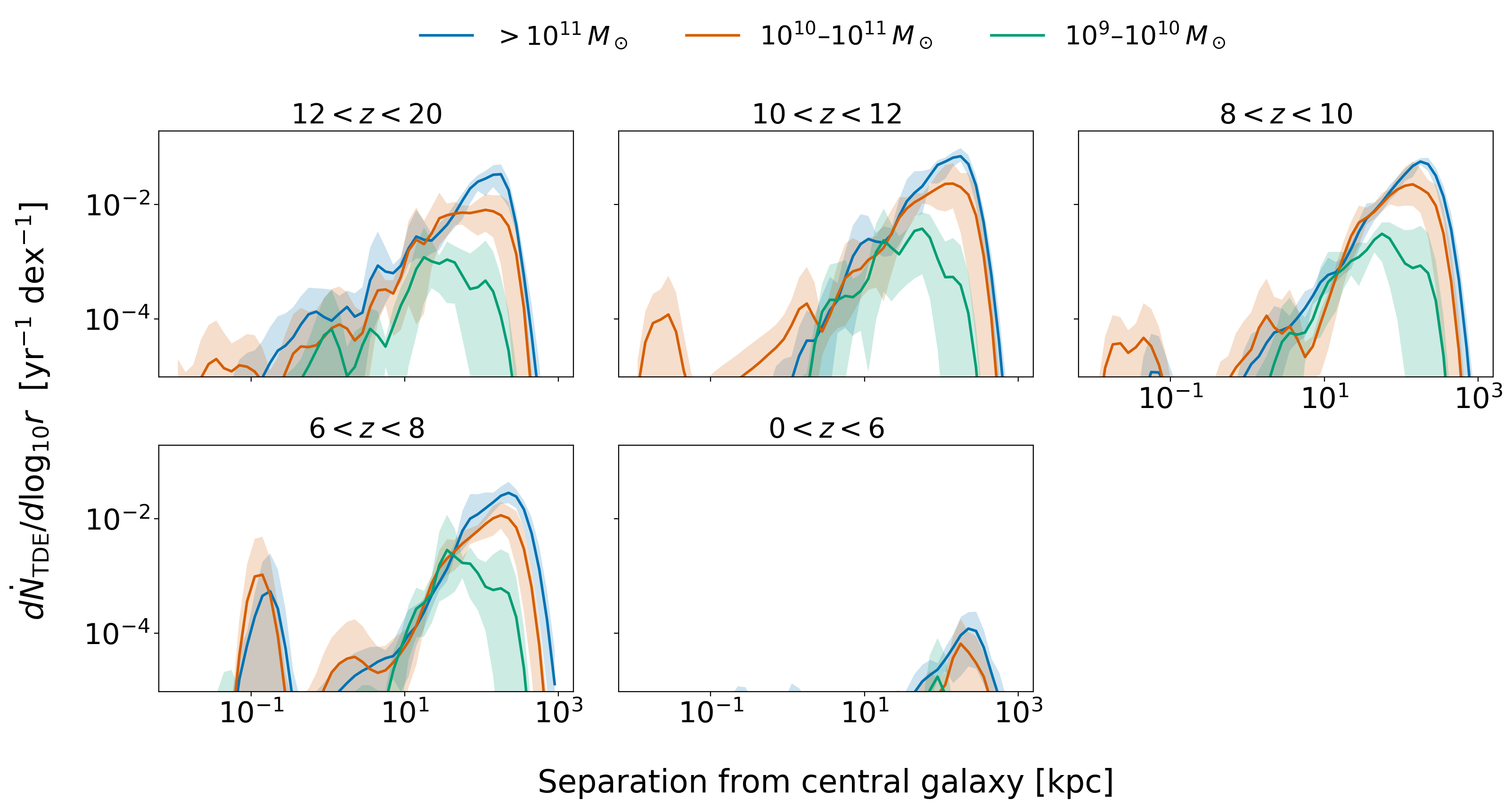}
    \caption{Radial distribution of off-nuclear TDEs about the central galaxy,
  in five redshift windows (panels).
  Each curve shows the rate of TDEs per dex of separation,
  ${\rm d}\dot N_{\rm TDE}/{\rm d}\log_{10} r$, where $r$ is the separation
  of the TDE-producing IMBH's host cluster from the main progenitor of the
  $z=0$ galaxy. For each halo, the rate is averaged over the time spanned by
  the redshift window and smoothed in $\log_{10} r$ with a Gaussian kernel of
  0.2\,dex FWHM. Colors denote bins of stellar mass at $z=0$,  solid lines
  are the mean over the five halos in each bin, and shaded regions span the
  full halo-to-halo range.}
    \label{fig:radial_maximal}
\end{figure*}

\begin{figure*}
    \centering
    \includegraphics[width=0.9\linewidth]{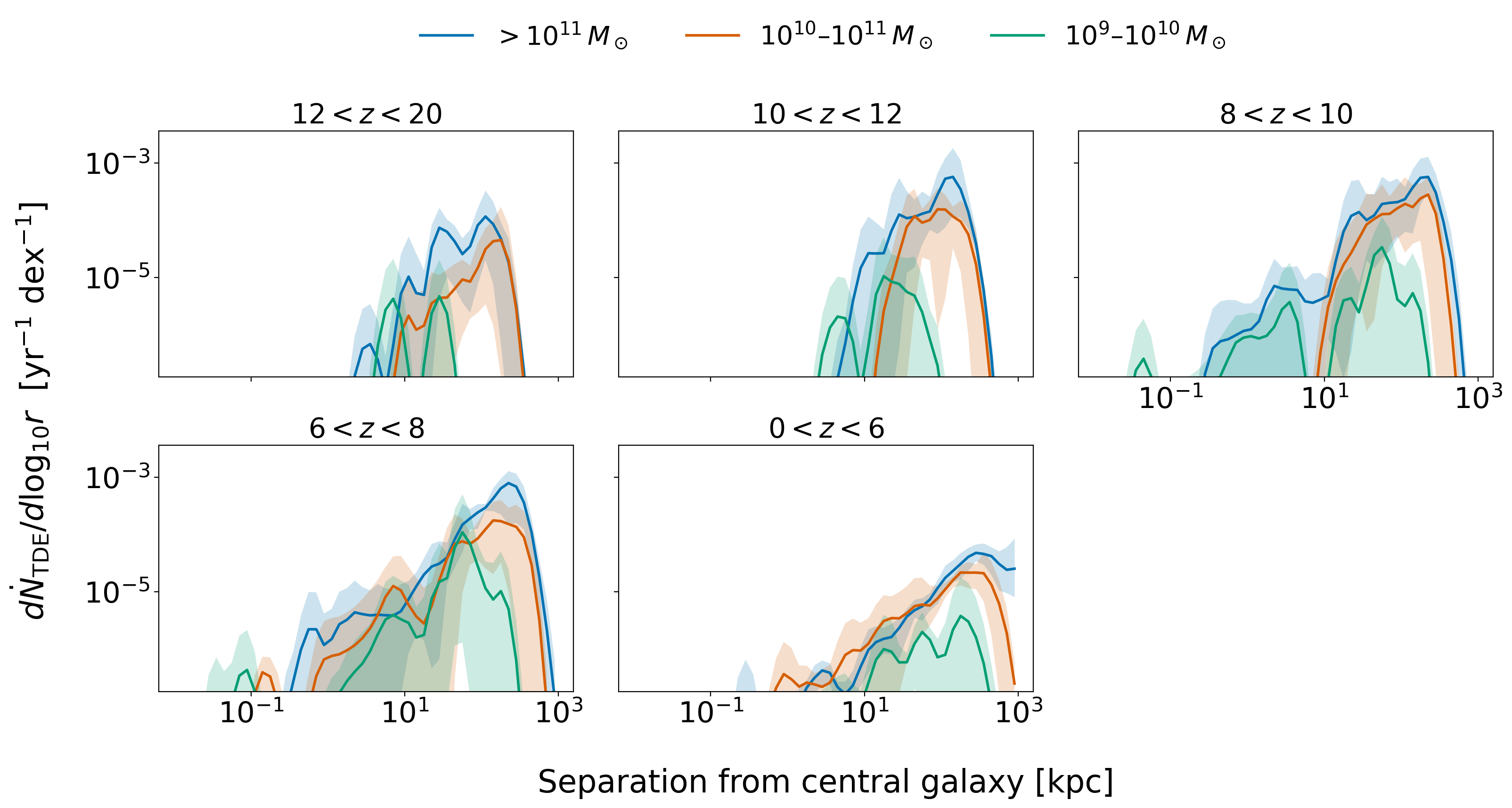}
    \caption{Same as \autoref{fig:radial_maximal}, but for the conservative star cluster model.}
    \label{fig:radial_minimal}
\end{figure*}

The radial distribution of off-nuclear TDE  events is shown for five time bins in \autoref{fig:radial_maximal} for the maximal model and \autoref{fig:radial_minimal} for the conservative model. 
The peak rate of off-nuclear TDEs is produced at $r\sim 10^2$ kpc at all times $z>6$. 
The tail towards smaller separations present at high redshift drops in the final redshift bin as those clusters are subsumed into the main galaxy, leaving clusters at large separation.


\begin{figure*}
    \centering
    \includegraphics[width=0.85\linewidth]{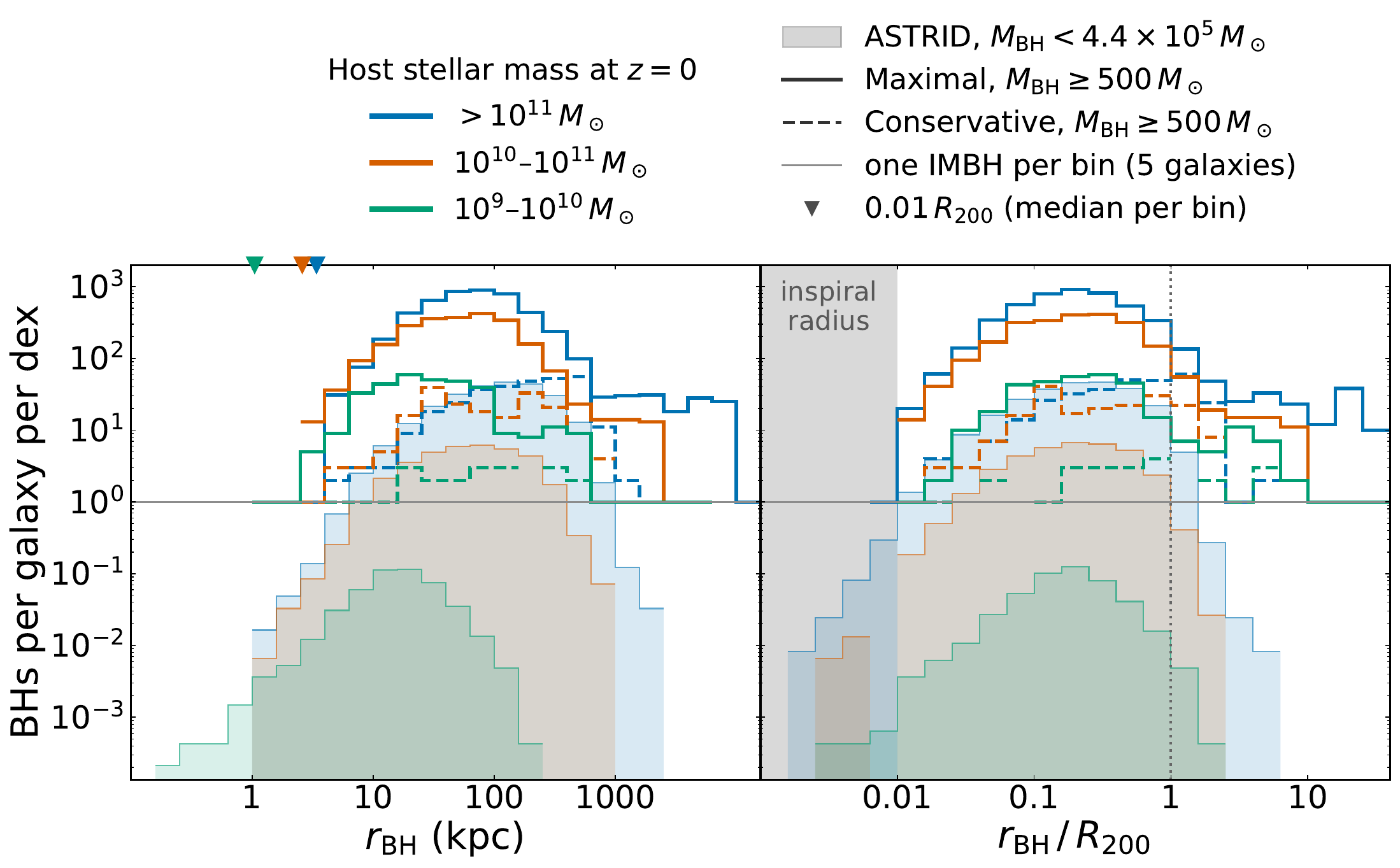}
    \caption{Radial distribution at $z=0$ of the IMBHs in our model, compared
    with the wandering black holes in ASTRID. Each panel shows the number of
    black holes per galaxy per dex of distance from the center of the central
    galaxy. The left column gives the physical distance $r_{\rm BH}$ and the
    right column the distance in units of the central's $R_{200c}$ at $z=0$.
    Colors give the $z=0$ stellar-mass bin.
    Thick solid and dashed lines are the maximal and conservative models,
    respectively, averaged over the five galaxies in each bin
(\autoref{tab:ellipticals}).They include every IMBH with
    $M_{\rm BH}\geq500\,M_\odot$ that forms before its cluster's orbit ends, but
    exclude IMBHs that reach the centre of the central galaxy (by inspiralling
    into, or remaining pinned in, the main progenitor). Shaded histograms show
    ASTRID's quenched, dispersion-dominated ($\kappa_{\rm rot}<0.5$) central
    galaxies in the same stellar-mass bins (23540, 763 and 611 galaxies, from
    the lowest to the highest bin). They include every black hole in each
    galaxy except its most massive one, with
    $M_{\rm BH}\leqslant 4.4\times10^{5}\,M_\odot$. Bins are 0.2~dex wide. The grey
    horizontal line marks one IMBH in a single bin across a mass bin's five
    galaxies, $1/(5\times0.23)\approx0.67$ per galaxy per dex, the lowest
    nonzero value our model can show. The grey band marks
    $r_{\rm BH}<0.01\,R_{200c}$, where orbit integration ends and a cluster is
    counted as having inspiralled, so our model contains no wandering IMBHs
    there by construction. Triangles mark the same radius in kpc for each bin's
    median $R_{200c}$. The dotted vertical line marks $R_{200c}$.}
    \label{fig:astrid}
\end{figure*}

\autoref{fig:astrid} shows the radial distribution of wandering IMBHs at $z=0$, versus galactocentric distance (left) and separation normalized to the parent host $R_{200}$ (right).
The shaded histograms show the wandering BH population in ASTRID at $z=0$. 
The sample from ASTRID contains thousands of galaxies, and thus is able to resolve counts far below our minimum, set by the 5 galaxies we have in each mass bin (this floor is shown as the grey horizontal line in both panels). 
Compared to ASTRID, we produce significantly more wandering black holes, especially in our lowest mass bin. 
The conservative model in the highest mass bin ($>10^{11}M_\odot$) is the only run where counts roughly agree between the two. 
These differences can be attributed to several effects. 
First, ASTRID does not extend to the low-mass IMBH regime that  our model follows. 
Due to this, the ASTRID simulations are missing  the IMBHs we produce below the simulation's minimum seed mass. 
Furthermore, ASTRID only seeds halos above $5\times10^9 h^{-1}M_\odot$ with MBHs.
By $z=0$, this means ASTRID halos in that mass bin have experienced very few mergers with MBH-hosting halos, whereas our catalog of seeds is populated in halos with  $\sim10^7M_\odot$, allowing lower-mass hierarchical progenitors to contribute additional black hole seeds. 
Finally, the seeding model is different--in ASTRID, a single MBH forms per halo, whereas our model allows many seeds to form. 
When implementing a star cluster-based seeding model, employing the same cluster to IMBH model as this work, \citet{zhou_acorn_2026} find that each halo hosts many individual seeds compared to the seeding model adopted in ASTRID. 

Additionally, our radial distribution extends further out than ASTRID's population, well beyond $3R_{200c}$, whereas ASTRID WBHs are typically located within $3R_{200c}$. 
Because we do not resolve the inspiral within $0.01R_{200 c}$, we have no tracked black holes in the inner regions. 
Additionally driving this effect is the fact that clusters merging into these host at separations comparable to a galaxy's $R_{200c}$ have dynamical friction times longer than the age of the Universe, and thus they typically live on very wide orbits and their black holes thus retain wide galactocentric separations to the present day. 
In ASTRID, structures with these low masses are not seeded with black holes, and thus do not contribute to the WBH population.


\section{Discussion}
\label{sec:discussion}

\subsection{Off-nuclear TDEs in the high-redshift Universe probe the initial conditions of star clusters}
\label{sec:disc_ics}

Dense star clusters that form at high redshift can host IMBHs formed through runaway stellar collisions.
Some of these clusters sink to the centers of their hosts through dynamical friction \citep[e.g.,][]{rantala_frost-clusters_2024}.
However, we find that a substantial population remains at large galactocentric distances for hundreds of Myr to Gyr.
These clusters remain at large separations because they form at large fractions of the virial radius of their birth halo (\autoref{fig:cluster_dist}),  and that halo is itself a low-mass progenitor far from the main branch of the merger tree.
Many can form even outside the virial radius of the birth halo, in a smaller halo. 
The dynamical friction timescale for a $\sim10^5$--$10^6\,M_\odot$ cluster orbiting at several $R_{200}$ of a $\gtrsim10^{10}\,M_\odot$ halo is far longer than a Hubble time \citep[e.g.,][]{Boylan-Kolchin+08}, so these clusters, and their IMBHs, remain wandering.

Within these clusters, the IMBH drives a burst of relaxation-driven TDEs that lasts until the bound stellar reservoir inside its sphere of influence is exhausted.
Because the reservoir is not replenished in our model, the time-integrated number of TDEs per cluster is bounded by the stellar mass inside $r_{\rm infl}$, i.e., $N_{\rm TDE}\lesssim M_{\rm BH}/M_*\sim10^{3}$--$10^{4}$ events.
The duration of the burst is set by the relaxation time. For a typical cluster from the high-resolution simulation ($M_{\rm cl}\approx2\times10^5\,M_\odot$, $r_{\rm h}\approx0.15$\,pc), the half-mass relaxation time is only $\approx5$\,Myr, so the TDEs from each cluster are emitted within $\sim10$--$100$\,Myr of the IMBH's formation.
The off-nuclear TDE rate of a galaxy at any time is therefore approximately the rate at which IMBH-hosting clusters form in its progenitors, multiplied by the IMBH mass each one carries. This is why the rate in the maximal model drops sharply once cluster seeding stops at $z=7$, falling below $10^{-5}$\,yr$^{-1}$ by $z\approx6$ (\autoref{fig:rates}).

In this model, the off-nuclear TDE rate is set by the host mass and the initial properties of the star cluster population.
Particularly, the host mass sets how many cluster-hosting progenitor halos there are.
The overasll rate scales monotonically with $z=0$ stellar mass, rising by roughly an order of magnitude from the $10^{9}$--$10^{10}~{\rm M}_\odot$ bin to the $>10^{11}~{\rm M}_\odot$ bin in both models (\autoref{fig:rates}).
The host galaxy mass does not change the IMBH mass that forms in a given cluster in our model.
The relation between IMBH mass and cluster mass is indistinguishable across the three galaxy mass bins (\autoref{fig:imbh-mass}) because it is determined by each cluster's internal density, not its environment. 
We note that we have limited our study to elliptical galaxies. 
The cluster abundance and seeding model
, by contrast, changes the rate substantially. The maximal and conservative models differ by $\approx2$\,dex at peak. Part of this reflects the number of clusters per halo, but much of it reflects their compactness: the clusters in the high-resolution simulation have $r_{\rm h}\approx0.1$--$0.2$\,pc, while the \citet{marks_inverse_2012} relation gives $r_{\rm h}\approx0.5$\,pc at the same mass.
Higher densities cause shorter relaxation and collision timescales, giving more massive IMBHs relative to their host cluster (typically $M_{\rm BH}/M_{\rm cl}\sim1\%$ in the maximal model, compared with a few$\times0.1\%$ in the conservative model) and shorter, more intense TDE bursts.
More of the clusters in the maximal model also exceed the $500\,{\rm M}_\odot$ threshold above which we compute TDEs.
Because the TDE rate is sensitive to cluster densities in this way, measuring the high-redshift off-nuclear TDE rate, or even an upper limit, would directly constrain the structure of the first star clusters, which is otherwise accessible only in rare, strongly lensed systems \citep{Adamo+24, Mowla+24}.

The runaway collision scenario has gained significant attention given recent {\it JWST} observations of extremely dense star clusters at high redshift \citep{Adamo+24, Mowla+24, claeyssens_tracing_2025, messa_jwst_2026} and of the ``Little Red Dots'' \citep[e.g.,][]{Labbe+23, kocevski_hidden_2023, matthee_little_2024, akins_2025_cosmosweb}, which may be related to dense cluster phenomena \citep[e.g.,][]{pacucci_little_2025, chisholm_little_2026,williams_cosmological_2026}.
We show here that, if such clusters form in the low-mass progenitors and satellites of massive galaxies, a necessary consequence is a population of off-nuclear TDEs at high redshift that peaks at $10\lesssim z\lesssim13$.

The observability of these events is uncertain.
First, cosmological time dilation reduces the observed-frame rate by a factor $(1+z)$. The peak intrinsic rate of $\sim4\times10^{-2}$\,yr$^{-1}$ in the most massive hosts of the maximal model therefore corresponds to $\sim3\times10^{-3}$ events per year in the observer frame for each such progenitor system, and the light curves are stretched by the same factor.
Second, at these redshifts the TDE-hosting clusters lie mostly at $\sim10^{2}$\,kpc from the main progenitor (\autoref{fig:radial_maximal}, \autoref{fig:radial_minimal}). This is roughly ten or more times the virial radius of a $10^{10}$--$10^{11}\,M_\odot$ halo at $z\approx10$ ($R_{200}\approx6$--$13$\,kpc) and corresponds to $\sim20''-45''$ on the sky.
At high redshift, ``off-nuclear'' therefore mostly means \emph{in a separate, low-mass halo that has not yet merged}. Such events would appear either hostless or associated with a faint neighbor of a bright galaxy, not as offset sources within a resolved host.
Unlike nuclear SMBH TDEs, IMBH TDEs of solar-type stars are expected to be short and highly super-Eddington, since the fallback timescale scales as $M_{\rm BH}^{1/2}$. Their detectability at $z\gtrsim8$ will depend on the emission mechanism and wavelength, which we leave to future work.

The off-nuclear TDE candidates observed so far are at low redshift ($z\approx0.03$--$0.06$) \citep{lin_luminous_2018, lin_multiwavelength_2020, jin_intermediate-mass_2025, yao_massive_2025, stein_tde_2026, grotova_population_2025, chang_multiwavelength_2025, maksym_tidal_2013, maksym_deep_2014, donato_tidal_2014}.
The four strong candidates compiled by \citet{guolo_demographics_2026} have projected offsets of $\approx1$--$34$\,kpc from hosts with $\log_{10}(M_*/M_\odot)\approx10.8$--$11.1$. These hosts are comparable to our most massive bin, where we predict the highest rates.
At least one of these events, 3XMM~J2150$-$05, is associated with a $\sim10^{4}$--$10^{5}\,{\rm M}_\odot$ black hole embedded in an off-center star cluster \citep{lin_luminous_2018, lin_multiwavelength_2020}, as the scenario modeled here would predict.
Because we limit our maximal model to high redshift, the two cluster models differ most in the low-redshift regime.
In the maximal model, which seeds clusters only at $z>7$, the cluster-IMBH channel contributes essentially no off-nuclear TDEs at $z\lesssim3$.
In the conservative model, clusters continue to form in halos at all redshifts, and the rate in the most massive hosts remains at $\sim10^{-5}$--$10^{-4}$\,yr$^{-1}$ down to $z=0$ (\autoref{fig:rates}). This is a non-negligible fraction of the optical nuclear TDE rate, $\approx3\times10^{-5}$\,galaxy$^{-1}$\,yr$^{-1}$ \citep{yao_tidal_2023}.
The late-time conservative rate is intermittent, reflecting the stochastic formation of individual IMBH-hosting clusters, and its radial distribution peaks at $\sim10^{2}$\,kpc, well outside the separations of the observed events (\autoref{fig:radial_minimal}).
Transient surveys associate a TDE with its host only within a limited projected radius, so events at $\gtrsim50$\,kpc would generally be classified as hostless or attributed to a different galaxy.
Part of the discrepancy also reflects the physics missing from our model (\S\ref{sec:disc_physics}): in particular, we do not form new clusters in situ in the main progenitor, and in situ clusters would naturally produce TDEs at smaller offsets.
If the low-redshift off-nuclear TDEs are produced by cluster-born IMBHs, our results suggest that the IMBHs either formed in clusters well after the epoch of reionization, or are being fed by a mechanism that refills the loss cone over Gyr timescales. A single high-redshift burst cannot supply them.

\subsection{Distant clusters hosting TDEs}
\label{sec:disc_unbound}

Some star clusters in our model end up outside $3R_{200c}$ of the $z=0$ galaxy, and can host TDEs at great distances from the galaxy whose progenitors they were born in.
As shown by \autoref{fig:rates}, these events are the majority of TDEs that originate in clusters formed in and around the progenitor halos of the main galaxy.
Separations from the main galaxy can reach Mpcs (\autoref{fig:radial_maximal} and \autoref{fig:radial_minimal}).

In the conservative model, every cluster starts deep inside ($R_{\rm e}=0.03R_{200}$) and bound to its birth halo. Escape in this model must therefore be driven by the hierarchical assembly itself.
Two processes can do this: the loss of the host subhalo's mass as it is tidally stripped, which lowers the binding energy of its clusters, and the orbital energy carried by the host subhalo when it is absorbed into a larger halo, which is inherited by its clusters at the hand-off.
In the maximal model, a large fraction of clusters additionally begin at 1--5\,$R_{200}$ of their minihalo (\autoref{fig:cluster_dist}), where they are only marginally bound, so this effect is stronger.

These TDEs, which can reach 1 Mpc distances, may not be associated with any luminous galaxy. Their only ``host'' would be the $\sim10^{5}$--$10^{6}\,M_\odot$ star cluster itself, likely too faint to detect at cosmological distances except perhaps under serendipitous strong lensing.
They would therefore be classified as hostless transients.
If confirmed, a population of hostless IMBH TDEs would reveal massive black holes that are not associated with any galaxy, a population that is invisible to dynamical searches and to most accretion-based searches.
We caution, however, that the escaped fraction is the quantity in our model most sensitive to the idealized potentials we adopt: untruncated NFW halos with fixed concentration, no large-scale (group or filament) potential beyond the main progenitor and the current host, and no tidal disruption of the clusters themselves (\S\ref{sec:disc_physics}).

\subsection{Comparison with ASTRID}
This work provides a complimentary view of the WBH population to the ASTRID simulations. 
Based on our comparison in \autoref{fig:astrid}, within similar galaxy selection criteria, we find that a large population of wandering IMBHs around seed mass is present in ASTRID. 
ASTRID captures WBHs produced by halo hierarchical assembly, but currently does not include multiple BHs formed in the same halo--these are followed in this work. 
Our work suggests that a large population of lower mass IMBHs than simulated by ASTRID may form due to the collapse of multiple star clusters in a single halo, and that the wandering MBH population can extend beyond a galaxy's $R_{200c}$. 
The differences between our model and the ASTRID simulations in this respect are largest for the lowest masses we consider ($10^{9}M_\odot<M_*<10^{10}M_\odot$). 
As discussed in \autoref{sec:results}, these have experienced very few mergers with BH-hosting halos in ASTRID. 
This implies that different seeding models would produce different wandering IMBH populations in the local universe, and thus affect observable TDE rates at low redshift. 
Higher-mass halos at low-$z$ may be less susceptible to this effect due to additional hierarchical assembly.

\subsection{Additional physical processes}
\label{sec:disc_physics}

\paragraph{Relaxation and refilling the loss cone}
Our analytic estimate assumes that stars are fed into the loss cone at the rate set by the local two-body relaxation time inside the sphere of influence, and that this reservoir is not replenished from outside.
However, the clusters extend beyond $r_{\rm infl}$ ($r_0\gg r_{\rm infl}$ for most of our clusters), so two-body relaxation throughout the cluster continues to scatter stars inward on the half-mass relaxation timescale. This supports a lower-level, long-lived TDE rate after the initial burst \citep[e.g.,][]{baumgardt_massive_2004}.
Our model therefore likely underestimates the late-time tail of the TDE rate from each cluster, which would partly fill in the  decline at $z\lesssim7$ in the maximal model.
However, after core collapse, clusters expand through two-body relaxation and heating by the central IMBH, and they lose mass to their host's tidal field. This serves to lower the central density and hence the refilling rate--thus we leave a more detailed calculation including these processes to future work.
In addition, we hold the stellar density profile fixed at $\alpha=1.2$ after IMBH formation. Around a central black hole, relaxation drives the profile toward a \citet{bahcall_star_1976} cusp with $\alpha=7/4$ within a relaxation time. Because $7/4>3/2$, this would move our calculation into the regime in which the TDE-supplying region lies \emph{inside} $r_{\rm crit}$, changing the time profile of the burst.
Finally, we assume a single stellar mass of $1\,M_\odot$. With a realistic mass function, mass segregation would concentrate stellar-mass black holes near the IMBH and partly shield it from low-mass stars.
Giant stars, which have large tidal radii, would contribute partial disruptions. 
A self-consistent treatment of these effects requires direct $N$-body or Monte Carlo cluster models, which we defer to future work.

\paragraph{IMBH growth through TDEs}
In each TDE, roughly half of the stellar debris is bound to the black hole \citep{rees_tidal_1988}.
Since the total mass disrupted per cluster is at most comparable to $M_{\rm BH}$, the IMBH could grow by up to $\sim50\%$ during the burst if this material is accreted. In practice the growth would be smaller, because much of the bound debris is expected to be lost to outflows when the fallback rate is highly super-Eddington.
This growth is not included in our model. It would slightly increase $r_{\rm infl}$ and the number of stars available for disruption, and would raise the IMBH masses of the wandering population by at most a factor of order unity.

\paragraph{Massive black hole binaries \& GW recoil}
Clusters that inspiral to the center of a halo (those tagged ``inspiraled'') deliver their IMBH to the nucleus of that halo. If the halo already contains a central black hole, or receives further IMBHs from other clusters, a black hole binary may form.
Such binaries are potential gravitational-wave sources in the deci-Hz and LISA bands, and the coalescence imparts a recoil kick to the remnant \citep[e.g.,][]{peres_classical_1962,bekenstein_gravitational-radiation_1973,fitchett_influence_1983,campanelli_large_2007,lousto_hangup_2011}.
The escape speeds of the minihalos that host most of our clusters are only $\sim10$--$60$\,km\,s$^{-1}$ (\autoref{fig:cluster_dist}), so even modest kicks can eject the merged black hole from a low-mass progenitor. This provides an additional channel for producing wandering IMBHs, one that is not tied to a surviving star cluster.
Within the clusters themselves, mergers between the IMBH and stellar-mass black holes can eject the IMBH unless the mass ratio is extreme \citep{holley-bockelmann_gravitational_2008}. This would end the TDE burst early, since an ejected IMBH carries only the stars tightly bound within its sphere of influence.
We do not model any of these processes here. The effect of recoil may thus be to move black holes from the ``inspiraled'' population into the wandering population and to truncate some bursts, while binary formation in nuclei is likely to enhance the nuclear, rather than the off-nuclear, TDE rate. 

\paragraph{Cluster tidal evolution and the host potential}
We hold each cluster's mass fixed during its orbit and do not follow tidal stripping or disruption of the cluster. We use the minimum Roche radius only to truncate the TDE-supplying region (\S\ref{sec:methods}).
A cluster that is fully disrupted by its host would leave behind a ``naked'' IMBH surrounded only by the stars tightly bound within its sphere of influence. That IMBH would remain a wandering black hole, but with a much lower TDE rate.
Conversely, dynamical friction weakens as the cluster loses mass, so stripped clusters would sink more slowly and remain at larger radii.
We also model every halo as an NFW profile with a fixed concentration $c=4$, and we do not include the stellar or gaseous components of the galaxies. Both choices mainly affect the inner few kpc of the orbits, and hence the inspiraled fraction and the small-separation tail of the radial distributions, rather than the $\sim10^2$\,kpc population that dominates the rates.

\paragraph{Continued star cluster formation}
In the maximal model, clusters are seeded only in halos that form at $z>7$ with $M_{\rm DM}<10^9\,M_\odot$, the regime resolved by the high-resolution simulation.
However, massive galaxies may continue to form dense star clusters well past this epoch: the bulk of present-day globular clusters formed at $z\sim2$--$6$, both in situ and in accreted satellites.
The conservative model includes later cluster formation, but only through relations calibrated on surviving $z=0$ globular cluster systems, with no redshift dependence.
Continued cluster formation would raise the off-nuclear TDE rate at $z\lesssim6$ relative to our maximal model. Clusters formed in situ within the main progenitor would also contribute TDEs at smaller separations ($\lesssim10$\,kpc), closer to the observed offsets.
On the other hand, clusters forming at later times form from metal-enriched gas. The stronger metallicity-dependent winds of the VMS \citep{vink_very_2018} would then reduce the IMBH mass formed per cluster, partially offsetting the larger number of clusters.
A cosmological model of cluster formation that spans $z\sim10$ to $z\sim1$ is needed to connect the high-redshift burst predicted here to the off-nuclear TDEs observed today.

\section{Summary}
\label{sec:summary}
Using a semi-analytic model, we explore the population of wandering IMBHs produced by dense star clusters in the early Universe and accreted onto elliptical galaxies.
We follow 15 TNG50 elliptical galaxies with $z=0$ stellar masses of $10^{9}$--$10^{11.3}\,M_\odot$ through their \textsc{SubLink} merger trees. We seed their progenitor halos with star clusters drawn either from a feedback-free, high-resolution cosmological simulation (``maximal'') or from local empirical scaling relations (``conservative''). We form IMBHs through runaway collisions and VMS collapse with the analytic model of \citet{williams_cosmological_2026}, and integrate each cluster's orbit under dynamical friction to $z=0$.
From this, we calculate the rate of off-nuclear TDEs produced in this scenario and predict their radial distribution. Our main findings are:
\begin{enumerate}
    \item Off-nuclear TDEs from cluster-born IMBHs occur in bursts. Once an IMBH forms, it disrupts the stars within its sphere of influence on the local relaxation timescale, typically within $\sim10$--$100$\,Myr. The number of events per cluster is bounded by $\sim M_{\rm BH}/M_*$, so the galaxy-wide rate traces the formation rate of IMBH-hosting clusters.
    \item The off-nuclear TDE rate per galaxy peaks at $9\lesssim z\lesssim13$ and increases with $z=0$ galaxy mass. In the maximal model, the most massive ellipticals ($M_*>10^{11}\,M_\odot$) reach $\sim2\times10^{-2}$\,yr$^{-1}$, compared with $\sim10^{-2}$ and $\sim10^{-3}$\,yr$^{-1}$ for the $10^{10}$--$10^{11}$ and $10^{9}$--$10^{10}\,M_\odot$ bins. In the conservative model the peak rate is $\approx1.5$\,dex lower, $\sim10^{-3}$\,yr$^{-1}$.
    (\autoref{fig:rates})
    \item The difference between the two models is driven largely by cluster compactness and the number of clusters seeded in each halo. The IMBH-to-cluster mass relation is set by the internal properties of the cluster and is independent of host galaxy mass (\autoref{fig:imbh-mass}). The high-redshift off-nuclear TDE rate depends on a combination of the cluster abundance, formation history, IMBH formation efficiency, and stellar replenishment, and  most strongly probes the densities and abundances of the first star clusters.
    \item At high redshift, the TDE-hosting clusters lie predominantly at $\sim10^{2}$\,kpc from the main progenitor. This is roughly ten virial radii, in low-mass halos that have not yet merged, so these events would appear hostless or associated with faint neighbors of bright galaxies. The small-separation tail is gradually removed as clusters sink into the main galaxy. (\autoref{fig:radial_maximal} and \autoref{fig:radial_minimal})
    \item The late-time rate depends on whether dense clusters continue to form after reionization. With cluster formation restricted to $z>7$, the rate falls below $10^{-5}$\,yr$^{-1}$ by $z\approx3$. With clusters formed at all epochs (conservative model), it remains at $\sim10^{-5}$--$10^{-4}$\,yr$^{-1}$ in and around massive ellipticals to $z=0$, a non-negligible fraction of the nuclear TDE rate. Even then, the predicted events lie at larger separations than the observed off-nuclear TDEs. (\autoref{fig:rates})
    \item A significant fraction of TDE-hosting clusters escape beyond $3R_{200c}$ of their $z=0$ galaxy, producing TDEs at Mpc distances at rates up to $\sim4\times10^{-2}$\,yr$^{-1}$ at high redshift in the maximal model. These would appear as hostless transients and would reveal massive black holes not associated with any galaxy.(\autoref{fig:rates})
\end{enumerate}
Future work including loss-cone refilling, cluster tidal evolution, gravitational-wave recoil, and continued cluster formation to low redshift will be needed to connect this high-redshift population to the off-nuclear TDEs now being discovered in the local Universe.

\section*{Acknowledgments}

 C.E.W. acknowledges the support of the National Science Foundation Graduate Research Fellowship, the University of California, Los Angeles (UCLA), the Mani L. Bhaumik Institute for Theoretical Physics, the Rudnick Abelmann Fellowship, and the UCLA Consortium for Developing Leadership in Science Fellowship. C.E.W., W.L., T.A-L., B.B., F.M., and M.V. thank the support of NASA grant No. 80NSSC24K0773 (ATP-23-ATP23-0149) and the XSEDE/ACCESS AST180056 allocation, as well as the UCLA cluster Hoffman2 for computational resources. C.E.W.,  T.A-L., and S.N. thank Howard and Astrid Preston for their generous support.
 N.Y. and C.E.W. acknowledge financial support from JSPS International Leading Research 23K20035.
 Y.Z. acknowledges the support from the NASA FINESST grant 80NSSC25K0318. 
 This material is based upon work supported by the National Science Foundation Graduate Research Fellowship Program under Grant No. DGE-2034835. Any opinions, findings, conclusions, or recommendations expressed in this material are those of the author(s) and do not necessarily reflect the views of the National Science Foundation.
This work used computational and storage services associated with the Hoffman2 Cluster, which is operated by the UCLA Office of Advanced Research Computing's Research Technology Group.

\software{astropy \citep{2013A&A...558A..33A,2018AJ....156..123A}, matplotlib \citep{Matplotlib},  numpy \citep{harris2020numpy}, scipy \citep{2020SciPy-NMeth}.}

\bibliography{cosmo}{}

@STRING{an      = "Astron. Nachr." }

@STRING{apjl    = "ApJL" }

@STRING{as      = "Appl. Spectrosc." }

@STRING{ga      = "Geomagn. Aehron." }

@STRING{ni      = "Nauchn. Inf." }

@STRING{on      = "Opt. News" }

@STRING{st      = "Sidereal Times" }

@ARTICLE{Boylan-Kolchin+08,
   author = {{Boylan-Kolchin}, M. and {Ma}, C.-P. and {Quataert}, E.},
    title = "{Dynamical friction and galaxy merging time-scales}",
  journal = {Monthly Notices of the Royal Astronomical Society},
archivePrefix = "arXiv",
   eprint = {0707.2960},
     year = 2008,
    month = jan,
   volume = 383,
    pages = {93-101},
      doi = {10.1111/j.1365-2966.2007.12530.x},
   adsurl = {http://adsabs.harvard.edu/abs/2008MNRAS.383...93B}
}

@ARTICLE{Chiaki+19,
       author = {{Chiaki}, Gen and {Wise}, John H.},
        title = "{Seeding the second star: enrichment from population III, dust evolution, and cloud collapse}",
      journal = {Monthly Notices of the Royal Astronomical Society},
         year = 2019,
        month = jan,
       volume = {482},
       number = {3},
        pages = {3933-3949},
          doi = {10.1093/mnras/sty2984},
archivePrefix = {arXiv},
       eprint = {1808.09515},
 primaryClass = {astro-ph.GA},
       adsurl = {https://ui.adsabs.harvard.edu/abs/2019MNRAS.482.3933C}
}

@ARTICLE{Marinacci+19,
       author = {{Marinacci}, Federico and {Sales}, Laura V. and {Vogelsberger}, Mark and {Torrey}, Paul and {Springel}, Volker},
        title = "{Simulating the interstellar medium and stellar feedback on a moving mesh: implementation and isolated galaxies}",
      journal = {Monthly Notices of the Royal Astronomical Society},
         year = 2019,
        month = nov,
       volume = {489},
       number = {3},
        pages = {4233-4260},
          doi = {10.1093/mnras/stz2391},
archivePrefix = {arXiv},
       eprint = {1905.08806},
 primaryClass = {astro-ph.GA},
       adsurl = {https://ui.adsabs.harvard.edu/abs/2019MNRAS.489.4233M}
}

@ARTICLE{NFW-97,
       author = {{Navarro}, Julio F. and {Frenk}, Carlos S. and {White}, Simon D.~M.},
        title = "{A Universal Density Profile from Hierarchical Clustering}",
      journal = {The Astrophysical Journal},
         year = 1997,
        month = dec,
       volume = {490},
       number = {2},
        pages = {493-508},
          doi = {10.1086/304888},
archivePrefix = {arXiv},
       eprint = {astro-ph/9611107},
 primaryClass = {astro-ph},
       adsurl = {https://ui.adsabs.harvard.edu/abs/1997ApJ...490..493N}
}

@ARTICLE{NFW-96a,
       author = {{Navarro}, Julio F. and {Frenk}, Carlos S. and {White}, Simon D.~M.},
        title = "{The Structure of Cold Dark Matter Halos}",
      journal = {The Astrophysical Journal},
         year = 1996,
        month = may,
       volume = {462},
        pages = {563},
          doi = {10.1086/177173},
archivePrefix = {arXiv},
       eprint = {astro-ph/9508025},
 primaryClass = {astro-ph},
       adsurl = {https://ui.adsabs.harvard.edu/abs/1996ApJ...462..563N}
}

@ARTICLE{Sales+12,
       author = {{Sales}, Laura V. and {Navarro}, Julio F. and {Theuns}, Tom and {Schaye}, Joop and {White}, Simon D.~M. and {Frenk}, Carlos S. and {Crain}, Robert A. and {Dalla Vecchia}, Claudio},
        title = "{The origin of discs and spheroids in simulated galaxies}",
      journal = {Monthly Notices of the Royal Astronomical Society},
         year = 2012,
        month = jun,
       volume = {423},
       number = {2},
        pages = {1544-1555},
          doi = {10.1111/j.1365-2966.2012.20975.x},
archivePrefix = {arXiv},
       eprint = {1112.2220},
 primaryClass = {astro-ph.CO},
       adsurl = {https://ui.adsabs.harvard.edu/abs/2012MNRAS.423.1544S}
}

@ARTICLE{Sakurai17,
       author = {{Sakurai}, Yuya and {Yoshida}, Naoki and {Fujii}, Michiko S. and {Hirano}, Shingo},
        title = "{Formation of intermediate-mass black holes through runaway collisions in the first star clusters}",
      journal = {Monthly Notices of the Royal Astronomical Society},
         year = 2017,
        month = dec,
       volume = {472},
       number = {2},
        pages = {1677-1684},
          doi = {10.1093/mnras/stx2044},
archivePrefix = {arXiv},
       eprint = {1704.06130},
 primaryClass = {astro-ph.GA},
       adsurl = {https://ui.adsabs.harvard.edu/abs/2017MNRAS.472.1677S}
}

@ARTICLE{Labbe+23,
       author = {{Labb{\'e}}, Ivo and {van Dokkum}, Pieter and {Nelson}, Erica and {Bezanson}, Rachel and {Suess}, Katherine A. and {Leja}, Joel and {Brammer}, Gabriel and {Whitaker}, Katherine and {Mathews}, Elijah and {Stefanon}, Mauro and {Wang}, Bingjie},
        title = "{A population of red candidate massive galaxies  600 Myr after the Big Bang}",
      journal = {Nature},
         year = 2023,
        month = apr,
       volume = {616},
       number = {7956},
        pages = {266-269},
          doi = {10.1038/s41586-023-05786-2},
archivePrefix = {arXiv},
       eprint = {2207.12446},
 primaryClass = {astro-ph.GA},
       adsurl = {https://ui.adsabs.harvard.edu/abs/2023Natur.616..266L}
}

@Article{Matplotlib,
  Author    = {Hunter, J. D.},
  Title     = {Matplotlib: A 2D graphics environment},
  Journal   = {Computing in Science \& Engineering},
  Volume    = {9},
  Number    = {3},
  Pages     = {90--95},
  publisher = {IEEE COMPUTER SOC},
  doi       = {10.1109/MCSE.2007.55},
  year      = 2007
}

@ARTICLE{Lake+23b,
       author = {{Lake}, William and {Naoz}, Smadar and {Marinacci}, Federico and {Burkhart}, Blakesley and {Vogelsberger}, Mark and {Williams}, Claire E. and {Chiou}, Yeou S. and {Chiaki}, Gen and {Nakazato}, Yurina and {Yoshida}, Naoki},
        title = "{The Supersonic Project: Star Formation in Early Star Clusters without Dark Matter}",
      journal = {\apjl},
         year = 2023,
        month = oct,
       volume = {956},
       number = {1},
          eid = {L7},
        pages = {L7},
          doi = {10.3847/2041-8213/acfa9b},
archivePrefix = {arXiv},
       eprint = {2306.01047},
 primaryClass = {astro-ph.GA},
       adsurl = {https://ui.adsabs.harvard.edu/abs/2023ApJ...956L...7L}
}

@ARTICLE{2018AJ....156..123A,
       author = {{Astropy Collaboration} and {Price-Whelan}, A.~M. and {Sip{\H{o}}cz}, B.~M. and {G{\"u}nther}, H.~M. and {Lim}, P.~L. and {Crawford}, S.~M. and {Conseil}, S. and {Shupe}, D.~L. and {Craig}, M.~W. and {Dencheva}, N. and {Ginsburg}, A. and {VanderPlas}, J.~T. and {Bradley}, L.~D. and {P{\'e}rez-Su{\'a}rez}, D. and {de Val-Borro}, M. and {Aldcroft}, T.~L. and {Cruz}, K.~L. and {Robitaille}, T.~P. and {Tollerud}, E.~J. and {Ardelean}, C. and {Babej}, T. and {Bach}, Y.~P. and {Bachetti}, M. and {Bakanov}, A.~V. and {Bamford}, S.~P. and {Barentsen}, G. and {Barmby}, P. and {Baumbach}, A. and {Berry}, K.~L. and {Biscani}, F. and {Boquien}, M. and {Bostroem}, K.~A. and {Bouma}, L.~G. and {Brammer}, G.~B. and {Bray}, E.~M. and {Breytenbach}, H. and {Buddelmeijer}, H. and {Burke}, D.~J. and {Calderone}, G. and {Cano Rodr{\'\i}guez}, J.~L. and {Cara}, M. and {Cardoso}, J.~V.~M. and {Cheedella}, S. and {Copin}, Y. and {Corrales}, L. and {Crichton}, D. and {D'Avella}, D. and {Deil}, C. and {Depagne}, {\'E}. and {Dietrich}, J.~P. and {Donath}, A. and {Droettboom}, M. and {Earl}, N. and {Erben}, T. and {Fabbro}, S. and {Ferreira}, L.~A. and {Finethy}, T. and {Fox}, R.~T. and {Garrison}, L.~H. and {Gibbons}, S.~L.~J. and {Goldstein}, D.~A. and {Gommers}, R. and {Greco}, J.~P. and {Greenfield}, P. and {Groener}, A.~M. and {Grollier}, F. and {Hagen}, A. and {Hirst}, P. and {Homeier}, D. and {Horton}, A.~J. and {Hosseinzadeh}, G. and {Hu}, L. and {Hunkeler}, J.~S. and {Ivezi{\'c}}, {\v{Z}}. and {Jain}, A. and {Jenness}, T. and {Kanarek}, G. and {Kendrew}, S. and {Kern}, N.~S. and {Kerzendorf}, W.~E. and {Khvalko}, A. and {King}, J. and {Kirkby}, D. and {Kulkarni}, A.~M. and {Kumar}, A. and {Lee}, A. and {Lenz}, D. and {Littlefair}, S.~P. and {Ma}, Z. and {Macleod}, D.~M. and {Mastropietro}, M. and {McCully}, C. and {Montagnac}, S. and {Morris}, B.~M. and {Mueller}, M. and {Mumford}, S.~J. and {Muna}, D. and {Murphy}, N.~A. and {Nelson}, S. and {Nguyen}, G.~H. and {Ninan}, J.~P. and {N{\"o}the}, M. and {Ogaz}, S. and {Oh}, S. and {Parejko}, J.~K. and {Parley}, N. and {Pascual}, S. and {Patil}, R. and {Patil}, A.~A. and {Plunkett}, A.~L. and {Prochaska}, J.~X. and {Rastogi}, T. and {Reddy Janga}, V. and {Sabater}, J. and {Sakurikar}, P. and {Seifert}, M. and {Sherbert}, L.~E. and {Sherwood-Taylor}, H. and {Shih}, A.~Y. and {Sick}, J. and {Silbiger}, M.~T. and {Singanamalla}, S. and {Singer}, L.~P. and {Sladen}, P.~H. and {Sooley}, K.~A. and {Sornarajah}, S. and {Streicher}, O. and {Teuben}, P. and {Thomas}, S.~W. and {Tremblay}, G.~R. and {Turner}, J.~E.~H. and {Terr{\'o}n}, V. and {van Kerkwijk}, M.~H. and {de la Vega}, A. and {Watkins}, L.~L. and {Weaver}, B.~A. and {Whitmore}, J.~B. and {Woillez}, J. and {Zabalza}, V. and {Astropy Contributors}},
        title = "{The Astropy Project: Building an Open-science Project and Status of the v2.0 Core Package}",
      journal = {\aj},
         year = 2018,
        month = sep,
       volume = {156},
       number = {3},
          eid = {123},
        pages = {123},
          doi = {10.3847/1538-3881/aabc4f},
archivePrefix = {arXiv},
       eprint = {1801.02634},
 primaryClass = {astro-ph.IM},
       adsurl = {https://ui.adsabs.harvard.edu/abs/2018AJ....156..123A}
}

@ARTICLE{2013A&A...558A..33A,
       author = {{Astropy Collaboration} and {Robitaille}, Thomas P. and
         {Tollerud}, Erik J. and {Greenfield}, Perry and {Droettboom}, Michael and
         {Bray}, Erik and {Aldcroft}, Tom and {Davis}, Matt and
         {Ginsburg}, Adam and {Price-Whelan}, Adrian M. and
         {Kerzendorf}, Wolfgang E. and {Conley}, Alexander and {Crighton}, Neil and
         {Barbary}, Kyle and {Muna}, Demitri and {Ferguson}, Henry and
         {Grollier}, Fr{\'e}d{\'e}ric and {Parikh}, Madhura M. and
         {Nair}, Prasanth H. and {Unther}, Hans M. and {Deil}, Christoph and
         {Woillez}, Julien and {Conseil}, Simon and {Kramer}, Roban and
         {Turner}, James E.~H. and {Singer}, Leo and {Fox}, Ryan and
         {Weaver}, Benjamin A. and {Zabalza}, Victor and {Edwards}, Zachary I. and
         {Azalee Bostroem}, K. and {Burke}, D.~J. and {Casey}, Andrew R. and
         {Crawford}, Steven M. and {Dencheva}, Nadia and {Ely}, Justin and
         {Jenness}, Tim and {Labrie}, Kathleen and {Lim}, Pey Lian and
         {Pierfederici}, Francesco and {Pontzen}, Andrew and {Ptak}, Andy and
         {Refsdal}, Brian and {Servillat}, Mathieu and {Streicher}, Ole},
        title = "{Astropy: A community Python package for astronomy}",
      journal = {\aap},
         year = "2013",
        month = "Oct",
       volume = {558},
          eid = {A33},
        pages = {A33},
          doi = {10.1051/0004-6361/201322068},
archivePrefix = {arXiv},
       eprint = {1307.6212},
 primaryClass = {astro-ph.IM},
       adsurl = {https://ui.adsabs.harvard.edu/abs/2013A&A...558A..33A}
}

@ARTICLE{Williams+24,
       author = {{Williams}, Claire E. and {Lake}, William and {Naoz}, Smadar and {Burkhart}, Blakesley and {Treu}, Tommaso and {Marinacci}, Federico and {Nakazato}, Yurina and {Vogelsberger}, Mark and {Yoshida}, Naoki and {Chiaki}, Gen and {Chiou}, Yeou S. and {Chen}, Avi},
        title = "{The Supersonic Project: Lighting Up the Faint End of the JWST UV Luminosity Function}",
      journal = {\apjl},
         year = 2024,
        month = jan,
       volume = {960},
       number = {2},
          eid = {L16},
        pages = {L16},
          doi = {10.3847/2041-8213/ad1491},
archivePrefix = {arXiv},
       eprint = {2310.03799},
 primaryClass = {astro-ph.GA},
       adsurl = {https://ui.adsabs.harvard.edu/abs/2024ApJ...960L..16W}
}

@ARTICLE{Lake+24a,
       author = {{Lake}, William and {Williams}, Claire E. and {Naoz}, Smadar and {Marinacci}, Federico and {Burkhart}, Blakesley and {Vogelsberger}, Mark and {Yoshida}, Naoki and {Chiaki}, Gen and {Chen}, Avi and {Chiou}, Yeou S.},
        title = "{The Supersonic Project: Early Star Formation with the Streaming Velocity}",
      journal = {The Astrophysical Journal},
         year = 2024,
        month = oct,
       volume = {973},
       number = {2},
          eid = {115},
        pages = {115},
          doi = {10.3847/1538-4357/ad6762},
archivePrefix = {arXiv},
       eprint = {2405.14938},
 primaryClass = {astro-ph.GA},
       adsurl = {https://ui.adsabs.harvard.edu/abs/2024ApJ...973..115L}
}

@ARTICLE{Adamo+24,
       author = {{Adamo}, Angela and {Bradley}, Larry D. and {Vanzella}, Eros and {Claeyssens}, Ad{\'e}la{\"\i}de and {Welch}, Brian and {Diego}, Jose M. and {Mahler}, Guillaume and {Oguri}, Masamune and {Sharon}, Keren and {Abdurro'uf} and {Hsiao}, Tiger Yu-Yang and {Xu}, Xinfeng and {Messa}, Matteo and {Lassen}, Augusto E. and {Zackrisson}, Erik and {Brammer}, Gabriel and {Coe}, Dan and {Kokorev}, Vasily and {Ricotti}, Massimo and {Zitrin}, Adi and {Fujimoto}, Seiji and {Inoue}, Akio K. and {Resseguier}, Tom and {Rigby}, Jane R. and {Jim{\'e}nez-Teja}, Yolanda and {Windhorst}, Rogier A. and {Hashimoto}, Takuya and {Tamura}, Yoichi},
        title = "{Bound star clusters observed in a lensed galaxy 460 Myr after the Big Bang}",
      journal = {Nature},
         year = 2024,
        month = aug,
       volume = {632},
       number = {8025},
        pages = {513-516},
          doi = {10.1038/s41586-024-07703-7},
archivePrefix = {arXiv},
       eprint = {2401.03224},
 primaryClass = {astro-ph.GA},
       adsurl = {https://ui.adsabs.harvard.edu/abs/2024Natur.632..513A}
}

@ARTICLE{Mowla+24,
       author = {{Mowla}, Lamiya and {Iyer}, Kartheik and {Asada}, Yoshihisa and {Desprez}, Guillaume and {Tan}, Vivian Yun Yan and {Martis}, Nicholas and {Sarrouh}, Ghassan and {Strait}, Victoria and {Abraham}, Roberto and {Brada{\v{c}}}, Maru{\v{s}}a and {Brammer}, Gabriel and {Muzzin}, Adam and {Pacifici}, Camilla and {Ravindranath}, Swara and {Sawicki}, Marcin and {Willott}, Chris and {Estrada-Carpenter}, Vince and {Jahan}, Nusrath and {Noirot}, Ga{\"e}l and {Matharu}, Jasleen and {Rihtar{\v{s}}i{\v{c}}}, Gregor and {Zabl}, Johannes},
        title = "{Formation of a low-mass galaxy from star clusters in a 600-million-year-old Universe}",
      journal = {Nature},
         year = 2024,
        month = dec,
       volume = {636},
       number = {8042},
        pages = {332-336},
          doi = {10.1038/s41586-024-08293-0},
       adsurl = {https://ui.adsabs.harvard.edu/abs/2024Natur.636..332M}
}

@article{smith_grackle_2017,
	title = {{GRACKLE}: a chemistry and cooling library for astrophysics},
	volume = {466},
	doi = {10.1093/mnras/stw3291},
	number = {2},
	journal = {Monthly Notices of the Royal Astronomical Society},
	author = {Smith, Britton D. and Bryan, Greg L. and Glover, Simon C. O. and Goldbaum, Nathan J. and Turk, Matthew J. and Regan, John and Wise, John H. and Schive, Hsi-Yu and Abel, Tom and Emerick, Andrew and O'Shea, Brian W. and Anninos, Peter and Hummels, Cameron B. and Khochfar, Sadegh},
	month = apr,
	year = {2017},
	note = {\_eprint: 1610.09591},
	pages = {2217--2234},
}

@article{pillepich_simulating_2018,
	title = {Simulating galaxy formation with the {IllustrisTNG} model},
	volume = {473},
	issn = {0035-8711},
	url = {https://doi.org/10.1093/mnras/stx2656},
	doi = {10.1093/mnras/stx2656},
	number = {3},
	urldate = {2025-01-16},
	journal = {Monthly Notices of the Royal Astronomical Society},
	author = {Pillepich, Annalisa and Springel, Volker and Nelson, Dylan and Genel, Shy and Naiman, Jill and Pakmor, Rüdiger and Hernquist, Lars and Torrey, Paul and Vogelsberger, Mark and Weinberger, Rainer and Marinacci, Federico},
	month = jan,
	year = {2018},
	pages = {4077--4106},
}

@Article{         harris2020numpy,
 title         = {Array programming with {NumPy}},
 author        = {Charles R. Harris and K. Jarrod Millman and St{\'{e}}fan J.
                 van der Walt and Ralf Gommers and Pauli Virtanen and David
                 Cournapeau and Eric Wieser and Julian Taylor and Sebastian
                 Berg and Nathaniel J. Smith and Robert Kern and Matti Picus
                 and Stephan Hoyer and Marten H. van Kerkwijk and Matthew
                 Brett and Allan Haldane and Jaime Fern{\'{a}}ndez del
                 R{\'{i}}o and Mark Wiebe and Pearu Peterson and Pierre
                 G{\'{e}}rard-Marchant and Kevin Sheppard and Tyler Reddy and
                 Warren Weckesser and Hameer Abbasi and Christoph Gohlke and
                 Travis E. Oliphant},
 year          = {2020},
 month         = sep,
 journal       = {Nature},
 volume        = {585},
 number        = {7825},
 pages         = {357--362},
 doi           = {10.1038/s41586-020-2649-2},
 publisher     = {Springer Science and Business Media {LLC}},
 url           = {https://doi.org/10.1038/s41586-020-2649-2}
}

@ARTICLE{2020SciPy-NMeth,
  author  = {Virtanen, Pauli and Gommers, Ralf and Oliphant, Travis E. and
            Haberland, Matt and Reddy, Tyler and Cournapeau, David and
            Burovski, Evgeni and Peterson, Pearu and Weckesser, Warren and
            Bright, Jonathan and {van der Walt}, St{\'e}fan J. and
            Brett, Matthew and Wilson, Joshua and Millman, K. Jarrod and
            Mayorov, Nikolay and Nelson, Andrew R. J. and Jones, Eric and
            Kern, Robert and Larson, Eric and Carey, C J and
            Polat, {\.I}lhan and Feng, Yu and Moore, Eric W. and
            {VanderPlas}, Jake and Laxalde, Denis and Perktold, Josef and
            Cimrman, Robert and Henriksen, Ian and Quintero, E. A. and
            Harris, Charles R. and Archibald, Anne M. and
            Ribeiro, Ant{\^o}nio H. and Pedregosa, Fabian and
            {van Mulbregt}, Paul and {SciPy 1.0 Contributors}},
  title   = {{{SciPy} 1.0: Fundamental Algorithms for Scientific
            Computing in Python}},
  journal = {Nature Methods},
  year    = {2020},
  volume  = {17},
  pages   = {261--272},
  adsurl  = {https://rdcu.be/b08Wh},
  doi     = {10.1038/s41592-019-0686-2},
}

@article{claeyssens_tracing_2025,
	title = {Tracing star formation across cosmic time at tens of parsec-scales in the lensing cluster field {Abell} 2744},
	volume = {537},
	copyright = {https://creativecommons.org/licenses/by/4.0/},
	issn = {0035-8711, 1365-2966},
	url = {https://academic.oup.com/mnras/article/537/3/2535/7954757},
	doi = {10.1093/mnras/staf058},
	language = {en},
	number = {3},
	urldate = {2025-02-21},
	journal = {Monthly Notices of the Royal Astronomical Society},
	author = {Claeyssens, Adélaïde and Adamo, Angela and Messa, Matteo and Dessauges-Zavadsky, Miroslava and Richard, Johan and Kramarenko, Ivan and Matthee, Jorryt and Naidu, Rohan P},
	month = feb,
	year = {2025},
	pages = {2535--2558},
}

@ARTICLE{Williams+25,
       author = {{Williams}, Claire E. and {Naoz}, Smadar and {Lake}, William and {Burkhart}, Blakesley and {Marinacci}, Federico and {Vogelsberger}, Mark and {Yoshida}, Naoki and {Menon}, Shyam H. and {Chen}, Avi and {Adamo}, Angela},
        title = "{{\ensuremath{\Lambda}}CDM Star Clusters at Cosmic Dawn: Stellar Densities, Environment, and Equilibrium}",
      journal = {The Astrophysical Journal},
         year = 2025,
        month = sep,
       volume = {990},
       number = {2},
          eid = {135},
        pages = {135},
          doi = {10.3847/1538-4357/adf19d},
archivePrefix = {arXiv},
       eprint = {2502.17561},
 primaryClass = {astro-ph.GA},
       adsurl = {https://ui.adsabs.harvard.edu/abs/2025ApJ...990..135W}
}

@ARTICLE{akins_2025_cosmosweb,
       author = {{Akins}, Hollis B. and {Casey}, Caitlin M. and {Lambrides}, Erini and {Allen}, Natalie and {Andika}, Irham T. and {Brinch}, Malte and {Champagne}, Jaclyn B. and {Cooper}, Olivia and {Ding}, Xuheng and {Drakos}, Nicole E. and {Faisst}, Andreas and {Finkelstein}, Steven L. and {Franco}, Maximilien and {Fujimoto}, Seiji and {Gentile}, Fabrizio and {Gillman}, Steven and {Gozaliasl}, Ghassem and {Harish}, Santosh and {Hayward}, Christopher C. and {Hirschmann}, Michaela and {Ilbert}, Olivier and {Kartaltepe}, Jeyhan S. and {Kocevski}, Dale D. and {Koekemoer}, Anton M. and {Kokorev}, Vasily and {Liu}, Daizhong and {Long}, Arianna S. and {McCracken}, Henry Joy and {McKinney}, Jed and {Onoue}, Masafusa and {Paquereau}, Louise and {Renzini}, Alvio and {Rhodes}, Jason and {Robertson}, Brant E. and {Shuntov}, Marko and {Silverman}, John D. and {Tanaka}, Takumi S. and {Toft}, Sune and {Trakhtenbrot}, Benny and {Valentino}, Francesco and {Zavala}, Jorge},
        title = "{COSMOS-Web: The Overabundance and Physical Nature of ``Little Red Dots''{\textemdash}Implications for Early Galaxy and SMBH Assembly}",
      journal = {The Astrophysical Journal},
         year = 2025,
        month = sep,
       volume = {991},
       number = {1},
          eid = {37},
        pages = {37},
          doi = {10.3847/1538-4357/ade984},
archivePrefix = {arXiv},
       eprint = {2406.10341},
 primaryClass = {astro-ph.GA},
       adsurl = {https://ui.adsabs.harvard.edu/abs/2025ApJ...991...37A}
}

@misc{pacucci_little_2025,
	title = {Little {Red} {Dots} {Are} {Nurseries} of {Massive} {Black} {Holes}},
	url = {http://arxiv.org/abs/2509.02664},
	doi = {10.48550/arXiv.2509.02664},
	urldate = {2025-10-14},
	publisher = {arXiv},
	author = {Pacucci, Fabio and Hernquist, Lars and Fujii, Michiko},
	month = sep,
	year = {2025},
	note = {arXiv:2509.02664 [astro-ph]},
}

@article{vink_very_2018,
	title = {Very massive stars: a metallicity-dependent upper-mass limit, slow winds, and the self-enrichment of globular clusters},
	volume = {615},
	copyright = {© ESO 2018},
	issn = {0004-6361, 1432-0746},
	shorttitle = {Very massive stars},
	url = {https://www.aanda.org/articles/aa/abs/2018/07/aa32773-18/aa32773-18.html},
	doi = {10.1051/0004-6361/201832773},
	language = {en},
	urldate = {2026-01-23},
	journal = {Astronomy \& Astrophysics},
	publisher = {EDP Sciences},
	author = {Vink, Jorick S.},
	month = jul,
	year = {2018},
	pages = {A119},
}

@misc{chisholm_little_2026,
	title = {Little {Red} {Dots} as {Globular} {Clusters} in {Formation}},
	url = {http://arxiv.org/abs/2602.15935},
	doi = {10.48550/arXiv.2602.15935},
	urldate = {2026-03-03},
	publisher = {arXiv},
	author = {Chisholm, John and Berg, Danielle A. and Boylan-Kolchin, Michael and Graaff, Anna de and Furtak, Lukas J. and Kokorev, Vasily and Matthee, Jorryt and Muñoz, Julian B. and Naidu, Rohan P. and Sander, Andreas A. C.},
	month = feb,
	year = {2026},
	note = {arXiv:2602.15935 [astro-ph]},
}

@article{vergara_rapid_2025,
	title = {Rapid formation of a very massive star ({\textgreater}50000 {M}$_{\textrm{⊙}}$ ), and subsequently, of an {IMBH}, from runaway collisions: {Direct} \textit{{N}} -body and {Monte} {Carlo} simulations of dense star clusters},
	volume = {704},
	copyright = {https://creativecommons.org/licenses/by/4.0},
	issn = {0004-6361, 1432-0746},
	shorttitle = {Rapid formation of a very massive star ({\textgreater}50000 {M}$_{\textrm{⊙}}$ ), and subsequently, of an {IMBH}, from runaway collisions},
	url = {https://www.aanda.org/10.1051/0004-6361/202555307},
	doi = {10.1051/0004-6361/202555307},
	language = {en},
	urldate = {2026-03-04},
	journal = {Astronomy \& Astrophysics},
	author = {Vergara, Marcelo C. and Askar, Abbas and Kamlah, Albrecht W. H. and Spurzem, Rainer and Flammini Dotti, Francesco and Schleicher, Dominik R. G. and Arca Sedda, Manuel and Hypki, Arkadiusz and Giersz, Mirek and Hurley, Jarrod and Berczik, Peter and Escala, Andres and Hoyer, Nils and Neumayer, Nadine and Pang, Xiaoying and Tanikawa, Ataru and Cen, Renyue and Naab, Thorsten},
	month = dec,
	year = {2025},
	pages = {A321},
}

@article{williams_observing_2026,
	title = {Observing {Compact} {Population} {III} {Star} {Clusters} and the {Presence} of {Cosmic} {Streaming}},
	volume = {997},
	issn = {0004-637X},
	url = {https://doi.org/10.3847/1538-4357/ae27c9},
	doi = {10.3847/1538-4357/ae27c9},
	language = {en},
	number = {2},
	urldate = {2026-03-04},
	journal = {The Astrophysical Journal},
	publisher = {The American Astronomical Society},
	author = {Williams, Claire E. and Naoz, Smadar and Yoshida, Naoki and Lake, William and Burkhart, Blakesley and Marinacci, Federico and Vogelsberger, Mark and Menon, Shyam H. and Chen, Avi and Kremer, Kyle},
	month = jan,
	year = {2026},
	pages = {202},
}

@article{gonzalez_prieto_intermediate-mass_2024,
	title = {Intermediate-mass {Black} {Hole} {Progenitors} from {Stellar} {Collisions} in {Dense} {Star} {Clusters}},
	volume = {969},
	issn = {0004-637X},
	url = {https://doi.org/10.3847/1538-4357/ad43d6},
	doi = {10.3847/1538-4357/ad43d6},
	language = {en},
	number = {1},
	urldate = {2026-03-04},
	journal = {The Astrophysical Journal},
	publisher = {The American Astronomical Society},
	author = {González Prieto, Elena and Weatherford, Newlin C. and Fragione, Giacomo and Kremer, Kyle and Rasio, Frederic A.},
	month = jun,
	year = {2024},
	pages = {29},
}

@article{kocevski_hidden_2023,
	title = {Hidden {Little} {Monsters}: {Spectroscopic} {Identification} of {Low}-mass, {Broad}-line {AGNs} at z {\textgreater} 5 with {CEERS}},
	volume = {954},
	issn = {2041-8205},
	shorttitle = {Hidden {Little} {Monsters}},
	url = {https://doi.org/10.3847/2041-8213/ace5a0},
	doi = {10.3847/2041-8213/ace5a0},
	language = {en},
	number = {1},
	urldate = {2026-03-04},
	journal = {The Astrophysical Journal Letters},
	publisher = {The American Astronomical Society},
	author = {Kocevski, Dale D. and Onoue, Masafusa and Inayoshi, Kohei and Trump, Jonathan R. and Arrabal Haro, Pablo and Grazian, Andrea and Dickinson, Mark and Finkelstein, Steven L. and Kartaltepe, Jeyhan S. and Hirschmann, Michaela and Aird, James and Holwerda, Benne W. and Fujimoto, Seiji and Juneau, Stéphanie and Amorín, Ricardo O. and Backhaus, Bren E. and Bagley, Micaela B. and Barro, Guillermo and Bell, Eric F. and Bisigello, Laura and Calabrò, Antonello and Cleri, Nikko J. and Cooper, M. C. and Ding, Xuheng and Grogin, Norman A. and Ho, Luis C. and Hutchison, Taylor A. and Inoue, Akio K. and Jiang, Linhua and Jones, Brenda and Koekemoer, Anton M. and Li, Wenxiu and Li, Zhengrong and McGrath, Elizabeth J. and Molina, Juan and Papovich, Casey and Pérez-González, Pablo G. and Pirzkal, Nor and Wilkins, Stephen M. and Yang, Guang and Yung, L. Y. Aaron},
	month = aug,
	year = {2023},
	pages = {L4},
}

@article{matthee_little_2024,
	title = {Little {Red} {Dots}: {An} {Abundant} {Population} of {Faint} {Active} {Galactic} {Nuclei} at z ∼ 5 {Revealed} by the {EIGER} and {FRESCO} {JWST} {Surveys}},
	volume = {963},
	issn = {0004-637X},
	shorttitle = {Little {Red} {Dots}},
	url = {https://doi.org/10.3847/1538-4357/ad2345},
	doi = {10.3847/1538-4357/ad2345},
	language = {en},
	number = {2},
	urldate = {2026-03-04},
	journal = {The Astrophysical Journal},
	publisher = {The American Astronomical Society},
	author = {Matthee, Jorryt and Naidu, Rohan P. and Brammer, Gabriel and Chisholm, John and Eilers, Anna-Christina and Goulding, Andy and Greene, Jenny and Kashino, Daichi and Labbe, Ivo and Lilly, Simon J. and Mackenzie, Ruari and Oesch, Pascal A. and Weibel, Andrea and Wuyts, Stijn and Xiao, Mengyuan and Bordoloi, Rongmon and Bouwens, Rychard and van Dokkum, Pieter and Illingworth, Garth and Kramarenko, Ivan and Maseda, Michael V. and Mason, Charlotte and Meyer, Romain A. and Nelson, Erica J. and Reddy, Naveen A. and Shivaei, Irene and Simcoe, Robert A. and Yue, Minghao},
	month = mar,
	year = {2024},
	pages = {129},
}

@article{messa_jwst_2026,
	title = {{JWST} spectroscopic confirmation of the {Cosmic} {Gems} arc at \textit{z} = 9.625: {Insights} into the small-scale structure of a post-burst system},
	volume = {705},
	copyright = {https://creativecommons.org/licenses/by/4.0},
	issn = {0004-6361, 1432-0746},
	shorttitle = {{JWST} spectroscopic confirmation of the {Cosmic} {Gems} arc at \textit{z} = 9.625},
	url = {https://www.aanda.org/10.1051/0004-6361/202556574},
	doi = {10.1051/0004-6361/202556574},
	language = {en},
	urldate = {2026-03-19},
	journal = {Astronomy \& Astrophysics},
	author = {Messa, M. and Vanzella, E. and Loiacono, F. and Adamo, A. and Oguri, M. and Sharon, K. and Bradley, L. D. and Christensen, L. and Claeyssens, A. and Richard, J. and {Abdurro’uf} and Bauer, F. E. and Bergamini, P. and Bolamperti, A. and Bradač, M. and Calura, F. and Coe, D. and Diego, J. M. and Grillo, C. and Hsiao, T. Y-Y. and Inoue, A. K. and Fujimoto, S. and Lombardi, M. and Meneghetti, M. and Resseguier, T. and Ricotti, M. and Rosati, P. and Welch, B. and Windhorst, R. A. and Xu, X. and Zackrisson, E. and Zanella, A. and Zitrin, A.},
	month = jan,
	year = {2026},
	pages = {A173},
}

@article{pechetti_detection_2022,
	title = {Detection of a 100,000 {M}$_{\textrm{⊙}}$ black hole in {M31}'s {Most} {Massive} {Globular} {Cluster}: {A} {Tidally} {Stripped} {Nucleus}},
	volume = {924},
	issn = {0004-637X, 1538-4357},
	shorttitle = {Detection of a 100,000 {M}$_{\textrm{⊙}}$ black hole in {M31}'s {Most} {Massive} {Globular} {Cluster}},
	url = {https://iopscience.iop.org/article/10.3847/1538-4357/ac339f},
	doi = {10.3847/1538-4357/ac339f},
	number = {2},
	urldate = {2026-03-20},
	journal = {The Astrophysical Journal},
	author = {Pechetti, Renuka and Seth, Anil and Kamann, Sebastian and Caldwell, Nelson and Strader, Jay and Den Brok, Mark and Luetzgendorf, Nora and Neumayer, Nadine and Voggel, Karina},
	month = jan,
	year = {2022},
	pages = {48},
}

@article{bahcall_star_1976,
	title = {Star distribution around a massive black hole in a globular cluster},
	volume = {209},
	issn = {0004-637X, 1538-4357},
	url = {http://adsabs.harvard.edu/doi/10.1086/154711},
	doi = {10.1086/154711},
	language = {en},
	urldate = {2026-03-26},
	journal = {The Astrophysical Journal},
	author = {Bahcall, J. N. and Wolf, R. A.},
	month = oct,
	year = {1976},
	pages = {214},
}

@article{zwart_runaway_2002,
	title = {The {Runaway} {Growth} of {Intermediate}-{Mass} {Black} {Holes} in {Dense} {Star} {Clusters}},
	volume = {576},
	issn = {0004-637X},
	url = {https://iopscience.iop.org/article/10.1086/341798},
	doi = {10.1086/341798},
	language = {en},
	number = {2},
	urldate = {2026-07-03},
	journal = {The Astrophysical Journal},
	publisher = {IOP Publishing},
	author = {Zwart, Simon F. Portegies and McMillan, Stephen L. W.},
	month = sep,
	year = {2002},
	pages = {899},
}

@article{glebbeek_evolution_2009,
	title = {The evolution of runaway stellar collision products},
	volume = {497},
	issn = {0004-6361, 1432-0746},
	url = {http://www.aanda.org/10.1051/0004-6361/200810425},
	doi = {10.1051/0004-6361/200810425},
	number = {1},
	urldate = {2026-07-13},
	journal = {Astronomy \& Astrophysics},
	author = {Glebbeek, E. and Gaburov, E. and De Mink, S. E. and Pols, O. R. and Portegies Zwart, S. F.},
	month = apr,
	year = {2009},
	pages = {255--264},
}

@misc{guolo_demographics_2026,
	title = {Demographics of {Wandering} {Black} {Holes} {Powering} {Off}-{Nuclear} {Tidal} {Disruption} {Events}},
	copyright = {Creative Commons Attribution 4.0 International},
	url = {https://arxiv.org/abs/2602.12970},
	doi = {10.48550/ARXIV.2602.12970},
	urldate = {2026-09-01},
	publisher = {arXiv},
	author = {Guolo, Muryel},
	year = {2026},
	note = {Version Number: 1},
}

@article{bellovary_wandering_2010,
	title = {Wandering {Black} {Holes} in {Bright} {Disk} {Galaxy} {Halos}},
	volume = {721},
	issn = {0004-637X},
	url = {https://ui.adsabs.harvard.edu/abs/2010ApJ...721L.148B},
	doi = {10.1088/2041-8205/721/2/L148},
	urldate = {2026-09-01},
	journal = {The Astrophysical Journal Letters},
	publisher = {IOP},
	author = {Bellovary, Jillian M. and Governato, Fabio and Quinn, Thomas R. and Wadsley, James and Shen, Sijing and Volonteri, Marta},
	month = oct,
	year = {2010},
	note = {ADS Bibcode: 2010ApJ...721L.148B},
	pages = {L148--L152},
}

@article{chandrasekhar_dynamical_1943,
	title = {Dynamical {Friction}. {I}. {General} {Considerations}: the {Coefficient} of {Dynamical} {Friction}.},
	volume = {97},
	issn = {0004-637X},
	shorttitle = {Dynamical {Friction}. {I}. {General} {Considerations}},
	url = {https://ui.adsabs.harvard.edu/abs/1943ApJ....97..255C},
	doi = {10.1086/144517},
	urldate = {2026-09-01},
	journal = {The Astrophysical Journal},
	publisher = {IOP},
	author = {Chandrasekhar, S.},
	month = mar,
	year = {1943},
	note = {ADS Bibcode: 1943ApJ....97..255C},
	pages = {255},
}

@article{ricarte_unveiling_2021,
	title = {Unveiling the {Population} of {Wandering} {Black} {Holes} via {Electromagnetic} {Signatures}},
	volume = {916},
	issn = {0004-637X},
	url = {https://ui.adsabs.harvard.edu/abs/2021ApJ...916L..18R},
	doi = {10.3847/2041-8213/ac1170},
	urldate = {2026-09-01},
	journal = {The Astrophysical Journal Letters},
	publisher = {IOP},
	author = {Ricarte, Angelo and Tremmel, Michael and Natarajan, Priyamvada and Quinn, Thomas},
	month = aug,
	year = {2021},
	note = {ADS Bibcode: 2021ApJ...916L..18R},
	pages = {L18},
}

@article{nelson_illustristng_2019,
	title = {The {IllustrisTNG} simulations: public data release},
	volume = {6},
	issn = {2197-7909},
	shorttitle = {The {IllustrisTNG} simulations},
	url = {https://link.springer.com/10.1186/s40668-019-0028-x},
	doi = {10.1186/s40668-019-0028-x},
	language = {en},
	number = {1},
	urldate = {2026-09-01},
	journal = {Computational Astrophysics and Cosmology},
	author = {Nelson, Dylan and Springel, Volker and Pillepich, Annalisa and Rodriguez-Gomez, Vicente and Torrey, Paul and Genel, Shy and Vogelsberger, Mark and Pakmor, Ruediger and Marinacci, Federico and Weinberger, Rainer and Kelley, Luke and Lovell, Mark and Diemer, Benedikt and Hernquist, Lars},
	month = dec,
	year = {2019},
	pages = {2},
}

@misc{williams_cosmological_2026,
	title = {A cosmological framework for stellar collisions at high redshift in proto-globular clusters, nuclear star clusters, and {Little} {Red} {Dots}},
	url = {http://arxiv.org/abs/2603.26872},
	doi = {10.48550/arXiv.2603.26872},
	urldate = {2026-06-08},
	publisher = {arXiv},
	author = {Williams, Claire E. and Naoz, Smadar and Rose, Sanaea C. and Burkhart, Blakesley and Yoshida, Naoki and Chen, Avi and Kremer, Kyle and Lake, William and Marinacci, Federico and Menon, Shyam H. and Vogelsberger, Mark},
	month = mar,
	year = {2026},
	note = {arXiv:2603.26872 [astro-ph.GA]},
}

@article{bird_astrid_2022,
	title = {The {ASTRID} simulation: galaxy formation and reionization},
	volume = {512},
	doi = {10.1093/mnras/stac648},
	number = {3},
	journal = {Monthly Notices of the Royal Astronomical Society},
	author = {Bird, Simeon and Ni, Yueying and Di Matteo, Tiziana and Croft, Rupert and Feng, Yu and Chen, Nianyi},
	month = may,
	year = {2022},
	pages = {3703--3716},
}

@article{di_matteo_vast_2023,
	title = {A vast population of wandering and merging {IMBHs} at cosmic noon},
	volume = {525},
	doi = {10.1093/mnras/stad2198},
	number = {1},
	journal = {Monthly Notices of the Royal Astronomical Society},
	author = {Di Matteo, Tiziana and Ni, Yueying and Chen, Nianyi and Croft, Rupert and Bird, Simeon and Pacucci, Fabio and Ricarte, Angelo and Tremmel, Michael},
	month = oct,
	year = {2023},
	pages = {1479--1497},
}

@article{ni_astrid_2022,
	title = {The {ASTRID} simulation: the evolution of supermassive black holes},
	volume = {513},
	doi = {10.1093/mnras/stac351},
	number = {1},
	journal = {Monthly Notices of the Royal Astronomical Society},
	author = {Ni, Yueying and Di Matteo, Tiziana and Bird, Simeon and Croft, Rupert and Feng, Yu and Chen, Nianyi and Tremmel, Michael and DeGraf, Colin and Li, Yin},
	month = jun,
	year = {2022},
	pages = {670--692},
}

@article{ricarte_origins_2021,
	title = {Origins and demographics of wandering black holes},
	volume = {503},
	doi = {10.1093/mnras/stab866},
	number = {4},
	journal = {Monthly Notices of the Royal Astronomical Society},
	author = {Ricarte, Angelo and Tremmel, Michael and Natarajan, Priyamvada and Zimmer, Charlotte and Quinn, Thomas},
	month = jun,
	year = {2021},
	pages = {6098--6111},
}

@article{springel_populating_2001,
	title = {Populating a cluster of galaxies - {I}. {Results} at z=0},
	volume = {328},
	doi = {10.1046/j.1365-8711.2001.04912.x},
	number = {3},
	journal = {Monthly Notices of the Royal Astronomical Society},
	author = {Springel, Volker and White, Simon D. M. and Tormen, Giuseppe and Kauffmann, Guinevere},
	month = dec,
	year = {2001},
	pages = {726--750},
}

@article{volonteri_assembly_2003,
	title = {The {Assembly} and {Merging} {History} of {Supermassive} {Black} {Holes} in {Hierarchical} {Models} of {Galaxy} {Formation}},
	volume = {582},
	doi = {10.1086/344675},
	number = {2},
	journal = {The Astrophysical Journal},
	author = {Volonteri, Marta and Haardt, Francesco and Madau, Piero},
	month = jan,
	year = {2003},
	pages = {559--573},
}

@article{zhou_astrid_2026,
	title = {The {ASTRID} {Simulation} at z = 0: {From} {Massive} {Black} {Holes} to {Large}-scale {Structure}},
	volume = {999},
	doi = {10.3847/1538-4357/ae3c08},
	number = {1},
	journal = {The Astrophysical Journal},
	author = {Zhou, Yihao and Di Matteo, Tiziana and Bird, Simeon and Croft, Rupert and Ni, Yueying and Yang, Yanhui and Chen, Nianyi and Lachance, Patrick and Zhang, Xiaowen and Hafezianzadeh, Fatemeh},
	month = mar,
	year = {2026},
	pages = {41},
}

@article{yao_massive_2025,
	title = {A {Massive} {Black} {Hole} 0.8 kpc from the {Host} {Nucleus} {Revealed} by the {Offset} {Tidal} {Disruption} {Event} {AT2024tvd}},
	volume = {985},
	issn = {2041-8205},
	url = {https://doi.org/10.3847/2041-8213/add7de},
	doi = {10.3847/2041-8213/add7de},
	language = {en},
	number = {2},
	urldate = {2026-09-29},
	journal = {The Astrophysical Journal Letters},
	publisher = {The American Astronomical Society},
	author = {Yao, Yuhan and Chornock, Ryan and Ward, Charlotte and Hammerstein, Erica and Sfaradi, Itai and Margutti, Raffaella and Kelley, Luke Zoltan and Lu, Wenbin and Liu, Chang and Wise, Jacob and Sollerman, Jesper and Alexander, Kate D. and Bellm, Eric C. and Drake, Andrew J. and Fremling, Christoffer and Gilfanov, Marat and Graham, Matthew J. and Groom, Steven L. and Hinds, K. R. and Kulkarni, S. R. and Miller, Adam A. and Miller-Jones, James C. A. and Nicholl, Matt and Perley, Daniel A. and Purdum, Josiah and Ravi, Vikram and Rich, R. Michael and Rehemtulla, Nabeel and Riddle, Reed and Smith, Roger and Stein, Robert and Sunyaev, Rashid and van Velzen, Sjoert and Wold, Avery},
	month = may,
	year = {2025},
	pages = {L48},
}

@article{burkert_high-precision_2020,
	title = {High-precision {Dark} {Halo} {Virial} {Masses} from {Globular} {Cluster} {Numbers}: {Implications} for {Globular} {Cluster} {Formation} and {Galaxy} {Assembly}},
	volume = {159},
	issn = {1538-3881},
	shorttitle = {High-precision {Dark} {Halo} {Virial} {Masses} from {Globular} {Cluster} {Numbers}},
	url = {https://doi.org/10.3847/1538-3881/ab5b0e},
	doi = {10.3847/1538-3881/ab5b0e},
	language = {en},
	number = {2},
	urldate = {2026-09-30},
	journal = {The Astronomical Journal},
	publisher = {The American Astronomical Society},
	author = {Burkert, Andreas and Forbes, Duncan A.},
	month = jan,
	year = {2020},
	pages = {56},
}

@article{weller_black_2026,
	title = {The black hole occupation fraction as a fossil record of seeding, dynamics, and galaxy assembly},
	volume = {1008},
	issn = {2041-8205, 2041-8213},
	url = {http://arxiv.org/abs/2607.09853},
	doi = {10.3847/2041-8213/ae9c38},
	number = {2},
	urldate = {2026-10-01},
	journal = {The Astrophysical Journal Letters},
	author = {Weller, Emma Jane and Natarajan, Priyamvada and Burke, Colin J.},
	month = sep,
	year = {2026},
	note = {arXiv:2607.09853 [astro-ph.GA]},
	pages = {L54},
}

@article{rodriguez-gomez_merger_2015,
	title = {The merger rate of galaxies in the {Illustris} simulation: a comparison with observations and semi-empirical models},
	volume = {449},
	issn = {0035-8711},
	shorttitle = {The merger rate of galaxies in the {Illustris} simulation},
	url = {https://doi.org/10.1093/mnras/stv264},
	doi = {10.1093/mnras/stv264},
	number = {1},
	urldate = {2026-10-01},
	journal = {Monthly Notices of the Royal Astronomical Society},
	author = {Rodriguez-Gomez, Vicente and Genel, Shy and Vogelsberger, Mark and Sijacki, Debora and Pillepich, Annalisa and Sales, Laura V. and Torrey, Paul and Snyder, Greg and Nelson, Dylan and Springel, Volker and Ma, Chung-Pei and Hernquist, Lars},
	month = may,
	year = {2015},
	pages = {49--64},
}

@article{chang_multiwavelength_2025,
	title = {Multiwavelength {Study} of a {Hyperluminous} {X}-{Ray} {Source} near {NGC} 6099: {A} {Strong} {IMBH} {Candidate}},
	volume = {983},
	doi = {10.3847/1538-4357/adbbee},
	number = {2},
	journal = {The Astrophysical Journal},
	author = {Chang, Yi-Chi and Soria, Roberto and Kong, Albert K. H. and Graham, Alister W. and Grishin, Kirill A. and Chilingarian, Igor V.},
	month = apr,
	year = {2025},
	pages = {109},
}

@article{donato_tidal_2014,
	title = {A {Tidal} {Disruption} {Event} in a nearby {Galaxy} {Hosting} an {Intermediate} {Mass} {Black} {Hole}},
	volume = {781},
	doi = {10.1088/0004-637X/781/2/59},
	number = {2},
	journal = {The Astrophysical Journal},
	author = {Donato, D. and Cenko, S. B. and Covino, S. and Troja, E. and Pursimo, T. and Cheung, C. C. and Fox, O. and Kutyrev, A. and Campana, S. and Fugazza, D. and Landt, H. and Butler, N. R.},
	month = feb,
	year = {2014},
	pages = {59},
}

@article{governato_fate_1994,
	title = {The fate of central black holes in merging galaxies},
	volume = {271},
	doi = {10.1093/mnras/271.2.317},
	journal = {Monthly Notices of the Royal Astronomical Society},
	author = {Governato, F. and Colpi, M. and Maraschi, L.},
	month = nov,
	year = {1994},
	pages = {317},
}

@article{grotova_population_2025,
	title = {The population of tidal disruption events discovered with {eROSITA}},
	volume = {697},
	doi = {10.1051/0004-6361/202553669},
	journal = {åp},
	author = {Grotova, I. and Rau, A. and Baldini, P. and Goodwin, A. J. and Liu, Z. and Merloni, A. and Salvato, M. and Anderson, G. E. and Arcodia, R. and Buchner, J. and Krumpe, M. and Malyali, A. and Masterson, M. and Miller-Jones, J. C. A. and Nandra, K. and Shirley, R.},
	month = may,
	year = {2025},
	pages = {A159},
}

@article{islam_massive_2004,
	title = {Massive black hole remnants of the first stars - {I}. {Abundance} in present-day galactic haloes},
	volume = {354},
	doi = {10.1111/j.1365-2966.2004.08201.x},
	number = {2},
	journal = {Monthly Notices of the Royal Astronomical Society},
	author = {Islam, Ranty R. and Taylor, James E. and Silk, Joseph},
	month = oct,
	year = {2004},
	pages = {427--442},
}

@article{jin_intermediate-mass_2025,
	title = {An {Intermediate}-mass {Black} {Hole} {Lurking} in {A} {Galactic} {Halo} {Caught} {Alive} during {Outburst}},
	doi = {10.48550/arXiv.2501.09580},
	journal = {arXiv e-prints},
	author = {Jin, C.-C. and Li, D.-Y. and Jiang, N. and Dai, L.-X. and Cheng, H.-Q. and Zhu, J.-Z. and Yang, C.-W. and Rau, A. and Baldini, P. and Wang, T.-G. and Zhou, H.-Y. and Yuan, W. and Zhang, C. and Shu, X.-W. and Shen, R.-F. and Wang, Y.-L. and Wen, S.-X. and Wu, Q.-Y. and Wang, Y.-B. and Thomsen, L. L. and Zhang, Z.-J. and Zhang, W.-J. and Coleiro, A. and Eyles-Ferris, R. and Fang, X. and Ho, L. C. and Hu, J.-W. and Jin, J.-J. and Li, W.-X. and Liu, B.-F. and Liu, F.-K. and Liu, M.-J. and Liu, Z. and Lu, Y.-J. and Merloni, A. and Qiao, E.-L. and Saxton, R. and Soria, R. and Wang, S. and Xue, Y.-Q. and Yang, H.-N. and Zhang, B. and Zhang, W.-D. and Cai, Z.-M. and Chen, F.-S. and Chen, H.-L. and Chen, T.-X. and Chen, W. and Chen, Y.-H. and Chen, Y.-F. and Chen, Y. and Cordier, B. and Cui, C.-Z. and Cui, W.-W. and Dai, Y.-F. and Ding, H.-C. and Fan, D.-W. and Fan, Z. and Feng, H. and Garcia, J. A. and Guan, J. and Han, D.-W. and Hou, D.-J. and Hu, H.-B. and Huang, M.-H. and Huo, J. and Jia, S.-M. and Jia, Z.-Q. and Jiang, B.-W. and Jin, G. and Kong, X. and Kuulkers, E. and Lei, W.-H. and Li, C.-K. and Li, J.-F. and Li, L.-H. and Li, M.-S. and Li, W. and Li, Z.-D. and Lian, T.-Y. and Ling, Z.-X. and Liu, C.-Z. and Liu, H.-Y and Liu, H.-Q. and Liu, J.-F. and Liu, Y. and Lu, F.-J. and Luo, L.-D. and Ma, J. and Mao, X. and Mu, H.-Y. and Nandra, K. and O'Brien, P. and Pan, H.-W. and Pan, X. and Qin, G.-J. and Rea, N. and Sanders, J. and Song, L.-M. and Sun, H. and Sun, S.-L. and Sun, X.-J. and Tan, Y.-Y. and Tang, Q.-J. and Tao, Y.-H. and Wang, B.-C. and Wang, J. and Wang, J.-F. and Wang, L. and Wang, W.-X. and Wang, Y.-S. and Wang, Z.-X. and Wu, Q.-W. and Wu, X.-F. and Xu, H.-T. and Xu, J.-J. and Xu, X.-P. and Xu, Y.-F. and Xu, Z. and Xue, C.-B. and Xue, S.-J. and Xue, Y.-L. and Yan, A.-L. and Yang, X.-T. and Yang, Y.-J. and Zhang, J. and Zhang, M. and Zhang, S.-N. and Zhang, Y.-H. and Zhang, Z. and Zhang, Z. and Zhang, Z.-L. and Zhao, D.-H. and Zhao, H.-S. and Zhao, X.-F. and Zhao, Z.-J. and Zheng, J. and Zhu, Q.-F. and Zhu, Y.-X. and Zhu, Z.-C. and Zou, H.},
	month = jan,
	year = {2025},
	pages = {arXiv:2501.09580},
}

@article{lin_multiwavelength_2020,
	title = {Multiwavelength {Follow}-up of the {Hyperluminous} {Intermediate}-mass {Black} {Hole} {Candidate} {3XMM} {J215022}.4-055108},
	volume = {892},
	doi = {10.3847/2041-8213/ab745b},
	number = {2},
	journal = {The Astrophysical Journal Letters},
	author = {Lin, Dacheng and Strader, Jay and Romanowsky, Aaron J. and Irwin, Jimmy A. and Godet, Olivier and Barret, Didier and Webb, Natalie A. and Homan, Jeroen and Remillard, Ronald A.},
	month = apr,
	year = {2020},
	pages = {L25},
}

@article{maksym_tidal_2013,
	title = {A tidal flare candidate in {Abell} 1795},
	volume = {435},
	doi = {10.1093/mnras/stt1379},
	number = {3},
	journal = {Monthly Notices of the Royal Astronomical Society},
	author = {Maksym, W. P. and Ulmer, M. P. and Eracleous, M. C. and Guennou, L. and Ho, L. C.},
	month = nov,
	year = {2013},
	pages = {1904--1927},
}

@article{maksym_deep_2014,
	title = {Deep spectroscopy of the {M}$_{\textrm{{V}}}$ {\textbackslash}ensuremath{\textbackslash}sim -14.8 host galaxy of a tidal disruption flare in {A1795}},
	volume = {444},
	doi = {10.1093/mnras/stu1485},
	number = {1},
	journal = {Monthly Notices of the Royal Astronomical Society},
	author = {Maksym, W. P. and Ulmer, M. P. and Roth, K. C. and Irwin, J. A. and Dupke, R. and Ho, L. C. and Keel, W. C. and Adami, C.},
	month = oct,
	year = {2014},
	pages = {866--873},
}

@article{perley_fast_2019,
	title = {The fast, luminous ultraviolet transient {AT2018cow}: extreme supernova, or disruption of a star by an intermediate-mass black hole?},
	volume = {484},
	doi = {10.1093/mnras/sty3420},
	number = {1},
	journal = {Monthly Notices of the Royal Astronomical Society},
	author = {Perley, Daniel A. and Mazzali, Paolo A. and Yan, Lin and Cenko, S. Bradley and Gezari, Suvi and Taggart, Kirsty and Blagorodnova, Nadia and Fremling, Christoffer and Mockler, Brenna and Singh, Avinash and Tominaga, Nozomu and Tanaka, Masaomi and Watson, Alan M. and Ahumada, Tomás and Anupama, G. C. and Ashall, Chris and Becerra, Rosa L. and Bersier, David and Bhalerao, Varun and Bloom, Joshua S. and Butler, Nathaniel R. and Copperwheat, Chris and Coughlin, Michael W. and De, Kishalay and Drake, Andrew J. and Duev, Dmitry A. and Frederick, Sara and González, J. Jesús and Goobar, Ariel and Heida, Marianne and Ho, Anna Y. Q. and Horst, John and Hung, Tiara and Itoh, Ryosuke and Jencson, Jacob E. and Kasliwal, Mansi M. and Kawai, Nobuyuki and Khanam, Tanazza and Kulkarni, Shrinivas R. and Kumar, Brajesh and Kumar, Harsh and Kutyrev, Alexander S. and Lee, William H. and Maeda, Keiichi and Mahabal, Ashish and Murata, Katsuhiro L. and Neill, James D. and Ngeow, Chow-Choong and Penprase, Bryan and Pian, Elena and Quimby, Robert and Ramirez-Ruiz, Enrico and Richer, Michael G. and Román-Zúñiga, Carlos G. and Sahu, D. K. and Srivastav, Shubham and Socia, Quentin and Sollerman, Jesper and Tachibana, Yutaro and Taddia, Francesco and Tinyanont, Samaporn and Troja, Eleonora and Ward, Charlotte and Wee, Jerrick and Yu, Po-Chieh},
	month = mar,
	year = {2019},
	pages = {1031--1049},
}

@article{rees_tidal_1988,
	title = {Tidal disruption of stars by black holes of 10⁶-10⁸ solar masses in nearby galaxies},
	volume = {333},
	doi = {10.1038/333523a0},
	number = {6173},
	journal = {Nature},
	author = {Rees, Martin J.},
	month = jun,
	year = {1988},
	pages = {523--528},
}

@article{seth_supermassive_2014,
	title = {A supermassive black hole in an ultra-compact dwarf galaxy},
	volume = {513},
	doi = {10.1038/nature13762},
	number = {7518},
	journal = {Nature},
	author = {Seth, Anil C. and van den Bosch, Remco and Mieske, Steffen and Baumgardt, Holger and Brok, Mark Den and Strader, Jay and Neumayer, Nadine and Chilingarian, Igor and Hilker, Michael and McDermid, Richard and Spitler, Lee and Brodie, Jean and Frank, Matthias J. and Walsh, Jonelle L.},
	month = sep,
	year = {2014},
	pages = {398--400},
}

@article{stein_tde_2026,
	title = {{TDE} 2025abcr: {A} {Tidal} {Disruption} {Event} in the {Outskirts} of a {Massive} {Galaxy}},
	volume = {1006},
	doi = {10.3847/2041-8213/ae77f3},
	number = {2},
	journal = {The Astrophysical Journal Letters},
	author = {Stein, Robert and Carney, Jonathan and Ward, Charlotte and Margutti, Raffaella and Hall, Xander J. and Sfaradi, Itai and Andreoni, Igor and Charalampopoulos, Panos and Chornock, Ryan and Gezari, Suvi and Mo, Geoffrey and Yao, Yuhan and Anumarlapudi, Akash and Bellm, Eric C. and Bloom, Joshua S. and Busmann, Malte and Caiazzo, Ilaria and Cenko, S. Bradley and Graham, Matthew J. and Groom, Steven L. and Gruen, Daniel and Hammerstein, Erica and Kaiser, Benjamin C. and Kasliwal, Mansi M. and O'Connor, Brendan and Palmese, Antonella and Purdum, Josiah and Rastinejad, Jillian C. and Riddle, Reed and Rusholme, Ben and Sollerman, Jesper and Somalwar, Jean J. and Veilleux, Sylvain},
	month = aug,
	year = {2026},
	pages = {L57},
}

@article{taylor_supermassive_2025,
	title = {A {Supermassive} {Black} {Hole} in a {Diminutive} {Ultracompact} {Dwarf} {Galaxy} {Discovered} with {JWST}/{NIRSpec}+{IFU}},
	volume = {991},
	doi = {10.3847/2041-8213/ae028e},
	number = {1},
	journal = {The Astrophysical Journal Letters},
	author = {Taylor, Matthew A. and Tahmasebzadeh, Behzad and Thompson, Solveig and Vasiliev, Eugene and Valluri, Monica and Drinkwater, Michael J. and Côté, Patrick and Ferrarese, Laura and Roediger, Joel and Baumgardt, Holger and Bentz, Misty C. and Dage, Kristen and Peng, Eric W. and Lapeer, Drew and Liu, Chengze and Sumners, Zach and Wang, Kaixiang and Baldassare, Vivienne and Blakeslee, John P. and Ko, Youkyung and Woods, Tyrone E.},
	month = sep,
	year = {2025},
	pages = {L24},
}

@article{tremmel_romulus_2017,
	title = {The {Romulus} cosmological simulations: a physical approach to the formation, dynamics and accretion models of {SMBHs}},
	volume = {470},
	doi = {10.1093/mnras/stx1160},
	number = {1},
	journal = {Monthly Notices of the Royal Astronomical Society},
	author = {Tremmel, M. and Karcher, M. and Governato, F. and Volonteri, M. and Quinn, T. R. and Pontzen, A. and Anderson, L. and Bellovary, J.},
	month = sep,
	year = {2017},
	pages = {1121--1139},
}

@article{yao_tidal_2023,
	title = {Tidal {Disruption} {Event} {Demographics} with the {Zwicky} {Transient} {Facility}: {Volumetric} {Rates}, {Luminosity} {Function}, and {Implications} for the {Local} {Black} {Hole} {Mass} {Function}},
	volume = {955},
	doi = {10.3847/2041-8213/acf216},
	number = {1},
	journal = {The Astrophysical Journal Letters},
	author = {Yao, Yuhan and Ravi, Vikram and Gezari, Suvi and van Velzen, Sjoert and Lu, Wenbin and Schulze, Steve and Somalwar, Jean J. and Kulkarni, S. R. and Hammerstein, Erica and Nicholl, Matt and Graham, Matthew J. and Perley, Daniel A. and Cenko, S. Bradley and Stein, Robert and Ricarte, Angelo and Chadayammuri, Urmila and Quataert, Eliot and Bellm, Eric C. and Bloom, Joshua S. and Dekany, Richard and Drake, Andrew J. and Groom, Steven L. and Mahabal, Ashish A. and Prince, Thomas A. and Riddle, Reed and Rusholme, Ben and Sharma, Yashvi and Sollerman, Jesper and Yan, Lin},
	month = sep,
	year = {2023},
	pages = {L6},
}

@article{boylan-kolchin_globular_2017,
	title = {The globular cluster-dark matter halo connection},
	volume = {472},
	doi = {10.1093/mnras/stx2164},
	number = {3},
	journal = {Monthly Notices of the Royal Astronomical Society},
	author = {Boylan-Kolchin, Michael},
	month = dec,
	year = {2017},
	pages = {3120--3130},
}

@article{harris_galactic_2017,
	title = {Galactic {Dark} {Matter} {Halos} and {Globular} {Cluster} {Populations}. {III}. {Extension} to {Extreme} {Environments}},
	volume = {836},
	doi = {10.3847/1538-4357/836/1/67},
	number = {1},
	journal = {The Astrophysical Journal},
	author = {Harris, William E. and Blakeslee, John P. and Harris, Gretchen L. H.},
	month = feb,
	year = {2017},
	pages = {67},
}

@article{hudson_correlation_2018,
	title = {The correlation between the sizes of globular cluster systems and their host dark matter haloes},
	volume = {477},
	issn = {0035-8711},
	url = {https://ui.adsabs.harvard.edu/abs/2018MNRAS.477.3869H},
	doi = {10.1093/mnras/sty844},
	urldate = {2026-10-01},
	journal = {Monthly Notices of the Royal Astronomical Society},
	publisher = {OUP},
	author = {Hudson, Michael J. and Robison, Bailey},
	month = jul,
	year = {2018},
	note = {ADS Bibcode: 2018MNRAS.477.3869H},
	pages = {3869--3885},
}

@article{marks_inverse_2012,
	title = {Inverse dynamical population synthesis. {Constraining} the initial conditions of young stellar clusters by studying their binary populations},
	volume = {543},
	issn = {0004-6361},
	url = {https://ui.adsabs.harvard.edu/abs/2012A&A...543A...8M},
	doi = {10.1051/0004-6361/201118231},
	urldate = {2026-10-01},
	journal = {Astronomy and Astrophysics},
	publisher = {EDP},
	author = {Marks, M. and Kroupa, P.},
	month = jul,
	year = {2012},
	note = {ADS Bibcode: 2012A\&A...543A...8M},
	pages = {A8},
}

@article{lim_next_2024,
	title = {The {Next} {Generation} {Virgo} {Cluster} {Survey} ({NGVS}). {XXVII}. {The} {Size} and {Structure} of {Globular} {Cluster} {Systems} and {Their} {Connection} to {Dark} {Matter} {Halos}},
	volume = {966},
	issn = {0004-637X},
	url = {https://ui.adsabs.harvard.edu/abs/2024ApJ...966..168L},
	doi = {10.3847/1538-4357/ad3444},
	urldate = {2026-10-01},
	journal = {The Astrophysical Journal},
	publisher = {IOP},
	author = {Lim, Sungsoon and Peng, Eric W. and Côté, Patrick and Ferrarese, Laura and Roediger, Joel C. and Liu, Chengze and Spengler, Chelsea and Sola, Elisabeth and Duc, Pierre-Alain and Sales, Laura V. and Blakeslee, John P. and Cuillandre, Jean-Charles and Durrell, Patrick R. and Emsellem, Eric and Gwyn, Stephen D. J. and Lançon, Ariane and Marleau, Francine R. and Mihos, J. Christopher and Müller, Oliver and Puzia, Thomas H. and Sánchez-Janssen, Rubén},
	month = may,
	year = {2024},
	note = {ADS Bibcode: 2024ApJ...966..168L},
	pages = {168},
}

@article{jordan_acs_2007,
	title = {The {ACS} {Virgo} {Cluster} {Survey}. {XII}. {The} {Luminosity} {Function} of {Globular} {Clusters} in {Early}-{Type} {Galaxies}},
	volume = {171},
	issn = {0067-0049},
	url = {https://ui.adsabs.harvard.edu/abs/2007ApJS..171..101J},
	doi = {10.1086/516840},
	urldate = {2026-10-01},
	journal = {The Astrophysical Journal Supplement Series},
	publisher = {IOP},
	author = {Jordán, Andrés and McLaughlin, Dean E. and Côté, Patrick and Ferrarese, Laura and Peng, Eric W. and Mei, Simona and Villegas, Daniela and Merritt, David and Tonry, John L. and West, Michael J.},
	month = jul,
	year = {2007},
	note = {ADS Bibcode: 2007ApJS..171..101J},
	pages = {101--145},
}

@article{prugniel_fundamental_1997,
	title = {The fundamental plane of early-type galaxies: non-homology of the spatial structure.},
	volume = {321},
	issn = {0004-6361},
	shorttitle = {The fundamental plane of early-type galaxies},
	url = {https://ui.adsabs.harvard.edu/abs/1997A&A...321..111P},
	urldate = {2026-10-01},
	journal = {Astronomy and Astrophysics},
	publisher = {EDP},
	author = {Prugniel, P. and Simien, F.},
	month = may,
	year = {1997},
	note = {ADS Bibcode: 1997A\&A...321..111P},
	pages = {111--122},
}

@article{lima_neto_specific_1999,
	title = {The specific entropy of elliptical galaxies: an explanation for profile-shape distance indicators?},
	volume = {309},
	issn = {0035-8711},
	shorttitle = {The specific entropy of elliptical galaxies},
	url = {https://ui.adsabs.harvard.edu/abs/1999MNRAS.309..481L},
	doi = {10.1046/j.1365-8711.1999.02849.x},
	urldate = {2026-10-01},
	journal = {Monthly Notices of the Royal Astronomical Society},
	publisher = {OUP},
	author = {Lima Neto, G. B. and Gerbal, D. and Márquez, I.},
	month = oct,
	year = {1999},
	note = {ADS Bibcode: 1999MNRAS.309..481L},
	pages = {481--495},
}

@article{rantala_frost-clusters_2024,
	title = {{FROST}-{CLUSTERS} – {I}. {Hierarchical} star cluster assembly boosts intermediate-mass black hole formation},
	volume = {531},
	issn = {0035-8711},
	url = {https://doi.org/10.1093/mnras/stae1413},
	doi = {10.1093/mnras/stae1413},
	number = {3},
	urldate = {2026-10-02},
	journal = {Monthly Notices of the Royal Astronomical Society},
	author = {Rantala, Antti and Naab, Thorsten and Lahén, Natalia},
	month = jul,
	year = {2024},
	pages = {3770--3799},
}

@article{abbott_gw190521_2020,
	title = {{GW190521}: {A} {Binary} {Black} {Hole} {Merger} with a {Total} {Mass} of 150 {M}{\textbackslash}ensuremathødot},
	volume = {125},
	doi = {10.1103/PhysRevLett.125.101102},
	number = {10},
	journal = {Physical Review Letters},
	author = {Abbott, R. and Abbott, T. D. and Abraham, S. and Acernese, F. and Ackley, K. and Adams, C. and Adhikari, R. X. and Adya, V. B. and Affeldt, C. and Agathos, M. and Agatsuma, K. and Aggarwal, N. and Aguiar, O. D. and Aich, A. and Aiello, L. and Ain, A. and Ajith, P. and Akcay, S. and Allen, G. and Allocca, A. and Altin, P. A. and Amato, A. and Anand, S. and Ananyeva, A. and Anderson, S. B. and Anderson, W. G. and Angelova, S. V. and Ansoldi, S. and Antier, S. and Appert, S. and Arai, K. and Araya, M. C. and Areeda, J. S. and Arène, M. and Arnaud, N. and Aronson, S. M. and Arun, K. G. and Asali, Y. and Ascenzi, S. and Ashton, G. and Aston, S. M. and Astone, P. and Aubin, F. and Aufmuth, P. and AultONeal, K. and Austin, C. and Avendano, V. and Babak, S. and Bacon, P. and Badaracco, F. and Bader, M. K. M. and Bae, S. and Baer, A. M. and Baird, J. and Baldaccini, F. and Ballardin, G. and Ballmer, S. W. and Bals, A. and Balsamo, A. and Baltus, G. and Banagiri, S. and Bankar, D. and Bankar, R. S. and Barayoga, J. C. and Barbieri, C. and Barish, B. C. and Barker, D. and Barkett, K. and Barneo, P. and Barone, F. and Barr, B. and Barsotti, L. and Barsuglia, M. and Barta, D. and Bartlett, J. and Bartos, I. and Bassiri, R. and Basti, A. and Bawaj, M. and Bayley, J. C. and Bazzan, M. and Bécsy, B. and Bejger, M. and Belahcene, I. and Bell, A. S. and Beniwal, D. and Benjamin, M. G. and Bentley, J. D. and Bergamin, F. and Berger, B. K. and Bergmann, G. and Bernuzzi, S. and Berry, C. P. L. and Bersanetti, D. and Bertolini, A. and Betzwieser, J. and Bhandare, R. and Bhandari, A. V. and Bidler, J. and Biggs, E. and Bilenko, I. A. and Billingsley, G. and Birney, R. and Birnholtz, O. and Biscans, S. and Bischi, M. and Biscoveanu, S. and Bisht, A. and Bissenbayeva, G. and Bitossi, M. and Bizouard, M. A. and Blackburn, J. K. and Blackman, J. and Blair, C. D. and Blair, D. G. and Blair, R. M. and Bobba, F. and Bode, N. and Boer, M. and Boetzel, Y. and Bogaert, G. and Bondu, F. and Bonilla, E. and Bonnand, R. and Booker, P. and Boom, B. A. and Bork, R. and Boschi, V. and Bose, S. and Bossilkov, V. and Bosveld, J. and Bouffanais, Y. and Bozzi, A. and Bradaschia, C. and Brady, P. R. and Bramley, A. and Branchesi, M. and Brau, J. E. and Breschi, M. and Briant, T. and Briggs, J. H. and Brighenti, F. and Brillet, A. and Brinkmann, M. and Brockill, P. and Brooks, A. F. and Brooks, J. and Brown, D. D. and Brunett, S. and Bruno, G. and Bruntz, R. and Buikema, A. and Bulik, T. and Bulten, H. J. and Buonanno, A. and Buscicchio, R. and Buskulic, D. and Byer, R. L. and Cabero, M. and Cadonati, L. and Cagnoli, G. and Cahillane, C. and Calderón Bustillo, J. and Callaghan, J. D. and Callister, T. A. and Calloni, E. and Camp, J. B. and Canepa, M. and Cannon, K. C. and Cao, H. and Cao, J. and Carapella, G. and Carbognani, F. and Caride, S. and Carney, M. F. and Carullo, G. and Casanueva Diaz, J. and Casentini, C. and Castañeda, J. and Caudill, S. and Cavaglià, M. and Cavalier, F. and Cavalieri, R. and Cella, G. and Cerdá-Durán, P. and Cesarini, E. and Chaibi, O. and Chakravarti, K. and Chan, C. and Chan, M. and Chandra, K. and Chao, S. and Charlton, P. and Chase, E. A. and Chassande-Mottin, E. and Chatterjee, D. and Chaturvedi, M. and Chatziioannou, K. and Chen, H. Y. and Chen, X.},
	month = sep,
	year = {2020},
	pages = {101102},
}

@article{campanelli_large_2007,
	title = {Large {Merger} {Recoils} and {Spin} {Flips} from {Generic} {Black} {Hole} {Binaries}},
	volume = {659},
	doi = {10.1086/516712},
	number = {1},
	journal = {The Astrophysical Journal Letters},
	author = {Campanelli, Manuela and Lousto, Carlos and Zlochower, Yosef and Merritt, David},
	month = apr,
	year = {2007},
	pages = {L5--L8},
}

@article{holley-bockelmann_gravitational_2008,
	title = {Gravitational {Wave} {Recoil} and the {Retention} of {Intermediate}-{Mass} {Black} {Holes}},
	volume = {686},
	doi = {10.1086/591218},
	number = {2},
	journal = {The Astrophysical Journal},
	author = {Holley-Bockelmann, Kelly and Gültekin, Kayhan and Shoemaker, Deirdre and Yunes, Nicolas},
	month = oct,
	year = {2008},
	pages = {829--837},
}

@article{lin_luminous_2018,
	title = {A luminous {X}-ray outburst from an intermediate-mass black hole in an off-centre star cluster},
	volume = {2},
	doi = {10.1038/s41550-018-0493-1},
	journal = {Nature Astronomy},
	author = {Lin, Dacheng and Strader, Jay and Carrasco, Eleazar R. and Page, Dany and Romanowsky, Aaron J. and Homan, Jeroen and Irwin, Jimmy A. and Remillard, Ronald A. and Godet, Olivier and Webb, Natalie A. and Baumgardt, Holger and Wijnands, Rudy and Barret, Didier and Duc, Pierre-Alain and Brodie, Jean P. and Gwyn, Stephen D. J.},
	month = jun,
	year = {2018},
	pages = {656--661},
}

@ARTICLE{baumgardt_massive_2004,
       author = {{Baumgardt}, Holger and {Makino}, Junichiro and {Ebisuzaki}, Toshikazu},
        title = "{Massive Black Holes in Star Clusters. I. Equal-Mass Clusters}",
      journal = {The Astrophysical Journal},
         year = 2004,
        month = oct,
       volume = {613},
       number = {2},
        pages = {1133-1142},
          doi = {10.1086/423298},
archivePrefix = {arXiv},
       eprint = {astro-ph/0406227},
 primaryClass = {astro-ph},
       adsurl = {https://ui.adsabs.harvard.edu/abs/2004ApJ...613.1133B}
}

@article{peres_classical_1962,
	title = {Classical {Radiation} {Recoil}},
	volume = {128},
	copyright = {http://link.aps.org/licenses/aps-default-license},
	issn = {0031-899X},
	url = {https://link.aps.org/doi/10.1103/PhysRev.128.2471},
	doi = {10.1103/PhysRev.128.2471},
	language = {en},
	number = {5},
	urldate = {2026-10-10},
	journal = {Physical Review},
	author = {Peres, Asher},
	month = dec,
	year = {1962},
	pages = {2471--2475},
}

@article{bekenstein_gravitational-radiation_1973,
	title = {Gravitational-{Radiation} {Recoil} and {Runaway} {Black} {Holes}},
	volume = {183},
	issn = {0004-637X, 1538-4357},
	url = {http://adsabs.harvard.edu/doi/10.1086/152255},
	doi = {10.1086/152255},
	language = {en},
	urldate = {2026-10-10},
	journal = {The Astrophysical Journal},
	author = {Bekenstein, Jacob D.},
	month = jul,
	year = {1973},
	pages = {657},
}

@article{fitchett_influence_1983,
	title = {The influence of gravitational wave momentum losses on the centre of mass motion of a {Newtonian} binary system},
	volume = {203},
	issn = {0035-8711, 1365-2966},
	url = {https://academic.oup.com/mnras/article-lookup/doi/10.1093/mnras/203.4.1049},
	doi = {10.1093/mnras/203.4.1049},
	language = {en},
	number = {4},
	urldate = {2026-10-10},
	journal = {Monthly Notices of the Royal Astronomical Society},
	author = {Fitchett, M. J.},
	month = aug,
	year = {1983},
	pages = {1049--1062},
}

@article{lousto_hangup_2011,
	title = {Hangup {Kicks}: {Still} {Larger} {Recoils} by {Partial} {Spin}-{Orbit} {Alignment} of {Black}-{Hole} {Binaries}},
	volume = {107},
	copyright = {http://link.aps.org/licenses/aps-default-license},
	issn = {0031-9007, 1079-7114},
	shorttitle = {Hangup {Kicks}},
	url = {https://link.aps.org/doi/10.1103/PhysRevLett.107.231102},
	doi = {10.1103/PhysRevLett.107.231102},
	language = {en},
	number = {23},
	urldate = {2026-10-10},
	journal = {Physical Review Letters},
	author = {Lousto, Carlos O. and Zlochower, Yosef},
	month = dec,
	year = {2011},
	pages = {231102},
}

@article{nicholl_at_2023,
	title = {{AT} 2022aedm and a {New} {Class} of {Luminous}, {Fast}-cooling {Transients} in {Elliptical} {Galaxies}},
	volume = {954},
	issn = {2041-8205, 2041-8213},
	url = {https://iopscience.iop.org/article/10.3847/2041-8213/acf0ba},
	doi = {10.3847/2041-8213/acf0ba},
	number = {1},
	urldate = {2026-10-12},
	journal = {The Astrophysical Journal Letters},
	author = {Nicholl, M. and Srivastav, S. and Fulton, M. D. and Gomez, S. and Huber, M. E. and Oates, S. R. and Ramsden, P. and Rhodes, L. and Smartt, S. J. and Smith, K. W. and Aamer, A. and Anderson, J. P. and Bauer, F. E. and Berger, E. and De Boer, T. and Chambers, K. C. and Charalampopoulos, P. and Chen, T.-W. and Fender, R. P. and Fraser, M. and Gao, H. and Green, D. A. and Galbany, L. and Gompertz, B. P. and Gromadzki, M. and Gutiérrez, C. P. and Howell, D. A. and Inserra, C. and Jonker, P. G. and Kopsacheili, M. and Lowe, T. B. and Magnier, E. A. and McCully, C. and McGee, S. L. and Moore, T. and Müller-Bravo, T. E. and Newsome, M. and Gonzalez, E. Padilla and Pellegrino, C. and Pessi, T. and Pursiainen, M. and Rest, A. and Ridley, E. J. and Shappee, B. J. and Sheng, X. and Smith, G. P. and Terreran, G. and Tucker, M. A. and Vinkó, J. and Wainscoat, R. J. and Wiseman, P. and Young, D. R.},
	month = sep,
	year = {2023},
	pages = {L28},
}

@article{vinko_luminous_2014,
	title = {A {LUMINOUS}, {FAST} {RISING} {UV}-{TRANSIENT} {DISCOVERED} {BY} {ROTSE}: {A} {TIDAL} {DISRUPTION} {EVENT}?},
	volume = {798},
	copyright = {http://iopscience.iop.org/info/page/text-and-data-mining},
	issn = {1538-4357},
	shorttitle = {A {LUMINOUS}, {FAST} {RISING} {UV}-{TRANSIENT} {DISCOVERED} {BY} {ROTSE}},
	url = {https://iopscience.iop.org/article/10.1088/0004-637X/798/1/12},
	doi = {10.1088/0004-637X/798/1/12},
	number = {1},
	urldate = {2026-10-12},
	journal = {The Astrophysical Journal},
	author = {Vinkó, J. and Yuan, F. and Quimby, R. M. and Wheeler, J. C. and Ramirez-Ruiz, E. and Guillochon, J. and Chatzopoulos, E. and Marion, G. H. and Akerlof, C.},
	month = dec,
	year = {2014},
	pages = {12},
}

@ARTICLE{DiMatteo2005,
       author = {{Di Matteo}, Tiziana and {Springel}, Volker and {Hernquist}, Lars},
        title = "{Energy input from quasars regulates the growth and activity of black holes and their host galaxies}",
      journal = {Nature},
         year = 2005,
        month = feb,
       volume = {433},
       number = {7026},
        pages = {604-607},
          doi = {10.1038/nature03335},
archivePrefix = {arXiv},
       eprint = {astro-ph/0502199},
 primaryClass = {astro-ph},
       adsurl = {https://ui.adsabs.harvard.edu/abs/2005Natur.433..604D}
}

@ARTICLE{Weinberger2017,
       author = {{Weinberger}, Rainer and {Springel}, Volker and {Hernquist}, Lars and {Pillepich}, Annalisa and {Marinacci}, Federico and {Pakmor}, R{\"u}diger and {Nelson}, Dylan and {Genel}, Shy and {Vogelsberger}, Mark and {Naiman}, Jill and {Torrey}, Paul},
        title = "{Simulating galaxy formation with black hole driven thermal and kinetic feedback}",
      journal = {Monthly Notices of the Royal Astronomical Society},
         year = 2017,
        month = mar,
       volume = {465},
       number = {3},
        pages = {3291-3308},
          doi = {10.1093/mnras/stw2944},
archivePrefix = {arXiv},
       eprint = {1607.03486},
 primaryClass = {astro-ph.GA},
       adsurl = {https://ui.adsabs.harvard.edu/abs/2017MNRAS.465.3291W}
}

@article{dimatteo_vast_2023,
	title = {A vast population of wandering and merging {IMBHs} at cosmic noon},
	volume = {525},
	copyright = {https://academic.oup.com/journals/pages/open\_access/funder\_policies/chorus/standard\_publication\_model},
	issn = {0035-8711, 1365-2966},
	url = {https://academic.oup.com/mnras/article/525/1/1479/7243407},
	doi = {10.1093/mnras/stad2198},
	language = {en},
	number = {1},
	urldate = {2026-10-12},
	journal = {Monthly Notices of the Royal Astronomical Society},
	author = {Di Matteo, Tiziana and Ni, Yueying and Chen, Nianyi and Croft, Rupert and Bird, Simeon and Pacucci, Fabio and Ricarte, Angelo and Tremmel, Michael},
	month = aug,
	year = {2023},
	pages = {1479--1497},
}

@misc{zhou_acorn_2026,
	title = {{ACORN} {I}. {Massive} {Black} {Hole} {Seeding} and {Tidal} {Disruption} {Events} from {Star} {Clusters} in {Cosmological} {Simulations}},
	copyright = {Creative Commons Attribution 4.0 International},
	url = {https://arxiv.org/abs/2610.12398},
	doi = {10.48550/ARXIV.2610.12398},
	urldate = {2026-10-12},
	publisher = {arXiv},
	author = {Zhou, Yihao and Di Matteo, Tiziana and Williams, Claire E. and Bhowmick, Aklant Kumar and Reina-Campos, Marta and Rodriguez, Carl L. and Prieto, Elena González and Blecha, Laura and Croft, Rupert},
	month = oct,
	year = {2026},
	note = {Version Number: 1},
}
\bibliographystyle{aasjournalv7}

\appendix

\section{Ancillary Data}
\begin{figure*}
    \centering
    \includegraphics[width=0.9\linewidth]{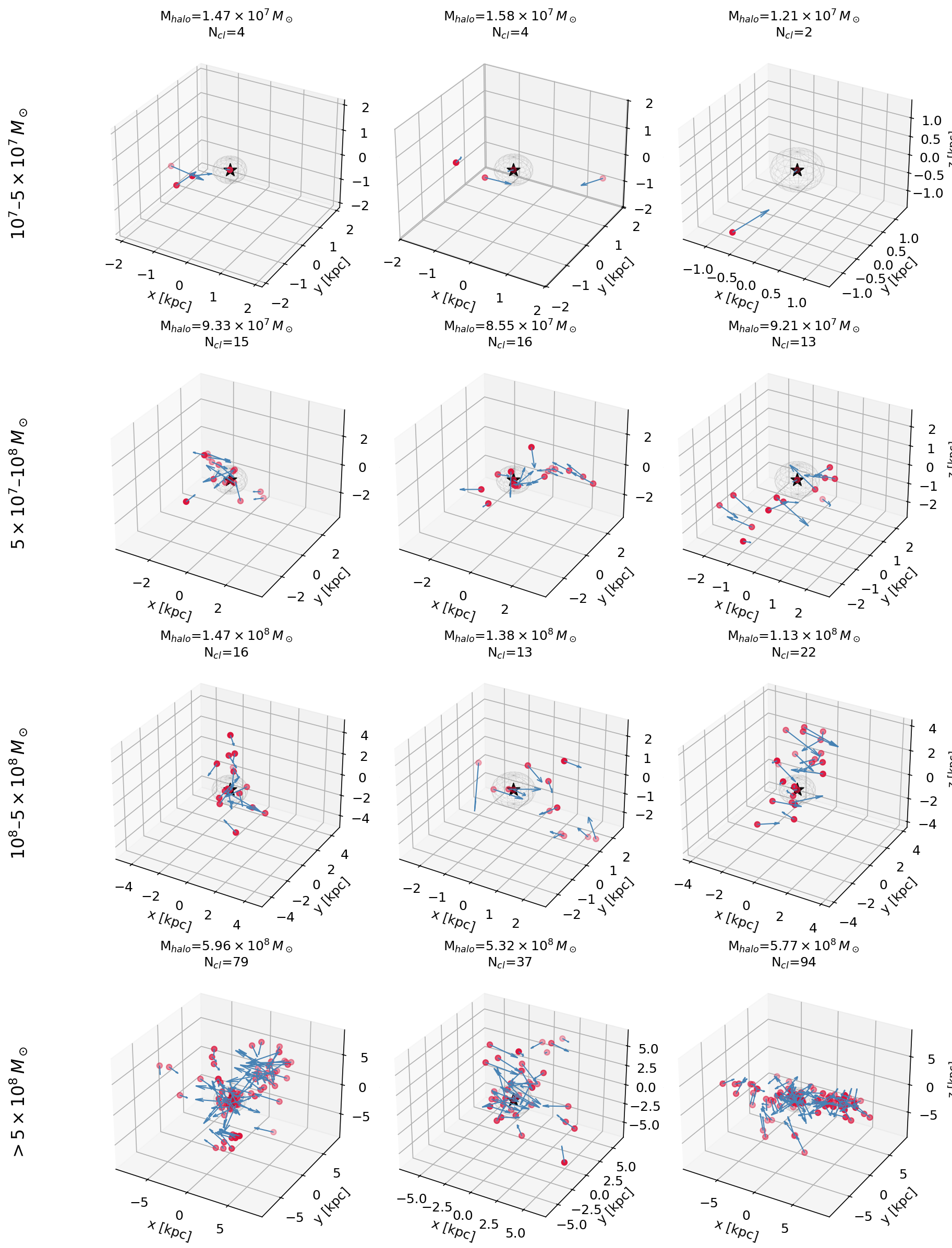}
    \caption{Three-dimensional distributions of star clusters around twelve example dark matter halos. Each row is a halo mass bin, labeled at left, and each column is a halo drawn at random from that bin among halos with at least one associated cluster. In each panel, the black star marks the halo center and the red points are the positions of the halo's associated clusters relative to the center, in physical kpc. The gray sphere shows the halo's virial radius, $R_{200}$. Blue arrows give each cluster's velocity relative to the halo. Each panel shows the halo mass and the number of associated clusters, $N_{\rm cl}$. Arrow lengths are scaled independently in each panel so that an arrow of the median velocity spans 30\% of the largest cluster–halo separation. 
    }
    \label{fig:example_halocluster}
\end{figure*}

\
\label{app:data}
\autoref{tab:ellipticals} gives the properties of TNG50 halos used in this work. 
\autoref{fig:example_halocluster} shows example clusters and halos in the high resolution cosmological simulation, from which the star cluster calibration catalog is compiled. 
Depicted are twelve example halos in the catalog (with centers of mass labeled with red stars) and the nearby clusters within $5R_{\rm vir}$. The clusters velocity and position are shown with respect to the halo (see caption) and the halo's virial radius is shown in grey. \autoref{tab:astrid_population} summarizes the ASTRID $z=0$ galaxies used in the analysis. 
\begin{table*}
\caption{Properties of the fifteen selected TNG50-1 galaxies at $z=0$: the five 
most dispersion-dominated galaxies (lowest $\kappa_{\rm rot}$) in each of three 
stellar-mass bins whose TNG50-1-Dark merger trees pass the checks described in 
\autoref{sec:methods}, ranked by $\kappa_{\rm rot}$ within each bin. $M_\star$ is the 
total bound stellar mass and SFR the instantaneous star formation rate of the 
subhalo; an SFR of 0 indicates no star-forming gas. $\kappa_{\rm rot}$, 
$f(\epsilon>0.7)$ and $f_{\rm sph}$ are computed from star particles within 30~kpc 
of the subhalo centre; $f_{\rm sph}$ is twice the mass fraction of stars with 
$\epsilon<0$. $M_{\rm DM}$ is the $z=0$ total mass of the matched subhalo in 
TNG50-1-Dark. All galaxies have $\kappa_{\rm rot}<0.5$ except TNG50-1 738736 
($\kappa_{\rm rot}=0.503$), which is marginally rotation-supported.}
\label{tab:ellipticals}
\begin{tabular}{rrcccccc}
\hline
TNG50-1 ID & TNG50-1-Dark ID & $M_\star$ & SFR & $\kappa_{\rm rot}$ & 
$f(\epsilon>0.7)$ & $f_{\rm sph}$ & $M_{\rm DM}$ \\
 & & ($10^{11}\,M_\odot$) & ($M_\odot\,{\rm yr}^{-1}$) & & & & ($10^{12}\,M_\odot$) \\
\hline
\multicolumn{8}{l}{$M_\star > 10^{11}\,M_\odot$} \\
434356 & 685512 & 1.32 & $7.5\times10^{-3}$ & 0.295 & 0.043 & 0.81 & 3.50 \\
450916 & 697044 & 1.31 & 0                  & 0.402 & 0.106 & 0.97 & 3.06 \\
379803 & 588075 & 1.75 & $1.2\times10^{-4}$ & 0.402 & 0.158 & 0.56 & 6.73 \\
418335 & 665702 & 1.85 & $4.9\times10^{-3}$ & 0.406 & 0.129 & 0.58 & 4.45 \\
447914 & 670354 & 1.45 & 0                  & 0.463 & 0.233 & 0.46 & 4.39 \\
\hline
\multicolumn{8}{l}{$10^{10} < M_\star/M_\odot < 10^{11}$} \\
509709 & 801308 & 0.736 & $4.0\times10^{-3}$ & 0.373 & 0.109 & 0.76 & 1.77 \\
495451 & 753345 & 0.697 & 0                  & 0.402 & 0.118 & 0.63 & 2.34 \\
481428 & 745415 & 0.831 & 0                  & 0.439 & 0.186 & 0.60 & 2.36 \\
513105 & 826784 & 0.456 & $9.5\times10^{-2}$ & 0.445 & 0.203 & 0.81 & 1.25 \\
524506 & 802209 & 0.457 & 0                  & 0.453 & 0.228 & 0.51 & 1.81 \\
\hline
\multicolumn{8}{l}{$10^{9} < M_\star/M_\odot < 10^{10}$} \\
799553 & 939095  & 0.0480 & $3.3\times10^{-3}$ & 0.350 & 0.085 & 0.64 & 0.531 \\
750258 & 1166615 & 0.0211 & $5.3\times10^{-2}$ & 0.421 & 0.131 & 0.79 & 0.0750 \\
715352 & 1110727 & 0.0201 & $9.5\times10^{-2}$ & 0.451 & 0.227 & 0.66 & 0.126 \\
711292 & 1099576 & 0.0217 & $8.6\times10^{-2}$ & 0.458 & 0.167 & 0.71 & 0.139 \\
738736 & 1136724 & 0.0360 & $5.4\times10^{-2}$ & 0.503 & 0.167 & 0.56 & 0.103 \\
\hline
\end{tabular}
\end{table*}

\clearpage

\movetabledown=4.5cm
\begin{table}
\begin{rotatetable}
\centering
\footnotesize
\setlength{\tabcolsep}{6pt}
\caption{Properties of the ASTRID $z=0$ quenched, dispersion-dominated centrals ($\kappa_{\rm rot}<0.5$) in the three stellar-mass bins: median, $1\sigma$ (16--84\%), $2\sigma$ (2.3--97.7\%) and $3\sigma$ (0.13--99.87\%) percentile ranges, and the full range.}
\label{tab:astrid_population}
\begin{tabular}{llrccccc}
\toprule
Parameter & $M_\star$ bin [$M_\odot$] & $N$ & median & $1\sigma$ & $2\sigma$ & $3\sigma$ & min--max \\
\midrule
$M_\star$ [$M_\odot$] & $10^{9}$--$10^{10}$ & 23540 & $1.34\times10^{9}$ & $1.07\times10^{9}$--$2.28\times10^{9}$ & $1.01\times10^{9}$--$5.50\times10^{9}$ & $1.00\times10^{9}$--$9.43\times10^{9}$ & $1.00\times10^{9}$--$9.97\times10^{9}$ \\
 & $10^{10}$--$10^{11}$ & 763 & $2.62\times10^{10}$ & $1.28\times10^{10}$--$7.55\times10^{10}$ & $1.03\times10^{10}$--$9.61\times10^{10}$ & $1.01\times10^{10}$--$9.99\times10^{10}$ & $1.00\times10^{10}$--$9.99\times10^{10}$ \\
 & $>10^{11}$ & 611 & $1.58\times10^{11}$ & $1.15\times10^{11}$--$2.24\times10^{11}$ & $1.04\times10^{11}$--$3.13\times10^{11}$ & $1.00\times10^{11}$--$3.98\times10^{11}$ & $1.00\times10^{11}$--$4.27\times10^{11}$ \\
\addlinespace[10pt]
SFR [$M_\odot\,{\rm yr^{-1}}$] & $10^{9}$--$10^{10}$ & 23540 & $0.048$ & $0$--$0.087$ & $0$--$0.0982$ & $0$--$0.0999$ & $0$--$0.1$ \\
 & $10^{10}$--$10^{11}$ & 763 & $0$ & $0$--$0.0435$ & $0$--$0.0891$ & $0$--$0.0992$ & $0$--$0.0999$ \\
 & $>10^{11}$ & 611 & $0.0313$ & $2.84\times10^{-3}$--$0.0761$ & $0$--$0.0953$ & $0$--$0.0987$ & $0$--$0.0995$ \\
\addlinespace[10pt]
$\kappa_{\rm rot}$ & $10^{9}$--$10^{10}$ & 23540 & $0.364$ & $0.306$--$0.445$ & $0.266$--$0.491$ & $0.223$--$0.499$ & $0.18$--$0.5$ \\
 & $10^{10}$--$10^{11}$ & 763 & $0.406$ & $0.33$--$0.466$ & $0.277$--$0.496$ & $0.22$--$0.499$ & $0.215$--$0.5$ \\
 & $>10^{11}$ & 611 & $0.409$ & $0.331$--$0.467$ & $0.248$--$0.495$ & $0.21$--$0.499$ & $0.194$--$0.5$ \\
\addlinespace[10pt]
$f(\epsilon>0.7)$ & $10^{9}$--$10^{10}$ & 23540 & $0.0914$ & $0.0442$--$0.187$ & $0.0285$--$0.249$ & $0.0151$--$0.286$ & $7.60\times10^{-3}$--$0.316$ \\
 & $10^{10}$--$10^{11}$ & 763 & $0.143$ & $0.0599$--$0.237$ & $0.03$--$0.314$ & $0.0168$--$0.363$ & $0.0122$--$0.388$ \\
 & $>10^{11}$ & 611 & $0.216$ & $0.092$--$0.318$ & $0.0251$--$0.375$ & $0.0146$--$0.406$ & $0.0141$--$0.408$ \\
\addlinespace[10pt]
$\sigma_{\star,\rm 1D}$ [km s$^{-1}$] & $10^{9}$--$10^{10}$ & 23540 & $35.5$ & $32.6$--$39.7$ & $29.8$--$48.2$ & $25.7$--$58.2$ & $19.4$--$66.9$ \\
 & $10^{10}$--$10^{11}$ & 763 & $75$ & $59.3$--$120$ & $54$--$134$ & $50.5$--$143$ & $50.2$--$148$ \\
 & $>10^{11}$ & 611 & $162$ & $143$--$185$ & $135$--$210$ & $126$--$227$ & $124$--$227$ \\
\addlinespace[10pt]
$R_{1/2,\star}$ [kpc] & $10^{9}$--$10^{10}$ & 23540 & $3.32$ & $2.81$--$4.17$ & $2.47$--$5.69$ & $2.21$--$8.01$ & $2.02$--$10.2$ \\
 & $10^{10}$--$10^{11}$ & 763 & $2.62$ & $2.2$--$3.26$ & $1.98$--$5.27$ & $1.89$--$8.33$ & $1.82$--$8.83$ \\
 & $>10^{11}$ & 611 & $3.25$ & $2.77$--$3.73$ & $2.5$--$4.84$ & $2.29$--$5.95$ & $2.21$--$6.15$ \\
\addlinespace[10pt]
$M_{\rm sub}$ [$M_\odot$] & $10^{9}$--$10^{10}$ & 23540 & $8.38\times10^{10}$ & $5.20\times10^{10}$--$1.13\times10^{11}$ & $2.57\times10^{10}$--$1.65\times10^{11}$ & $9.38\times10^{9}$--$2.69\times10^{11}$ & $4.80\times10^{9}$--$6.06\times10^{11}$ \\
 & $10^{10}$--$10^{11}$ & 763 & $7.89\times10^{11}$ & $2.40\times10^{11}$--$4.09\times10^{12}$ & $1.44\times10^{11}$--$7.77\times10^{12}$ & $6.63\times10^{10}$--$1.10\times10^{13}$ & $4.81\times10^{10}$--$1.90\times10^{13}$ \\
 & $>10^{11}$ & 611 & $1.10\times10^{13}$ & $6.81\times10^{12}$--$1.59\times10^{13}$ & $4.75\times10^{12}$--$2.19\times10^{13}$ & $3.60\times10^{12}$--$2.85\times10^{13}$ & $2.99\times10^{12}$--$3.02\times10^{13}$ \\
\addlinespace[10pt]
$M_{\rm FOF}$ [$M_\odot$] & $10^{9}$--$10^{10}$ & 23540 & $9.17\times10^{10}$ & $6.13\times10^{10}$--$1.30\times10^{11}$ & $3.00\times10^{10}$--$2.20\times10^{11}$ & $1.07\times10^{10}$--$4.88\times10^{11}$ & $5.77\times10^{9}$--$1.98\times10^{12}$ \\
 & $10^{10}$--$10^{11}$ & 763 & $9.16\times10^{11}$ & $2.93\times10^{11}$--$4.54\times10^{12}$ & $1.67\times10^{11}$--$1.02\times10^{13}$ & $6.86\times10^{10}$--$1.94\times10^{13}$ & $4.93\times10^{10}$--$2.78\times10^{13}$ \\
 & $>10^{11}$ & 611 & $1.28\times10^{13}$ & $7.65\times10^{12}$--$1.95\times10^{13}$ & $5.01\times10^{12}$--$2.91\times10^{13}$ & $3.71\times10^{12}$--$5.41\times10^{13}$ & $3.05\times10^{12}$--$6.85\times10^{13}$ \\
\addlinespace[10pt]
$N_{\rm BH}$ & $10^{9}$--$10^{10}$ & 23540 & $1$ & $1$--$2$ & $1$--$3$ & $1$--$5$ & $0$--$8$ \\
 & $10^{10}$--$10^{11}$ & 763 & $6$ & $2$--$27$ & $1$--$54$ & $1$--$87.3$ & $1$--$115$ \\
 & $>10^{11}$ & 611 & $70$ & $44$--$103$ & $30$--$141$ & $23.8$--$185$ & $23$--$206$ \\
\addlinespace[10pt]
$M_{\rm BH,max}$ [$M_\odot$] & $10^{9}$--$10^{10}$ & 23516 & $1.13\times10^{7}$ & $4.70\times10^{6}$--$1.97\times10^{7}$ & $9.85\times10^{5}$--$3.40\times10^{7}$ & $9.12\times10^{4}$--$6.16\times10^{7}$ & $4.70\times10^{4}$--$9.64\times10^{7}$ \\
 & $10^{10}$--$10^{11}$ & 763 & $2.59\times10^{8}$ & $4.21\times10^{7}$--$7.16\times10^{8}$ & $8.99\times10^{6}$--$1.13\times10^{9}$ & $2.86\times10^{5}$--$1.31\times10^{9}$ & $2.12\times10^{5}$--$1.39\times10^{9}$ \\
 & $>10^{11}$ & 611 & $1.24\times10^{9}$ & $9.19\times10^{8}$--$1.48\times10^{9}$ & $7.08\times10^{8}$--$1.73\times10^{9}$ & $5.54\times10^{8}$--$2.40\times10^{9}$ & $4.54\times10^{8}$--$2.52\times10^{9}$ \\
\addlinespace[10pt]
$N_{\rm wand}$ & $10^{9}$--$10^{10}$ & 23540 & $0$ & $0$--$1$ & $0$--$2$ & $0$--$4$ & $0$--$7$ \\
 & $10^{10}$--$10^{11}$ & 763 & $5$ & $1$--$26$ & $0$--$53$ & $0$--$86.3$ & $0$--$114$ \\
 & $>10^{11}$ & 611 & $69$ & $43$--$102$ & $29$--$140$ & $22.8$--$184$ & $22$--$205$ \\
\addlinespace[10pt]
$N_{\rm wand}\,(r<30\,{\rm kpc})$ & $10^{9}$--$10^{10}$ & 23540 & $0$ & $0$--$1$ & $0$--$1$ & $0$--$3$ & $0$--$5$ \\
 & $10^{10}$--$10^{11}$ & 763 & $2$ & $0$--$5$ & $0$--$9$ & $0$--$14$ & $0$--$14$ \\
 & $>10^{11}$ & 611 & $8$ & $5$--$11$ & $3$--$16$ & $1.79$--$18.4$ & $1$--$20$ \\
\bottomrule
\end{tabular}
\end{rotatetable}
\end{table}
\clearpage

\section{Impact of model choices on TDE rates}
\label{app:modelchoices}

\begin{figure}
    \centering
    \includegraphics[width=0.95\linewidth]{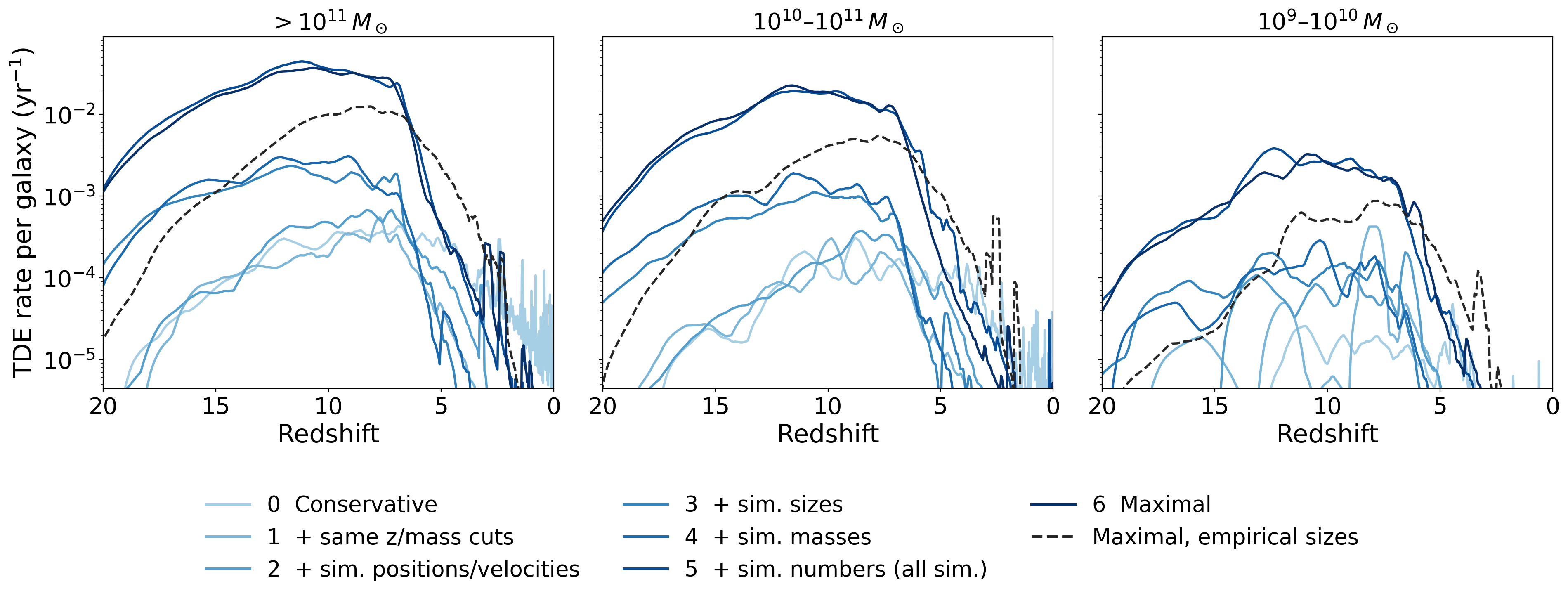}
    \caption{Average TDE rate per galaxy as a function of redshift, as the 
    conservative model (Section~\ref{sec:obs_sampler}) is converted into the 
    maximal model (Section~\ref{sec:cosmoseeding}) one ingredient at a time. 
    Each panel is one $z=0$ stellar-mass bin (\autoref{tab:ellipticals}). Each 
    line is the mean over the bin's five galaxies of the rate per galaxy, 
    smoothed over 50~Myr, and includes all TDEs regardless of where they occur. 
    Line colour runs from light to dark blue as ingredients are switched, 
    cumulatively, from the empirical relations to the cluster-formation 
    simulation: 
    (0)~the conservative model; 
    (1)~the same cuts on host-halo mass and formation redshift as the maximal 
    model; 
    (2)~initial positions and velocities relative to the host taken from the 
    simulation in place of Equation~\eqref{eq:ps}; 
    (3)~half-mass radii from the simulation, with 0.1~dex log-normal scatter, in 
    place of Equation~\eqref{eq:mk12}; 
    (4)~cluster masses from the simulation in place of Equation~\eqref{eq:gclf}; 
    (5)~the number of clusters per host from the simulation in place of 
    Equation~\eqref{eq:ncl}, so that every ingredient comes from the simulation; 
    and (6)~the maximal model. Steps~5 and~6 differ only in how the clusters are 
    drawn, so they should agree to within sampling noise. The dashed grey line is 
    the reverse test: the maximal model with the empirical radii of 
    Equation~\eqref{eq:mk12}. }
    \label{fig:model_compare}
\end{figure}

\begin{figure}
    \centering
    \includegraphics[width=0.95\linewidth]{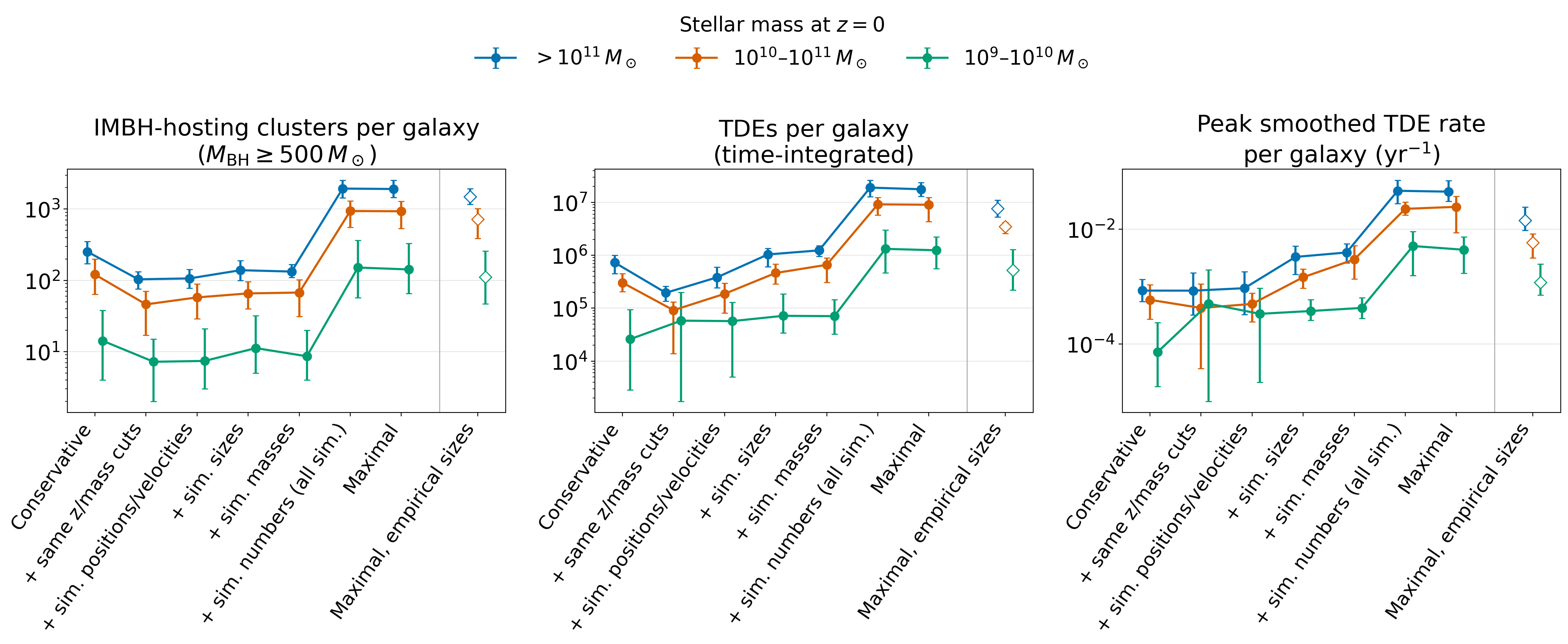}
    \caption{Summary of the model sequence in \autoref{fig:model_compare}. From 
    left to right, the panels show three quantities per galaxy as a function of 
    model step: the number of clusters that host an IMBH of 
    $M_{\rm BH}\geq500\,M_\odot$, the time-integrated number of TDEs, and the 
    peak of the 50~Myr-smoothed TDE rate. Colours give the $z=0$ stellar-mass 
    bin. Points are the mean over the five galaxies in each bin, and error bars 
    span the full range among them, so they show halo-to-halo variation rather 
    than a statistical uncertainty. Points within a step are offset horizontally 
    for clarity. Open diamonds, to the right of the vertical line, show the 
    reverse test (the maximal model with the empirical radii of 
    Equation~\eqref{eq:mk12}).}
    \label{fig:model_summary}
\end{figure}

The conservative and maximal models differ in every ingredient of the cluster
population: the number of clusters per host, their masses, half-mass radii and
initial phase-space coordinates, and the cuts on host-halo mass and formation
redshift. To find out which of these differences sets the gap between the two
models' TDE rates, we convert the conservative model into the maximal model
one ingredient at a time, keeping each change in the steps that follow. The
steps are:
\begin{enumerate}
\setcounter{enumi}{-1}
  \item the conservative model (Section~\ref{sec:obs_sampler});
  \item the same cuts on host-halo mass and cluster formation redshift as the
        maximal model (Section~\ref{sec:cosmoseeding});
  \item initial positions and velocities relative to the host drawn from the
        cluster-formation simulation instead of Equation~\eqref{eq:ps};
  \item half-mass radii drawn from simulated clusters of similar mass, with
        0.1~dex log-normal scatter, instead of Equation~\eqref{eq:mk12};
  \item cluster masses drawn from the simulation instead of
        Equation~\eqref{eq:gclf};
  \item the number of clusters per host drawn from the simulation instead of
        Equation~\eqref{eq:ncl}, so that every ingredient now comes from the
        simulation; and
  \item the maximal model itself.
\end{enumerate}
Steps~5 and~6 use the same ingredients and differ only in the code path that draws them. 
We also run the
maximal model with the empirical radii of Equation~\eqref{eq:mk12} (the
``reverse test''). 
Every run uses all fifteen galaxies of
\autoref{tab:ellipticals}, the same orbit integration, and no cluster orbit was discarded in any run.
\autoref{fig:model_compare} shows the resulting TDE rates per galaxy as a
function of redshift, \autoref{fig:model_summary} summarizes them, and
\autoref{tab:ladder} lists the numbers.

\begin{table*}
\caption{Cluster populations and TDEs per galaxy along the model sequence of
\autoref{fig:model_compare}, averaged over the five galaxies in each $z=0$
stellar-mass bin. $N_{\rm cl}$ is the number of clusters seeded over cosmic
time, $f_{500}$ the fraction of them that form an IMBH with
$M_{\rm BH}\geq500\,M_\odot$, $N_{\rm TDE}$ the time-integrated number of TDEs,
and $\dot N_{\rm peak}$ the peak of the 50~Myr-smoothed TDE rate.}
\label{tab:ladder}
\begin{tabular}{lcccccc}
\hline
 & \multicolumn{4}{c}{$M_\star>10^{11}\,M_\odot$} & $10^{10}$--$10^{11}\,M_\odot$ & $10^{9}$--$10^{10}\,M_\odot$ \\
Step & $N_{\rm cl}$ & $f_{500}$ & $N_{\rm TDE}$ ($10^6$) & $\dot N_{\rm peak}$ ($10^{-3}\,{\rm yr^{-1}}$) & $N_{\rm TDE}$ ($10^6$) & $N_{\rm TDE}$ ($10^6$) \\
\hline
0 Conservative                & 336  & 0.70 & 0.80 & 1.0  & 0.43 & 0.065 \\
1 + same $z$/mass cuts        & 150  & 0.67 & 0.34 & 0.77 & 0.16 & 0.027 \\
2 + sim.\ positions/velocities & 160 & 0.67 & 0.39 & 1.1  & 0.19 & 0.043 \\
3 + sim.\ sizes               & 147  & 0.91 & 1.06 & 3.5  & 0.54 & 0.087 \\
4 + sim.\ masses              & 151  & 0.90 & 1.73 & 10.5 & 0.80 & 0.046 \\
5 + sim.\ numbers (all sim.)  & 2179 & 0.88 & 19.0 & 47   & 8.35 & 1.41  \\
6 Maximal                     & 2206 & 0.88 & 18.3 & 44   & 9.00 & 1.48  \\
\hline
Maximal, empirical sizes      & 2167 & 0.69 & 8.65 & 14   & 3.78 & 0.64  \\
\hline
\end{tabular}
\end{table*}

The maximal model produces 21--23 times as many TDEs per galaxy as the
conservative model in all three mass bins, and its peak rate is 25--43 times
higher. 
The gap can be attributed mostly to the number of clusters (step~5) and the cluster sizes (step~3). With
the same cuts applied to both, the simulation forms about fifteen times as
many clusters per galaxy as the seed-halo normalization of
Equation~\eqref{eq:ncl} predicts (e.g.\ $\approx2200$ rather than
$\approx150$ for $M_\star>10^{11}\,M_\odot$).
The number of TDEs rises by a
similar factor, 10--11 in the two upper mass bins. 
Since the TDE count
scales almost linearly with the number of clusters, the normalization of the
cluster number--halo mass relation at high redshift is the main uncertainty
in our predicted rates.

The cluster sizes are the other major factor (step~3). The median
half-mass radius in the simulation is $\approx0.15$~pc, about three times
smaller than the $\approx0.5$~pc of Equation~\eqref{eq:mk12}. The relaxation
time scales as $r_{\rm h}^{3/2}$, so it is about six times shorter in the
simulated clusters. As a result, more clusters form an IMBH of at least
$500\,M_\odot$ (90 rather than 70 per cent), the median mass of these IMBHs
triples (from $\approx1.7\times10^{3}$ to
$\approx5.4\times10^{3}\,M_\odot$), and the number of TDEs per galaxy rises
by a factor of 2--3. 
Giving the maximal model's clusters the
empirical radii lowers $f_{500}$ back to 0.70 and the number of TDEs by a
factor of 2.1--2.4, moves the peak back to $z\approx8$ and raises the
late-time fraction back to $\approx45$ per cent. The size effect is thus
nearly the same whether it is applied to the conservative or the maximal
population, so the ingredients act close to independently and their effects
roughly multiply.

The other ingredients change the TDE count by less than a factor of
$\approx2$. The cuts on halo mass and formation redshift (step~1) remove
clusters that would form in low-mass hosts or after $z=7$. This more than
halves the number of clusters and TDEs, and mostly removes late TDEs: the
fraction after $z=7$ falls from 0.7 to 0.5. Taking the initial positions and
velocities from the simulation (step~2) changes the number of TDEs by a
factor of 1.1--1.6, within the halo-to-halo spread. Taking the cluster
masses from the simulation (step~4) raises the number of TDEs by a factor of
1.5--1.6 in the two upper mass bins. The median cluster mass barely changes
($\approx2.4\times10^{5}\,M_\odot$ in both), so this increase must come from
differences in the shape of the mass distribution rather than its typical
value. In the lowest mass bin, steps~1--4 involve only $\approx10$--15
clusters per galaxy, so the factor for any single step is dominated by
sampling noise (for example, the apparent drop at step~4). Steps~4 and~5
together, however, give a factor of 15--18 in all three bins.

Finally, steps~5 and~6 agree to within 4--8 per cent in the number of TDEs
and to within 7--13 per cent in the peak rate. This is comparable to the
seed-to-seed scatter of the maximal model, and it confirms that the
comparison measures the effect of the ingredients rather than of the
sampling procedure. In summary, the factor of $\sim20$ between our two models
comes mostly from the number of clusters formed per unit halo mass at high
redshift (a factor of $\sim10$) and their compactness (a factor of
2--3). The halo-mass and redshift cuts partly offset these increases (a
factor of $\approx0.4$), and the other ingredients matter only at the level
of tens of per cent.

\section{Impact of random draw on TDE rates}
\label{app:randomdraw}

In order to investigate the impact of scatter between model realizations, we run five additional maximal model integrations for each of the 15 halos, varying the random seed during the drawing of star clusters. 
\autoref{tab:seed_clusters} shows the results of this investigation for each mass bin. 
The seed-to-seed scatter in all three mass bins is consistent with
Poisson fluctuations, $\sim N_{\rm cl}^{-1/2}$.
The 16\% in the lowest bin comes from 1166615, which has only about 76 clusters (60–90 across seeds). 
For all the other 14 halos, the dominant contribution to the overall scatter is the halo-to-halo scatter, rather than random fluctuations between seeds. 

\begin{figure}
    \centering
    \includegraphics[width=0.95\linewidth]{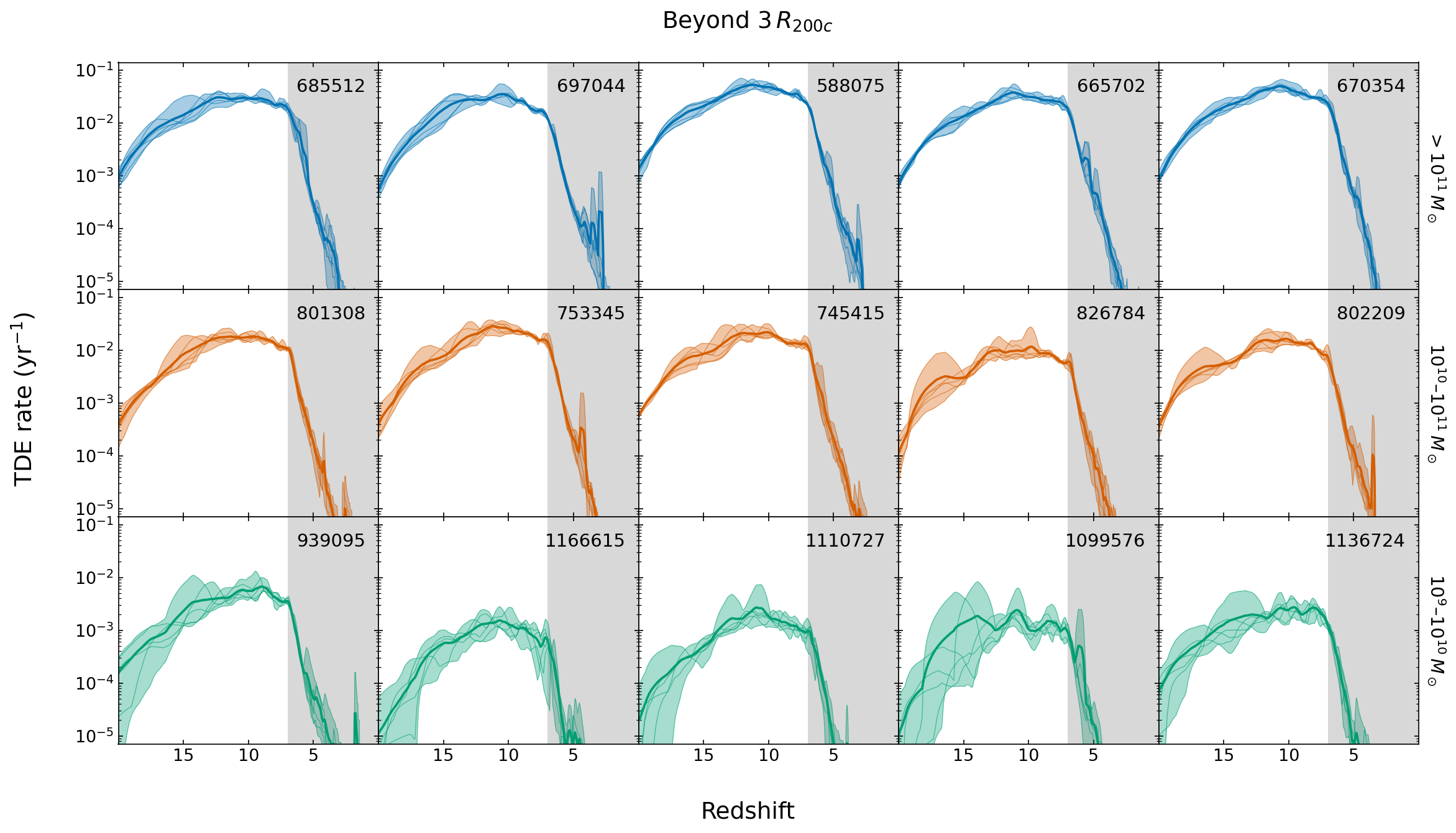}
    \caption{Smoothed TDE rate per galaxy versus redshift depicting six independent realizations for each halo in our sample. All runs are the maximal model, and only the TDEs that occur beyond $3R_{200c}$ are shown (see \autoref{fig:seeds_within}). For each galaxy, the solid line shows the mean over the realizations, the shaded band shows the spread of the realizations, and the thin lines show each realization's mean rate. }
    \label{fig:seeds_beyond}
\end{figure}

\begin{figure}
    \centering
    \includegraphics[width=0.95\linewidth]{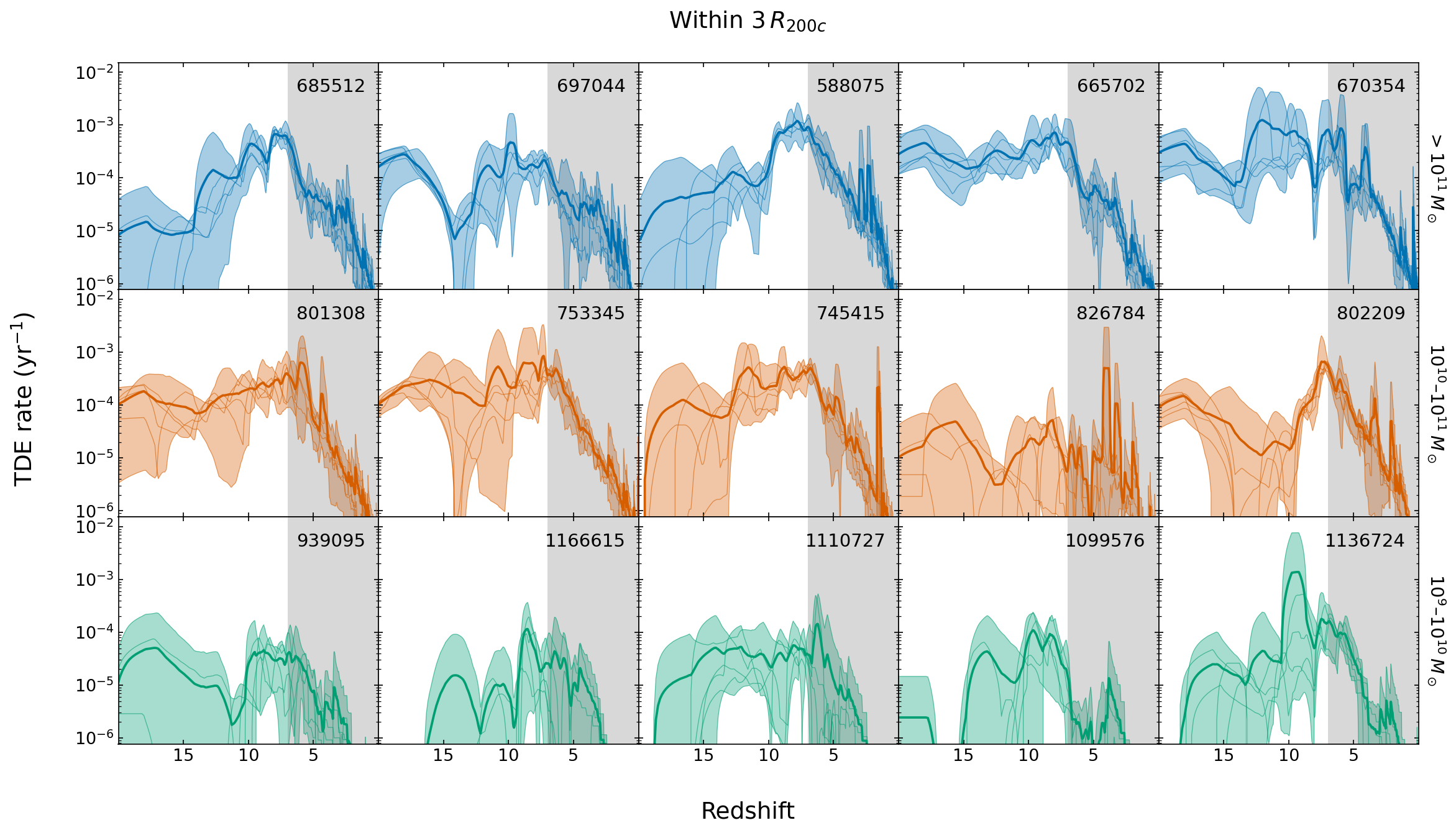}
    \caption{Same as \autoref{fig:seeds_beyond}, but for TDEs that occur within $3R_{200c}$.}
    \label{fig:seeds_within}
\end{figure}

In \autoref{fig:seeds_beyond}, we show the smoothed TDE rate per galaxy versus redshift for every halo, including only the TDEs that occur outside of $3R_{200c}$. 
Since the majority of TDEs occur beyond this radius, the rates reflect the overall robustness of the analysis to random scatter (contributing a few percent in each mass bin, similar to the overall counts). 
In \autoref{fig:seeds_within}, we show the same plot of TDE rate versus redshift for TDEs occurring within $3R_{200c}$. 
Here, the rates vary by $15-20\%$ in the upper mass bins. 
This increased scatter is due to the fact that only a handful of clusters reach the main progenitor while the IMBH-driven burst of TDEs is still occurring in each realization. 
Thus, the large scatter comes from which clusters are drawn and where their orbits take them in a given realization. 
In the lowest mass bin, few clusters are drawn, so the small-number statistics  to affect scatter in the rate of TDE events at $>20\%$ .

\begin{table}
\centering
\caption{Number of clusters seeded per galaxy in the maximal model, from six
independent realizations (the original run and five additional random
seeds). $\langle N_{\rm cl}\rangle$ is the mean over the five galaxies in each
$z=0$ stellar-mass bin and over the realizations. The halo range spans the
realization-averaged values of the individual galaxies. The seed-to-seed
scatter is the standard deviation of $N_{\rm cl}$ across realizations divided
by its mean, computed separately for each galaxy; the column gives the range
of this quantity over the galaxies in the bin. }
\label{tab:seed_clusters}
\begin{tabular}{lccc}
\hline
$M_\star(z=0)$ & $\langle N_{\rm cl}\rangle$ & Halo range & Seed-to-seed scatter \\
\hline
$>10^{11}\,M_\odot$           & 2180 & 1703--2818 & 0.9--2.8\% \\
$10^{10}$--$10^{11}\,M_\odot$ & 1059 & 620--1473  & 1.5--2.8\% \\
$10^{9}$--$10^{10}\,M_\odot$  & 165  & 76--368    & 3.5--16\%  \\
\hline
\end{tabular}
\end{table}

\end{CJK*}
\end{document}